\documentclass{aa}  

\usepackage[]{hyperref}
\usepackage{xcolor}
\usepackage{graphicx}
\usepackage{txfonts}
\usepackage{lipsum}
\usepackage{subcaption}         % necessary for continued figures, example in section 3
\usepackage{lscape}             % to rotate a single page table, example in appendix.
\usepackage{placeins}           % useful with \FloatBarrier, to keep 

\newcommand{\dn}{$\rm D4000_n$}
\newcommand{\hb}{H$\beta$}
\newcommand{\hd}{H$\delta_A$}
\newcommand{\hg}{H$\gamma_A$}
\newcommand{\hdg}{H$\delta_A$+H$\gamma_A$}
\newcommand{\mgfep}{$\rm [MgFe]^\prime$}

\newcommand{\mgtwofe}{$\rm [Mg_2Fe]$}
\newcommand{\mgfef}{$\rm [MgFe50]^\prime$}
\newcommand{\mgfeft}{$\rm [MgFe52]^\prime$}
\newcommand{\micron}{$\mu m$}

\newcommand{\silver}{{\it silver}}
\newcommand{\golden}{{\it golden}}

\newcommand{\mli}[1]{\mathit{#1}}

\begin{document}
\title{LEGA-C stellar population scaling relations}
\subtitle{I Chemo-archaeological downsizing trends at $z\sim0.7$}

\author{Anna R. Gallazzi\inst{1}\thanks{anna.gallazzi@inaf.it}
\and Stefano Zibetti\inst{1}
\and Arjen van der Wel\inst{2}
\and Angelos Nersesian\inst{2,3}
\and Yasha Kaushal\inst{4}
\and Rachel Bezanson\inst{4}
\and Francesco D'Eugenio\inst{5,6}
\and Eric F. Bell\inst{7}
\and Joel Leja\inst{8,9,10}
\and Laura Scholz-Diaz\inst{1}
\and Po-Feng Wu\inst{11,12,13}
\and Camilla Pacifici\inst{14}
\and Michael Maseda\inst{15}
\and Daniele Mattolini\inst{1,16}}
\institute{INAF-Osservatorio Astrofisico di Arcetri, Largo Enrico Fermi 5, 50126, Firenze, Italy
\and Sterrenkundig Observatorium Universiteit Gent, Krijgslaan 281 S9, B-9000 Gent, Belgium
\and STAR Institute, Université de Liège, Quartier Agora, Allée du six Aout 19c, B-4000 Liege, Belgium
\and Department of Physics and Astronomy and PITT PACC, University of Pittsburgh, Pittsburgh, PA 15260, USA
\and Kavli Institute for Cosmology, University of Cambridge, Madingley Road, Cambridge CB3 0HA, UK 
\and Cavendish Laboratory – Astrophysics Group, University of Cambridge, 19 JJ Thomson Avenue, Cambridge CB3 0HE, UK 
\and Department of Astronomy, University of Michigan, 1085 South University Avenue, Ann Arbor, MI 48109, USA
\and Department of Astronomy and Astrophysics, 525 Davey Lab, The Pennsylvania State University, University Park, PA 16802, USA
\and Institute for Gravitation and the Cosmos, The Pennsylvania State University, University Park, PA 16802, USA
\and Institute for Computational and Data Sciences, The Pennsylvania State University, University Park, PA 16802, USA
\and Graduate Institute of Astrophysics, National Taiwan University, Taipei, 10617, Taiwan
\and Department of Physics and Center for Theoretical Physics, National Taiwan University, Taipei, 10617, Taiwan
\and Physics Division, National Center for Theoretical Sciences, Taipei, 10617, Taiwan
\and Space Telescope Science Institute, 3700 San Martin Drive, Baltimore, MD 21218, USA
\and Department of Astronomy, University of Wisconsin-Madison, 475 N. Charter St., Madison, WI, 53706, USA
\and Dipartimento di Fisica, Universit\`a di Trento, Via Sommarive 14, I-38123 Povo (TN), Italy}

\abstract{
We analyse the stellar population properties of a well-defined sample of 552 galaxies at redshift $0.6<z<0.77$ drawn from the LEGA-C spectroscopic survey. This paper is the first of a series, and it is aimed at i) presenting the catalog of revised absorption indices for LEGA-C DR3 and of the inferred physical parameter estimates, describing their systematic uncertainties, and at ii) deriving benchmark scaling relations for the general massive galaxy population at intermediate redshift. We estimate light-weighted mean ages and stellar metallicities by a careful analysis of key absorption features in the galaxies’ stellar continuum spectra coupled with photometry. The observables are interpreted in a Bayesian framework with a comprehensive library of model spectra based on stochastic star formation histories, metallicity histories and dust attenuations. We discuss various sources of systematic uncertainties within our method, as well as systematic differences with results from other spectral fitting approaches.
We derive volume-weighted scaling relations connecting light-weighted mean ages and stellar metallicities with galaxy stellar mass for the general galaxy population at $\left<z\right>=0.7$~and masses $>10^{10}M_\odot$. We find the downsizing trends observed in the local Universe to be already in place 6 Gyr ago. We also observe a bimodal distribution of light-weighted ages as a function of mass, transitioning around $10^{11}M_\odot$. Such a bimodality is not observed in the stellar metallicity-mass relation, which changes from a steep to a flat regime across $M_\ast\sim10^{10.8}M_\odot$. Similar trends in age and metallicity also emerge as a function of velocity dispersion, but with a sharper transition from young to old around $\log\sigma_\ast=2.3$. Differences with respect to the trends as a function of stellar mass suggest that age is primarily dependent on velocity dispersion below and above the transition regime, while both the stellar mass and the depth of the total gravitational potential well (as traced by the velocity dispersion) contribute to stellar metallicity.
The catalogs of revised absorption index measurements for LEGA-C DR3 used in this work and of the inferred stellar population physical parameters are released to public repositories.
}
   \keywords{galaxies: evolution -- galaxies: fundamental parameters -- galaxies: high-redshift
               }

   \maketitle

\section{Introduction} \label{sec:intro}

Galaxy evolution is governed by a complex and non-linear interplay between assembly processes, regulated by the matter distribution in the Universe, and baryonic processes, regulated by star formation and feedback and the matter cycle within galaxies \citep{SomervilleDave15}. Despite the wealth and complexity of these processes and their interaction, present-day galaxies are observed to follow several scaling relations that connect the physical properties of their stellar and gaseous components to global properties such as their total mass, their size and structure \citep[e.g.][]{kauffmann03,shen03, Tremonti04,bernardi05,gallazzi05,cappellari06,Saintonge11,mcdermid15,Catinella18}.
In particular, among these, the relations between the age and chemical composition of stellar populations in galaxies are fossil records of the past star formation and chemical enrichment histories of galaxies that reach different final masses. It has long been established that the stellar populations in present-day passive galaxies are older, metal-richer and more $\alpha$-enhanced in more massive systems than in their less massive bretheren \citep[e.g.][]{trager2000,Thomas05,gallazzi06,Renzini06,graves09}. These trends are referred to as an archaeological manifestation of star-formation downsizing \citep{fontanot09,Cowie96}, such that star formation and metal enrichment have been more efficient in the past for more massive galaxies.
These relations extend to star-forming galaxies \citep{gallazzi05,panter08,Zahid17,Trussler20}, which generally show similar properties to passive galaxies in the high-mass regime, but increasingly deviate from the passive trends toward lower masses. The transition from the passive to the star-forming regime in the scaling relations is suggested to be associated to a mass scale where the gas accretion mode and the dominant quenching mechanism change \citep[e.g.][]{DekelBirnboim06}. 
While stellar population properties are observed to scale to first order with galaxy mass (with a debate whether it is the stellar or the dynamical mass the more relevant quantity), 
other parameters are observed to contribute to the scatter in the scaling relations, such as the current star formation activity \citep[e.g.][]{Looser24}, size and structure \citep{cappellari06,mcdermid15}, the galaxy hierarchy \citep[e.g.][]{pasquali10,Trussler20,Gallazzi21}, host halo mass and environment \citep{baldry06,Thomas10,pasquali09,ScholzDiaz22,oyarzun22}.

Scaling relations are important benchmarks for hydrodynamical simulations in cosmological context, semi-analytic models as well as empirical models. All these models and approaches adopt different implementations of astrophysical processes (star formation efficiency, the production and cycle of metals, stellar and SuperNova feedback, growth and evolution of super-massive Black Holes and Active Galactic Nuclei feedback, gas cycling within the interstellar medium and to/from the inter-galactic medium). These processes regulate the growth of galaxies and their chemical enrichment, and the epoch and timescale of quenching, shaping the resulting distribution of galaxy populations in physical properties, hence their scaling relations. 

The interpretation of the observed scaling relations in the local Universe is limited by the fact that these archaeological tracers tell us when the stellar populations were formed but not directly when they assembled, hence they are less sensitive to galaxy assembly histories. Moreover, scaling relations are not evolutionary tracks but represent a snapshot in time of the population evolution. For these reasons, the archaeological approach and the direct look-back approach, tracing the cosmic evolution of star formation activity and galaxy number densities \citep{MD14}, need to complement each other by tracing the evolution of scaling relations \citep[e.g.][]{Lilly13,Chen20}.
The way forward is to apply the archaeological approach at different cosmic epochs to constrain galaxy star formation and assembly histories and connect progenitors and descendants in a statistical sense. 

Because of the time variation of spectral diagnostics, moving to higher redshifts (hence to younger cosmic ages), presents the advantage of allowing us to better resolve the early phases of galaxy formation, which are very challanging to disentangle in the old populations of present-day galaxies. Moreover, we would catch galaxies at epochs that are progressively closer to their peak activity and subsequent quenching as depicted by the cosmic star formation rate density \citep{MD14}. However, studies of the stellar component in higher-redshift galaxies have lagged behind studies of the ionized gas component \citep[e.g.][]{Erb06,maiolino08,moustakas11,zahid13}, because they require deep observations to reach a sufficient S/N in the stellar continuum sampling the rest-frame optical (thus moving into the observed NIR). It is important to fill this gap and complement information on the metallicity of the ISM with that on the chemical composition of stellar populations, as it can give constraints to the mechanisms that regulate the baryon and metal cycle in galaxies \citep[e.g][]{Garcia24,Stanton24}.

The potential of applying the archaeological approach at earlier cosmic times has been shown by a number of pioneering studies at intermediate and high redshift. Until about a decade ago, studies of stellar ages and chemical composition from the rest-frame optical were limited to a few tens of galaxies or coadded spectra up to $z\sim0.8$, mostly targeting red quiescent galaxies in clusters \citep{Jorgensen05,SB09,jorgensen13} or in the field \citep{Schiavon06,ferreras09,choi14,Leetho19}, and a mass-selected sample of $\sim70$~galaxies including both quiescent and star-forming in our previous work \cite{gallazzi14} (hereafter G14).

At higher redshift $z>1-3$, the stellar metallicities of limited samples of the brightest objects or of co-added spectra were studied from rest-frame UV absorption features \citep{Rix04, Halliday08,Sommariva12,cullen19}, sampling the younger stellar component. To sample the bulk mass in the rest-frame optical, NIR spectroscopic data have been obtained for few galaxies with deep medium-resolution from the ground \citep{Toft12,Stockman20,Onodera15,Lonoce15,Kriek19,Saracco20,Carnall22} and from space \citep{Beverage25,Carnall24}, or for relatively larger samples with low-resolution slitless spectroscopy \citep{Ferreras19,EC19}. 

Currently, the highest redshift available for the necessary deep stellar continuum spectroscopy for large and representative samples of galaxies is $0.6<z<1$, with the $\sim200$ night Large Early Galaxy Astrophysics Census (LEGA-C) VIMOS/VLT survey \citep{vdWel16,straatman18,DR3} which provides spectra of the required stellar continuum quality and for near-mass-selected samples at intermediate redshifts ($0.6<z<1$). The deep 20-hr observations of LEGA-C spectra allow robust measurement and fitting of rest-frame optical absorption features, enabling the analysis of ages, metallicities and star formation histories \citep{Chauke18,Chauke19, Wu18a,Wu18b,DEugenio20,Barone22,Woodrum22,Cappellari23, Beverage23,kaushal24,Nersesian25}, as well as stellar kinematics \citep{Bezanson18,deGraaff21,DEugenio23b,DEugenio23a} and dynamical modeling \citep{vanHoudt21,vdW22}.

In this work we leverage the high spectral quality, large statistics and well-defined target selection of LEGA-C to characterize the volume-weighted distributions of light-weighted mean ages and stellar metallicities as a function of stellar mass and velocity dispersion for massive galaxies at $0.6<z<0.77$. We rely on optimally-selected key stellar absorption features, sensitive to age and total metallicity, in combination to rest-frame optical photometry. These diagnostics are interpreted with our Bayesian Stellar population Analysis ({\tt BaStA}) framework, by comparison with a Monte Carlo library of model spectra encompassing a range of star formation histories (SFH), chemical enrichment histories (CEH) and dust attenuations \citep{gallazzi05,Zibetti17,Zibetti22}. 
This approach is complementary to full spectral fitting methods in that it focuses on selected, well-understood spectral features that are chiefly sensitive to age and metallicity \citep{worthey94,wo97,Vazdekis99,thomas04}. Moreover, thanks to the direct and full sampling of the prior as determined by the precomputed model library, the resulting posterior provides uncertainties that fully capture the underlying degeneracies.

Our sample of $\sim$550 galaxies, carefully selected to have reliable measurements of stellar absorption features, used as constraints to age and metallicity, is representative of the whole galaxy population down to $10^{10.4}M_\odot$ and of the star-forming population down to $10^{10}M_\odot$. With this we explore the stellar population properties of massive galaxies at $\left<z\right>=0.7$, and how their scaling as a function of stellar mass or stellar velocity dispersion differs. 

In the second paper of this series \citep[][hereafter Paper II]{Gallazzi25_paperII} we further explore how the $z=0.7$ global age and stellar metallicity scaling relations differentiate between quiescent and star-forming galaxies, and to what extent the stellar metallicity, averaged over the SFH, is sensitive to ongoing star formation activity. We then combine our LEGA-C results with a consistent analysis, in terms of observational constraints and modeling assumptions, of a volume- and completeness-corrected sample of galaxies from SDSS DR7 presented in \cite{Mattolini25}, to quantify the evolution of the scaling relations between $z=0.7$~and $z=0.1$.

As a product of the work presented here, we provide two catalogs with revised absorption index measurements for the whole LEGA-C DR3, one with measurements derived from individual spectra and one with measurements obtained by combining duplicate observations when available. We also provide the catalog of inferred stellar population physical properties for the full LEGA-C DR3 sample. The catalog includes stellar population properties (mean ages, stellar metallicities, stellar mass and dust attenuation) derived in this work with {\tt BaStA} based on duplicate-combined index measurements for the whole DR3, as well as those derived with independent spectral fitting analysis performed within the LEGA-C team, specifically from spectrum+photometry fits with {\tt Prospector} presented in \cite{Nersesian25} and with {\tt BAGPIPES} presented in \cite{kaushal24}.

The paper is organized as follows. In Sec.~\ref{sec:data} we describe the LEGA-C spectroscopic data used in this work, the measurement of stellar absorption features and the galaxy sample analysed, along with sample statistical corrections. In Sec.~\ref{sec:stelpop_method} we introduce our {\tt BaStA} stellar population fitting approach, building on \cite{gallazzi05} and \cite{Zibetti17}, and we describe the adopted model library together with the spectroscopic and photometric constraints used. The resulting stellar population physical parameter estimates and uncertainties are presented in Sec.~\ref{sec:stelpop_results}. Our main results on the stellar population scaling relations in LEGA-C are presented in Sec.~\ref{sec:scaling_legac} where we characterize the volume-weighted relations for the global massive galaxy population connecting light-weighted ages and stellar metallicity to stellar mass and velocity dispersion. We discuss the observed downsizing trends in Sec.~\ref{sec:discussion} and summarize our results in Sec.~\ref{sec:summary}.
Readers interested in more technical details can find them in the Appendices on: the revised catalog of absorption indices (Appendix~\ref{Appendix:index_measures}) and the treatment of LEGA-C duplicate observations (Appendix~\ref{Appendix:duplicates}); an assessment of the impact of dust modeling on the age and metallicity inference (Appendix~\ref{sec:appendix_model_prior}); an assessment of the impact of modeling assumptions within our method in comparison to our previous work (Appendix~\ref{sec:appendix_basta}), as well as a comparison with parameters estimated from {\tt Prospector} and {\tt BAGPIPES} that were presented in \cite{Nersesian25} and \cite{kaushal24}, respectively, and from {\tt pPXF} in \cite{Cappellari23} (Appendix E).
Throughout the paper we assume a $\Lambda CDM$ cosmology with $H_0=70 \rm km/s/Mpc$, $\Omega_M=0.3$, $\Omega_\Lambda=0.7$, and a solar stellar metallicity of $Z_\odot=0.0154$ and we adopt a \cite{chabrier03} IMF.

\section{Data and sample}\label{sec:data}

\subsection{The LEGA-C data}
The LEGA-C survey has obtained deep spectra for a total sample of 4081 galaxies, 3029 of which are primary targets selected from the UltraVISTA catalog \citep{Muzzin13a} with redshift $0.6<z<1$~and $K_S$ magnitude brighter than $K_{S,lim}=20.7-7.5 \log((1+z)/1.8)$. Observations were conducted with the VIMOS spectrograph on VLT, integrating $\sim20$~hr on source,  with an effective spectral resolution of $R\sim3500$ over the observed wavelength range $6300<\lambda<8800$\AA. Details on the survey goals and design, and the observational set-up are extensively presented in \cite{vdWel16} and \cite{straatman18}.

In this work we use the 1-d spectra from the LEGA-C third data release (DR3). Details on the data reduction, spectra extraction and spectral measurements in DR3 can be found in \cite{DR3}. The LEGA-C $8"\times1"$~slits typically encompass a representative fraction of the galaxy light. The 1-d spectra are obtained from optimal extraction using the HST-based Sersic light profile. In this work we are particularly interested in the metal-sensitive features around rest-frame 5000\AA, which for the LEGA-C galaxies fall in the $>7000\AA$~observer-frame range affected by telluric absorption. The revision of the telluric absorption correction devised for DR3 is thus of particular importance for the accuracy of our measurements. Spectrophotometric calibration has been obtained comparing the uncalibrated LEGA-C spectra with best-fitting templates from {\it BVrizYJ} SED fits. 
\subsection{Measurement of stellar absorption features}
We measure stellar absorption features using the index definitions of the Lick system \citep{worthey94}, in addition to the 4000\AA-break \dn~\citep{balogh99} and the high-order Balmer lines \citep{wo97}. Note that indices are not translated to the Lick system, but are measured on the rest-frame observed spectra at their native resolution, thus including instrumental resolution and broadening due to the intrinsic velocity dispersion of the galaxy. The strength of stellar absorption indices is measured off of the rest-frame emission-line-subtracted spectra. As described in \cite{DR3}, the decoupling of the stellar continuum from the emission lines is performed with a coordinated fit with {\tt\string pPXF} \citep[which provides a bestfit continuum consisting of a combination of MILES stellar spectra][and 15th order multiplicative Legendre polynomial]{miles11} and {\tt\string platefit} (which determines emission line fluxes and EW from the continuum subtracted spectrum). 
Because in the index measurement pixels are assigned equal weight regardless of their noise, to avoid being strongly biased by high-noise wavelength elements, we flag pixels that deviate by more than $2\sigma$~(and with a minimum of 10\%) from the bestfit {\tt\string pPXF} continuum. This model only serves to identify highly deviant pixels but does not enter in the index measurement. The flagged pixels are linearly interpolated over. They are in total about 7\% of all pixels used in the index measurement. Details on this procedure are described in \cite{DR3}. We consider the index measurement reliable only if less than 1/3 of the pixels in any of the central or pseudo-continuum side bands are flagged.
After the DR3 was released we discovered that a fraction of $\sim15$\% of LEGA-C spectra did not receive a measurement of the absorption indices, despite the good quality of the spectrum and the lack of artifacts. This was tracked down to be related to a silent bug in the run of {\tt\string pPXF} associated with the rebinning of the spectra. We thus re-processed the emission-line subtracted spectra for cleaning and index measurement. We then measure absorption indices with our routine {\tt\string BaStA\_index}.  

Formal errors are estimated in first place via standard error propagation based on the noise spectrum. In a second step, based on repeated LEGA-C observations we estimate scaling factors to apply to these formal errors and obtain the actual errors adopted in the stellar population analysis. This factor is estimated to be 1.3 for all indices, except for \dn~(2) and \hb~(1.8) owing to larger uncertainties associated to spectro-photometric calibration and emission line infill, respectively \citep{DR3}.
We release the revised catalog of index measurements for the whole DR3 (see Appendix~\ref{Appendix:index_measures}), both for individual spectra and combining duplicate observations when available (see Sec.~\ref{Sec:sample_definition}).

\subsection{Spectroscopic requirements and samples definition}\label{Sec:sample_definition}
The sample used in this work to derive scaling relations comprises galaxies belonging to the primary sample (${\tt primary} = 1$), with no sign of AGN contamination in the LEGA-C spectra (${\tt flag\_spec} = 0$), with a measurement of stellar velocity dispersion from the {\tt\string pPXF} fit (${\tt sigma\_star} > 0$) and spectroscopic redshift $0.55 < {\tt z\_spec} < 1.1$. This represents a parent sample of 2864 spectra. 

Some of the objects in the sample have duplicate observations. These are useful not only to assess more reliably the uncertainties on spectral measurements that are not captured by the sole noise spectrum but also could be used to increase the wavelength coverage for a given galaxy. Because of the different position of the slit on the mask in different observing runs, the actual observed wavelength range can vary between different observations of the same object. We take advantage of this to maximize the number of absorption features covered for each galaxy. For each object in the parent sample we check whether there are more than one usable observation (typically two or three). If so, for each absorption index we compute the error-weighted mean of the index measurements available for that objects. Objects with duplicate observations are thus included only once in our final, parent sample, using the duplicate-combined index measurements. After accounting for duplicate observations, the parent sample consists of 2588 unique galaxies. In Appendix~\ref{Appendix:duplicates} we assess the reliability and improved quality of the parameter estimates based on combined duplicate observations with respect to individual ones.

As described in Sec.~\ref{sec:stelpop_method_data}, for our stellar population analysis we further require the measurement of at least one of the Balmer absorption indices (\hb, \hd, \hg) and at least one of the composite metal-sensitive indices (among \mgfep, \mgtwofe, \mgfeft, \mgfef). After combining the measurements for duplicate observations, this requirement reduces the sample to 575 unique galaxies. The large reduction of the sample from 2588 to 575 unique galaxies mainly results from the wavelength coverage of the LEGA-C spectra which limits the observability of the Mg, Fe5270 and Fe5335 features up to $z\sim0.77$, and only a minor fraction of galaxies are excluded because of large spectroscopic uncertainties preventing an index measurement (as explained in the previous section). Considering only the sets of indices that we deem reliable for stellar population fitting (see Sec.~\ref{sec:stelpop_method}), our analysis is finally reduced to a {\it silver} sample of 552 galaxies and a {\it golden} subsample of 323 galaxies with spectral $S/N>20/\rm pixel$\footnote{As a reference S/N we take the LEGA-C DR3 catalog entry ${\tt SN\_RF\_4000}$ which is the S/N per pixel at 4000\AA~rest-frame. In the case of duplicate observations we take the squared sum of the individual S/N values.}.
The number of galaxies in each subsample are summarized in Table~\ref{Tab:mean_errors}. We notice that if we did not combine duplicate observations, but simply took the observation with the highest S/N for each object with repeated observations, the \silver/\golden~samples would reduce to 491/236 unique galaxies. 

\begin{figure}
\includegraphics[width=0.5\textwidth]{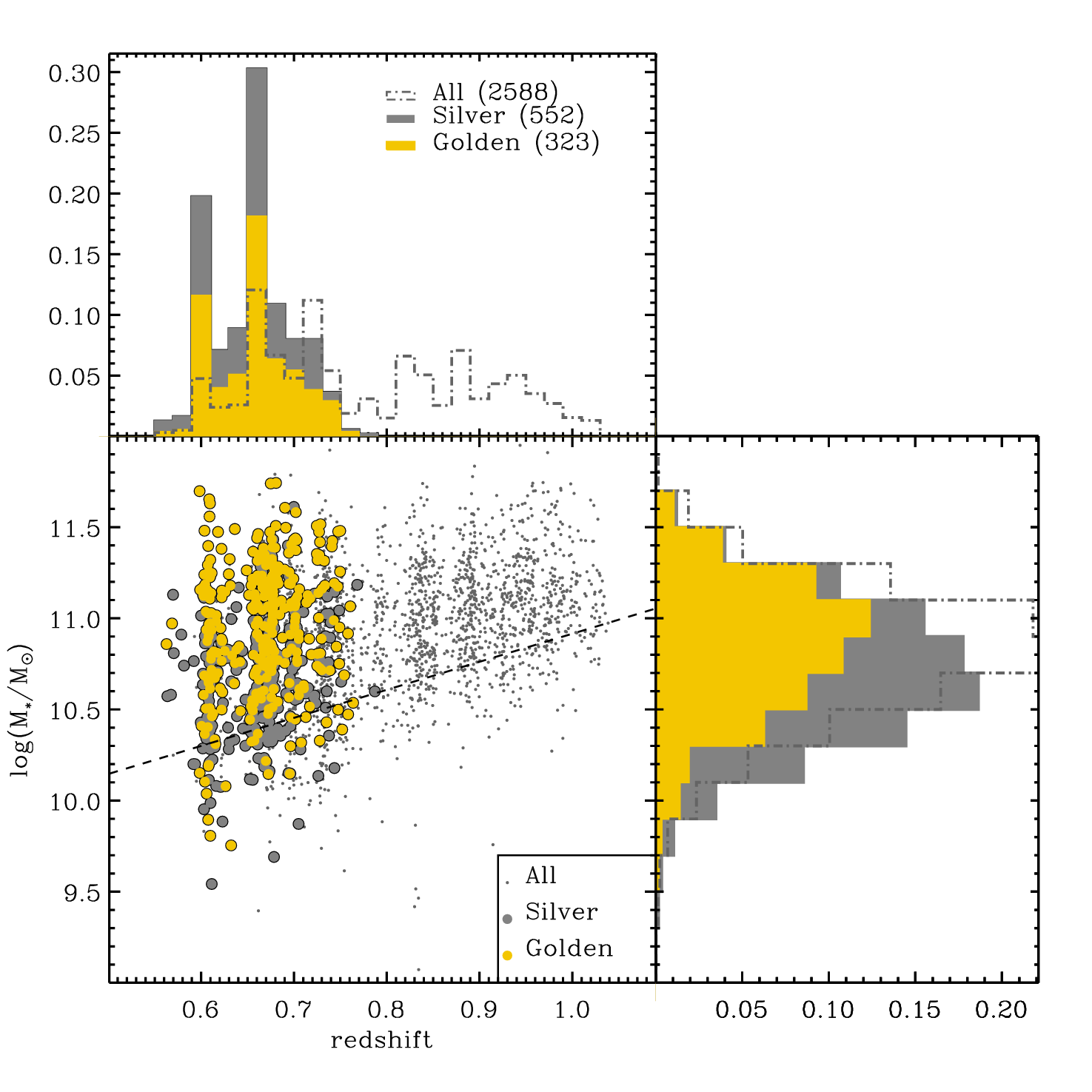}
\caption{Distribution in stellar mass and redshift for LEGA-C primary sample (grey dots and dot-dashed histograms, normalized to unit integral) and for the subsamples used in the
analysis: \silver~sample (grey symbols and histograms); \golden~sample (golden symbols and histograms). The histograms for the \silver~and \golden~samples are normalized by the number of galaxies in the \silver~sample. The number of galaxies in each subsample is given in parenthesis. Redshifts come from the LEGA-C DR3 catalog, while stellar masses are those computed in this work.}\label{fig:mass_redshift}
\end{figure}

\begin{figure}
\includegraphics[width=0.5\textwidth]{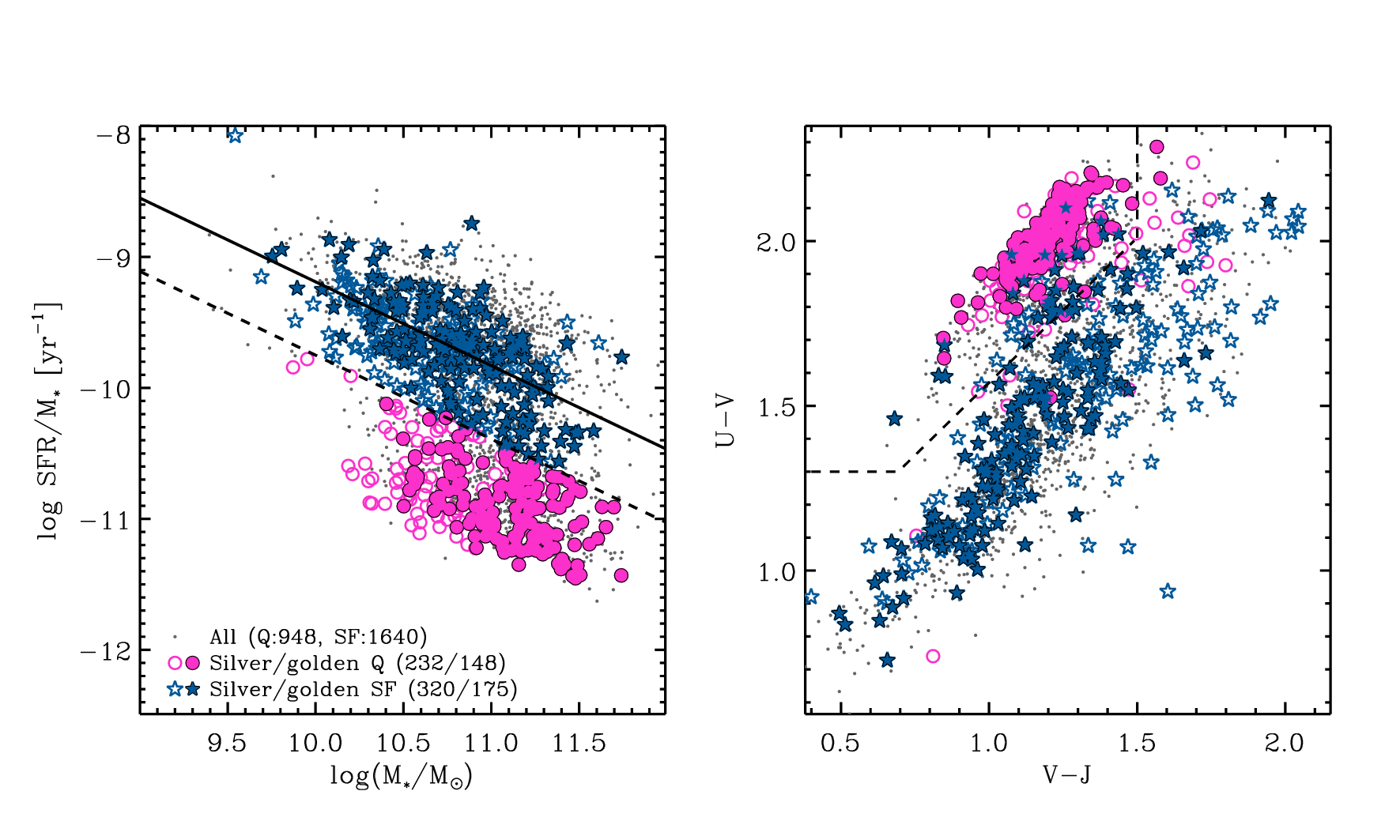}
\caption{{\it Left}:
Distribution in specific SFR versus stellar mass. {\it Right}: Distribution in U-V, V-J rest-frame colors. We separate galaxies into quiescent (Q, magenta circles) and star-forming (SF, blue stars) based on their location with respect to the Star Forming Main Sequence (solid line in the left panel): Q galaxies are those deviating by more than $2\sigma$ below the MS (i.e. they lie below the dashed line in the left panel). The right-hand panel illustrates how this selection compares with the UVJ quiescent zone outlined by the dashed segment \citep{Muzzin13b}. Empty/filled symbols indicate galaxies in the \silver/\golden~samples. The numbers in the left panel give the number of SF and Q galaxies in the \golden, \silver~and parent LEGA-C samples. All quantities in this plot come from the LEGA-C DR3 catalog, except stellar masses which are those computed in this work.\label{fig:UVJ_SSFR}}
\end{figure}

Our final sample spans a range in physical properties, in particular mass and star formation activity. Figure~\ref{fig:mass_redshift} shows the distribution in stellar mass versus redshift for the whole LEGA-C sample and for the sample analysed here. As already mentioned, the redshift limit $z\lesssim0.77$~is imposed by the wavelength coverage. Stellar mass extends over the range $\rm 10^{10}M_\odot\lesssim M_\ast \lesssim 10^{11.5}M_\odot$. The requirement of $S/N>20$ preferentially selects against lower mass galaxies ($M_\ast<10^{11}M_\odot$). 
Figure~\ref{fig:UVJ_SSFR} illustrates how our sample is representative of galaxies with different levels of current star formation, by showing the distribution in specific star formation rate versus stellar mass ($\rm SSFR-M_\ast$) and in $U-V$ versus $V-J$ (UVJ) diagrams. We define as quiescent (Q) those galaxies that lie more the $2\sigma$~ below the Main Sequence of star formation (i.e. below the dashed line in the left-hand panel)\footnote{To fit the star forming Main Sequence we use galaxies that would be classified as star-forming according to their location in the UVJ diagram (rightward of the dashed segment in the right panel).} and as star-forming (SF) galaxies consistent within $2\sigma$~with the MS (i.e. above the dashed line). Since our sample is selected in K band, the requirement of high S/N in the optical restframe (roughly corresponding to the R and I bands) disfavours the tail of the galaxy distribution with the reddest optical-NIR colours. As we can see from Fig.~\ref{fig:UVJ_SSFR}, this tail is largely composed of heavily dust-attenuated star-forming galaxies.

In this paper, we separate Q and SF galaxies in Sec.\ref{sec:stelpop_method} and~\ref{sec:stelpop_results} in order to illustrate the quality of spectroscopic measurements for the two classes of objects, the different accuracy of parameter estimates and the different systematic uncertainties that pertain to Q and SF galaxy spectra. This is useful for anyone wishing to use our catalog for different analysis. For convenience, we opt to use the definition of Q/SF that we adopt in Paper II where we analyse the scaling relations separately for the two classes of galaxies and their evolution with respect to local galaxies classified with an equivalent criterion.

\begin{figure*}
\includegraphics[width=\textwidth]{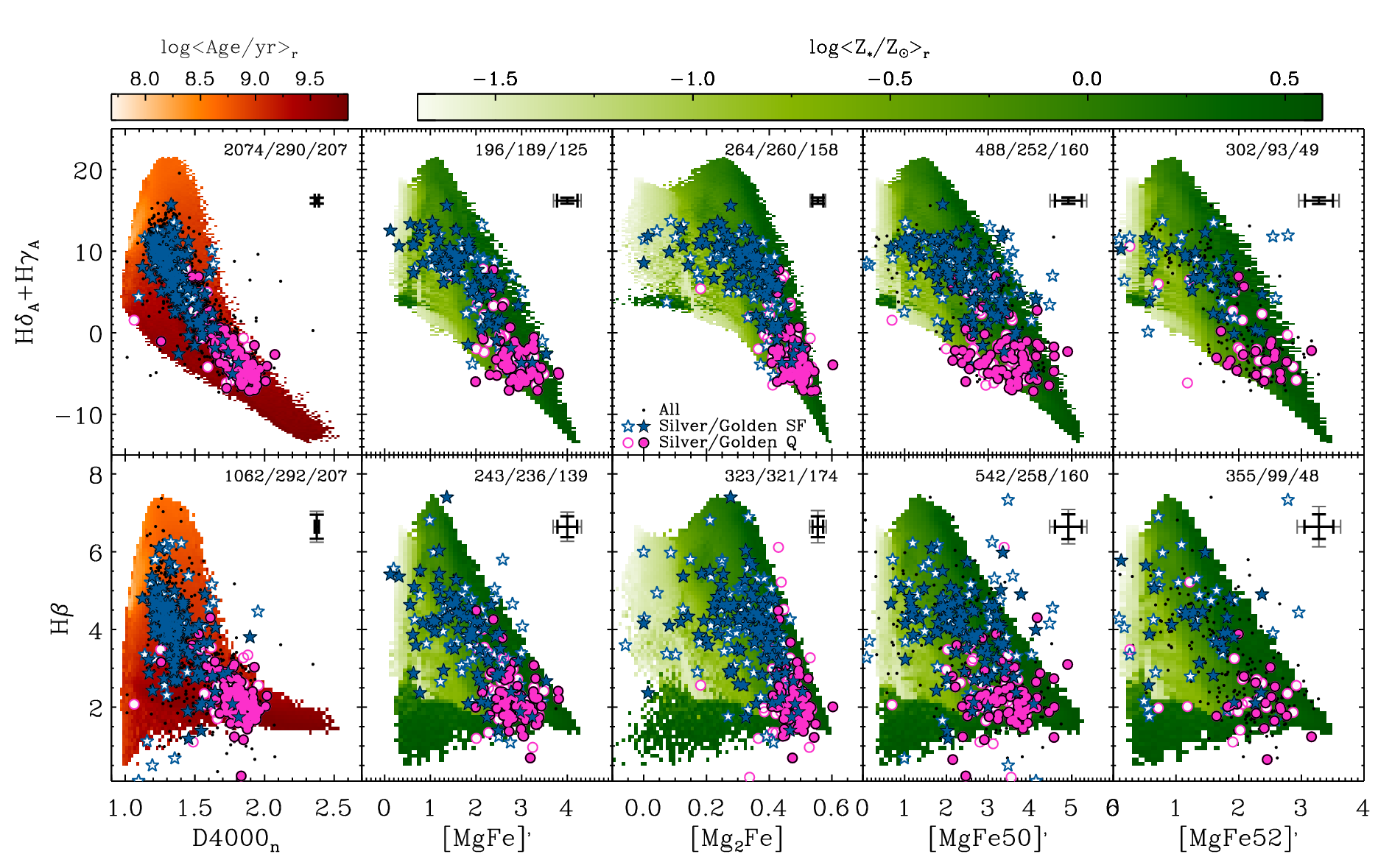}
\caption{Distribution in absorption indices diagnostic diagrams, showing the combinations of indices considered in this work for their different sensitivity to age and metallicity. The colored region shows the distribution of the adopted model library of complex star formation and metal enrichment histories (see Sec.~\ref{sec:stelpop_method_library}), color-coded for r-band light-weighted age (left-most panels) and for r-band light-weighted stellar metallicity. Symbols are as in Fig.\ref{fig:UVJ_SSFR}: galaxies are distinguished into quiescent (magenta) and star-forming (blue) based on a mass-dependent threshold in specific SFR; empty symbols indicate galaxies belonging to the {\it silver} sample, i.e. with optimal combinations of indices, while filled symbols refer to the {\it golden} subsample (with $\rm S/N>20$); dots display all the galaxies in the LEGA-C sample with measurements of the two indices in each panel. In each panel only \silver/\golden~galaxies for which the corresponding couple of indices has been {\it used} in the fit are shown (therefore all the \silver/\golden~ galaxies shown lie within $3\sigma$ of the model region). In each panel we report the number of galaxies shown among those in the full LEGA-C/\silver/\golden~samples. The error bars display the median uncertainty on each index for the \golden~(thick black) and the \silver~galaxies (thin grey).}\label{fig:idx_plane_5panel}
\label{default}
\end{figure*}

\subsection{Sample statistical corrections}\label{sec:stats_corr}
Our sample is drawn from the $K_s$-band selected LEGA-C parent sample according to the availability of key spectral absorption features and to their spectral S/N. Therefore some statistical corrections are needed to compute volume-limited quantities accounting for incompleteness and selection bias. First, we consider i) the volume correction factor {\tt Vcor} which accounts for the fact that the LEGA-C sample is magnitude-limited, ii) the completeness correction factor {\tt Scor} which accounts for mass-dependent incompleteness in the observed sample with respect to the parent UltraVISTA sample \citep[see][]{DR3}. The product of the two corrections is the {\tt Tcor} parameter in the DR3 catalog. The full LEGA-C sample is representative of the whole galaxy population at $\left<z\right>=0.7$~down to $\log M_\ast/M_\odot=10.4$. The completeness limit for blue galaxies only extends to masses 0.2~dex smaller.

We then quantify and correct for biases introduced by our sample selection for studying age and metallicity scaling relations. In particular, we wish to account for incompleteness as a function of stellar population properties at a given stellar mass. This is quantified by the fraction of galaxies that enter our sample definitions (specifically the \silver~and the \golden~samples which we use for the statistical analysis) with respect to the parent sample that would be available to us. As parent sample we consider the main LEGA-C sample ({\tt PRIMARY$=$1, FLAG\_SPEC$=$0, SIGMA\_STARS\_PRIME $> $0}) restricted to the redshift range over which at least one of the metal-sensitive composite indices would be available (i.e. the redshift out to which Fe5335 would be observable, $z < 0.79$). We opt to measure incompleteness in a space of the observables that trace stellar population properties independently from our modeling, namely we consider the rest-frame $U-V$ color as proxy for stellar populations physical parameters and the absolute g-band luminosity $L_g$ as proxy for galaxy stellar mass. For each galaxy in the \silver~(\golden) sample we consider the 20 nearest neighbors in the $\rm (U-V)-L_g$ plane in the parent sample. The inverse of the fraction of \silver~(\golden) galaxies among these 20 is then taken as the statistical weight {\tt w\_spec\_silver} ({\tt w\_spec\_golden}) to apply to correct for incompleteness. The weights {\tt w\_spec\_silver} range from 1 to 5 quite uniformly over the color-luminosity plane, with the tendency for bluer galaxies at fixed luminosity to have slightly larger weights. For the golden sample {\tt w\_spec\_golden} range from 1 to 10 with a stronger gradient with stellar mass.
Because of the larger variations and uncertainties in {\tt w\_spec\_golden} due to smaller number statistics in the low-mass/low-luminosity regime, in the analysis presented in Sec.~\ref{sec:scaling_legac} we will rely on the weighted \silver~sample as our reference for population statistical trends, while we will use the unweighted \golden~sample as a control with higher-quality measurements.

\section{Stellar population modelling}\label{sec:stelpop_method}
In order to estimate stellar population physical parameters, such as mean weighted ages and metallicities, in addition to stellar mass, we use our code BaStA \citep{Zibetti22} which adopts and builds upon the bayesian approach described in \cite{gallazzi05} and \cite{gallazzi14}. The code compares a selected set of spectral diagnostics and photometric data to those predicted by a large library of model spectra constructed by convolving SSP models with randomly generated SFHs, metallicity histories and dust attenuation parameters. The posterior probability distribution function (PDF) of selected physical parameters, marginalized over all the other parameters, is computed by weighing each model by the likelihood of the data for that model. We then take the median of the PDFs and half of the $84^{th}-16^{th}$ percentile range as fiducial value of the parameter and its associated uncertainty, respectively. We stress that these uncertainties include not only the error associated to observational uncertainties, but also degeneracies within the assumed prior distribution of SFHs, metallicity and dust.

\subsection{The library of spectral models}\label{sec:stelpop_method_library}
The model library used here is constructed following \cite{Zibetti17} and \cite{Zibetti22}, and comprises 500000 models obtained by convolving Simple Stellar Population models with Monte Carlo SFHs. In this work we adopt the SSPs of the 2019 version of \cite{BC03} population synthesis code (CB19) which are based on the MILES stellar library \citep{miles11} and on the Parsec evolutionary tracks \citep{parsec15}. We adopt the \cite{chabrier03} IMF. For a more detailed description of the ingredients adopted in CB19 we refer to Appendix A of \cite{sanchez22}.
Following \cite{plat19} and \cite{gutkin16} for the solar scale adopted in the Parsec evolutionary tracks, we normalize the stellar metallicities to the present-day photospheric solar metallicity of $\rm Z_\odot=0.01524$.\footnote{Note that this is lower than the most recent estimate of $\rm Z_\odot=0.0225$ \citep{Magg22}.}

In our model library, the SFHs are described as the superposition of a smooth secular component and random bursts. The secular component is given in the form of a delayed gaussian \citep[][see also \cite{Gavazzi02a}]{Sandage1986}, which allows for both a rising and a declining phase. On top of this continuum SFH, bursts of star formation can occur in random number (up to six), age and intensity.

A major change over our previous works and others using precomputed model libraries in a bayesian approach \citep[e.g.][]{gallazzi05,Gallazzi21,daCunha08} is the introduction of a Chemical Enrichment History (CEH) through a parametrization for the metallicity evolution along the SFH, rather than assuming a constant metallicity value. The chosen parametrization is reminiscent of a leaking box model \citep{Erb08}, having as parameters the initial and the final metallicity and a parameter describing the swiftness of enrichment. The metallicity increases with the stellar mass formed following this parametrization. Bursts are allowed to have a metallicity drawn from a gaussian centered on the value of the secular metallicity at the same epoch, thus introducing some scatter in the metal enrichment history.
Finally, the emergent spectra are attenuated according to a \cite{CF00} dust attenuation model separating the contribution from the birth clouds to that of the diffuse interstellar medium. A summary of the adopted priors is provided in Table~\ref{tab:SFHinpars} and a full description is provided in Appendix A of \cite{Mattolini25}. Notice that for each galaxy we then only consider models that have a formation age $t_{form}$~ younger than the Universe age at the galaxy's redshift (with an allowance of 0.2 dex).

Beside the parameters that generate the SFH, CEH and dust attenuation (as summarized in Table~\ref{tab:SFHinpars}), for each model spectrum we compute a number of derived parameters which describe in a non-parametric way the resulting properties averaged over the SFH, namely: i) the r-band-weighted average age and stellar metallicity\footnote{See Equations 4 and 6 in \cite{Zibetti17} for the exact definition of light-weighted mean age. The analogous definition applies to the light-weighted mean metallicity.}, $\log\left<Age/yr\right>_r$ and $\log \left<Z_\ast/Z_\odot\right>_r$; ii) the mass-weighted average age and stellar metallicity,  $\log\left<Age/yr\right>_M$ and $\log \left<Z_\ast/Z_\odot\right>_M$; iii) the total present-day stellar mass, $\log(M_\ast/M_\odot)$; iv) the effective dust attenuation in the g-band, $A_g$. These are the parameters of which we are interested in deriving the posterior PDF for each galaxy, by marginalizing over all the other model parameters.

\begin{table*}[]
    \caption{Summary table of the parameters used to generate the composite stellar populations library.}
    \label{tab:SFHinpars}
    \centering
    \begin{tabular}{p{0.15\textwidth}|
    p{0.35\textwidth}|
    p{0.40\textwidth}}
    
         {{\bf \small Parameters}} & {{\bf \small Description}} & {{\bf \small Prior PDF}}  \\
         \hline \hline
                   \noalign{\vspace{3pt}} 
        \multicolumn{3}{c}{{\it Secular SFH:} $\mli{SFR}_\tau(t)\propto\frac{(t-t_{\mathrm{form}})}{\tau}\exp\left(-\frac{(t-t_{\mathrm{form}})^2}{\tau^2}\right)$}\\
         \hline\hline
          {$t_\mathrm{form}$} & Time elapsed since the beginning of the SFH & Log-uniform between $5\cdot10^8$ and $2\cdot10^{10}\,yr^{ (a)}$  \\
          \hline
          {$\tau$} & Time scale/peak & From log-uniform of $\tau/t_\mathrm{form}$ between $1/20$ and $2$  \\
          
          \hline\hline
          \noalign{\vspace{3pt}} 

         \multicolumn{3}{c}{{\it Secular Chemical Enrichment History:}}\\
         \multicolumn{3}{c}{$Z_*(t)=Z_*\left(M(t)\right)=Z_{*\,\text{final}}-\left(Z_{*\,\text{final}}-Z_{*\,\text{0}}\right)\left(1-\frac{M(t)}{M_\text{final}}\right)^\alpha$}\\
         \hline \hline
         $\left<Z_{*\,\rm{fm-w}}\right>$ & Formed-mass-weighted metallicity (non-generative, constraint for generative pars.)& $\tanh$ distribution of $P(\log \left<Z_{*\,\rm{fm-w}}\right>)$: $P(1/50\, \mathrm{Z_\odot})=0$, $P(0.1\, \mathrm{Z_\odot})=0.9$, $P(3.8\, \mathrm{Z_\odot})=1$, $\left<Z_{*\,\rm{fm-w}}\right> <3.8\, \mathrm{Z_\odot}$ \\
         \hline
         {$\alpha$} & Swiftness of chemical enrichment & $0.25 \leq \alpha \leq 19$,  from a uniform distribution in $\beta\equiv\frac{1}{1+\alpha},\, 0.05 \leq \beta \leq 0.80$  \\
         \hline
         $Z_{*\,\text{0}}$ & Initial metallicity & Log-uniform between $\log (Z_{\rm{min}}/Z_\odot) \equiv \log(1/50)$ and $\min(\log Z_{\rm{0\,max}}/Z_\odot \equiv \log(0.05), \left<Z_{*\,\rm{fm-w}}\right>)$ \\
         \hline
         $Z_{*\,\text{final}}$ & Final metallicity & Derived from  $Z_{\rm{min}}$, $\alpha$, and $\left<Z_{*\,\rm{fm-w}}\right>$, with the constraint $Z_{*\,\text{final}} \leq 4 Z_{*\,\odot}$ \\

         \hline \hline
          \noalign{\vspace{3pt}} 
         \multicolumn{3}{c}{{\it Bursts}}\\
         \hline \hline
         $N_\mathrm{burst}$ & Number of bursts & $1/3$ of models without bursts; for the remaining $2/3$: $N_\mathrm{burst}\leq 6$, $P(N_\mathrm{burst})\propto e^{-N_\mathrm{burst}}$\\
         \hline
         $t_{\mathrm{burst},i}$ & Age of the $i$th burst (instantaneous) & Log-uniform distribution between $10^5$\, yr and the $t_\mathrm{form}$ of the secular component\\
         \hline
         $Z_{*\,\mathrm{burst},i}$ & Metallicity of the $i$th burst & Gaussian distribution in log-$Z$, with mean equal to the metallicity value of the secular component at $t=t_{\mathrm{burst},i}$ and width $\sigma_{Z_{*}\,\mathrm{burst}} = 0.2$\,dex\\
         \hline
         $\mathcal{M}_{\mathrm{burst},i}$ & Mass fraction formed in the $i$th burst relative to the mass formed in secular mode & Log-uniform distribution between $10^{-3}$ and $2$. More restrictive upper limits applied for $t_{\mathrm{burst},i} < 10^8$\,yr \\
         \hline \hline
          \noalign{\vspace{3pt}} 
         \multicolumn{3}{c}{{\it Dust attenuation} \citep{CF00}}\\
         \hline
         $\tau_V$ & Total optical depth due to diffuse ISM and birth cloud (affects stars younger than $10^7$\,yr) & $P(\tau_V)\propto1-\tanh(1.5\tau_V-6.7)$, constant at low values, exponential drop to $0$ between $\tau_V=4$ and $\tau_V=6$ \\
         \hline
         $\mu$ & Fraction of total optical depth in diffuse ISM & $P(\mu)\propto1-\tanh(8\mu-6)$, exponential drop to $0$ between $\mu=0.5$ and $\mu=1$\\
         \hline \hline
    \end{tabular}
\tablefoot{$^{(a)}$In the fit, we restrict the range to the age of the Universe at the redshift of each galaxy, with an allowance of 0.2~dex.}
\end{table*}

\subsection{Spectroscopic and photometric constraints}\label{sec:stelpop_method_data}
The observational constraints that we adopt combine a selected set of stellar absorption features with photometric data. All the results shown in the following are obtained fitting both spectroscopic and photometric data simultaneously. As stated in Sec.~\ref{sec:data}, the observed absorption indices are measured at the original spectral resolution (including instrumental and velocity dispersion broadening). For each galaxy, these are compared to indices measured on the model spectra convolved at the same effective resolution and broadening of the galaxy. The absorption features are chosen among those showing distinct sensitivity to age and metallicity, while having at the same time a negligible dependence on element abundance ratios, since our models are based on scaled-solar SSPs and do not allow for variable abundance ratios for individual elements. Building on our previous work (G05, G14), the ideal set of indices combines age-sensitive features such as \dn, \hb, \hdg, with metal-sensitive features such as \mgtwofe~\citep{BC03}, \mgfep~\citep{Thomas03}. Because of the wavelength coverage of LEGA-C spectra, there are only 57 galaxies with this complete set of five indices. We consider other two combinations of Mg and bluer Fe features that show a negligible dependence on [$\alpha$/Fe], namely \mgfef~and \mgfeft~as defined in \cite{Kuntschner10}\footnote{For convenience we report the definition of the composite indices:\\
\mgfep \(= \sqrt{\mathrm{Mgb}\times (0.72\times \mathrm{Fe5270}+0.28\times \mathrm{Fe5335})}\)\\
\mgtwofe \(= 0.6\times \mathrm{Mg}_2+0.4\times \log(\mathrm{Fe4531}+\mathrm{Fe5015})\)\\
\mgfef \(= 0.5\times (0.69\times \mathrm{Mgb} + \mathrm{Fe5015})\)\\
\mgfeft \(= 0.5\times (0.64\times \mathrm{Mgb} + \mathrm{Fe5270})\).}. 

We thus consider the following sets of indices:
\begin{itemize}
\item[s1)] any index available among \dn, \hb, \hd, \hg~(or, preferred, \hdg~if both Balmer indices are available)
\end{itemize}
combined with the set s2a if possible or, alternatively, with set s2b, where: 
\begin{itemize}
\item[s2a)] \mgtwofe~in combination with \mgfep~or \mgfeft~or alone;
\item[s2b)] \mgfef~in combination with \mgfep~or \mgtwofe~or \mgfeft~or alone.
\end{itemize}

Our minimum requirement for reliable age and metallicity estimates is to have at least one of the Balmer absorption lines and at least one of the Mg-Fe composite indices listed above. Galaxies meeting this requirement and that can be fit by one of the index sets above constitute our \silver~sample of 552 galaxies.\footnote{Note that we run our fitting procedure on all the LEGA-C DR3 galaxies. We provide a catalog of derived stellar population parameters for the whole DR3, since this may be useful for some science applications that are not interested in stellar metallicity (that critically require the availability of Mg and Fe features).}

The index sets listed above were chosen after checking that stellar population parameters inferred from different indices do not show a bias, with respect to the optimal set of 5 indices (\dn, \hb, \hdg, \mgtwofe, \mgfep). For this test, we considered the 52 galaxies for which all the age and metal-sensitive indices are available, and we fit them with all the different index combinations allowed. We then compare the resulting ages and metallicities with those obtained with the reference fit based on the optimal index set. We find that all the chosen index combinations with which LEGA-C galaxies are fit do not show systematic bias (within few hundreths of dex) with respect to the reference fit, with a scatter comparable or smaller than the mean parameter uncertainties. 
More than 40\% of the \silver~sample is fit (in addition to s1) with both \mgfep~and \mgtwofe, about 30\% with \mgfef, 10\% with \mgfef~and \mgfeft, 7\% with \mgtwofe~and \mgfeft, the other index combinations apply to a few percent each.

In Fig.\ref{fig:idx_plane_5panel} we show the distribution of LEGA-C galaxies in the index-index diagnostic diagrams combining the Balmer absorption lines with \dn~and with the red composite Mg-Fe indices. In each panel, dots display all LEGA-C galaxies for which the two indices are measured, while empty/filled symbols indicate galaxies in the \silver/\golden~sample, distinguished into Q (magenta circles) and SF (blue stars).
The coverage of the model library is shown by the colored map, where the color reflects the mean light-weighted age (left-most panels of Balmer lines versus \dn) and the mean light-weighted stellar metallicity (right panels of Balmer lines versus Mg-Fe composite indices). This figure serves as an illustration of how the physical parameters map onto the observational diagnostics, but we remind that the fit is performed in the multi-dimensional space of indices and photometry. 
The figure shows that our model library provides an adequate coverage of the observational space of LEGA-C galaxies. In turn, we note that the LEGA-C sample analysed here spans a large range of physical parameters and underlying SFHs, spreading over a substantial fraction of the model parameter space. From Fig.~\ref{fig:idx_plane_5panel} we see that most of the observations lie within the parameter space defined by our models, with a few exceptions due to large observational errors. To control for possible outliers, in the fit we discard an absorption index if its measurement lies more than $3\sigma$ away from the models (where $\sigma$~is the observational error on the index). All the \silver/\golden~data points shown in Fig.\ref{fig:idx_plane_5panel} indicate galaxies whose measurement is within $3\sigma$~of the model grid.

\begin{figure}
\begin{center}
\includegraphics[width=0.23\textwidth]{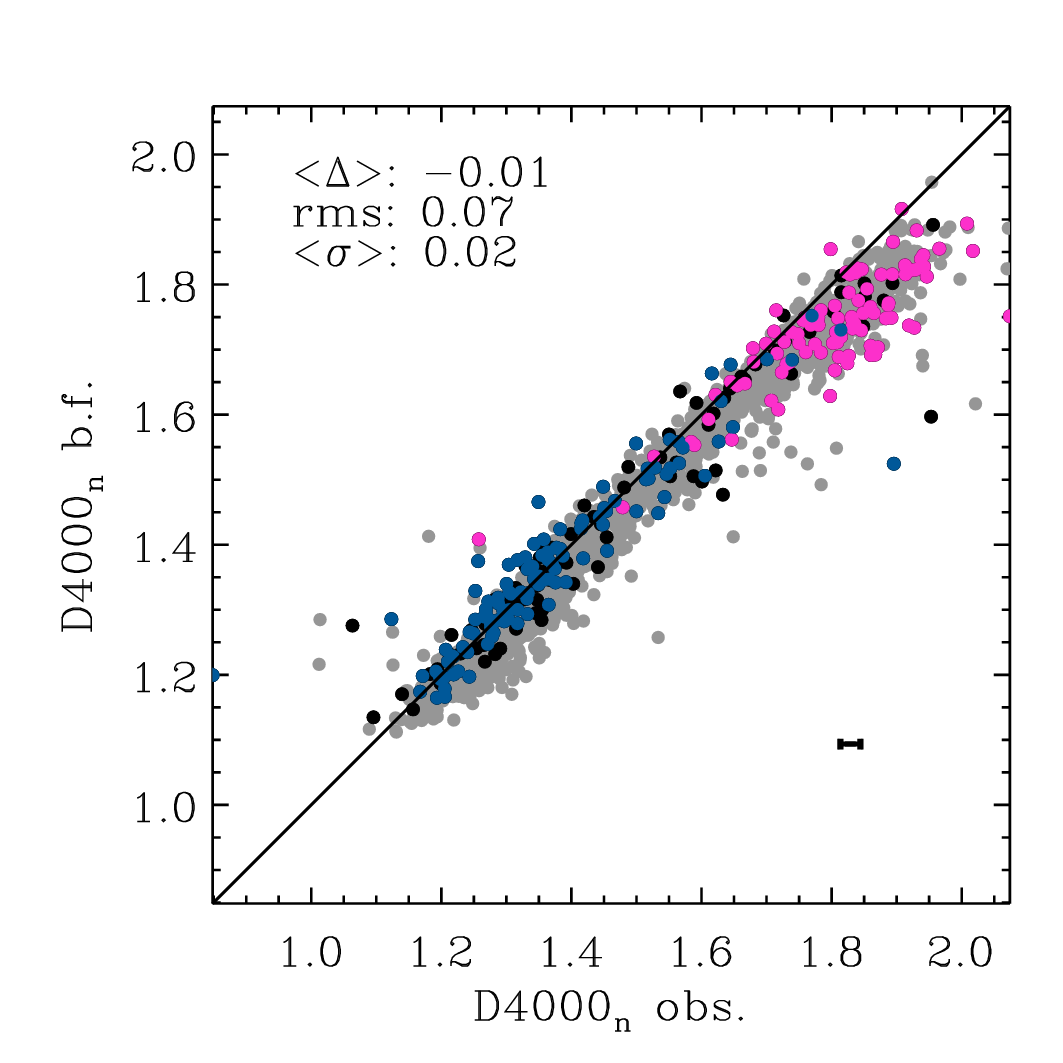}
\includegraphics[width=0.23\textwidth]{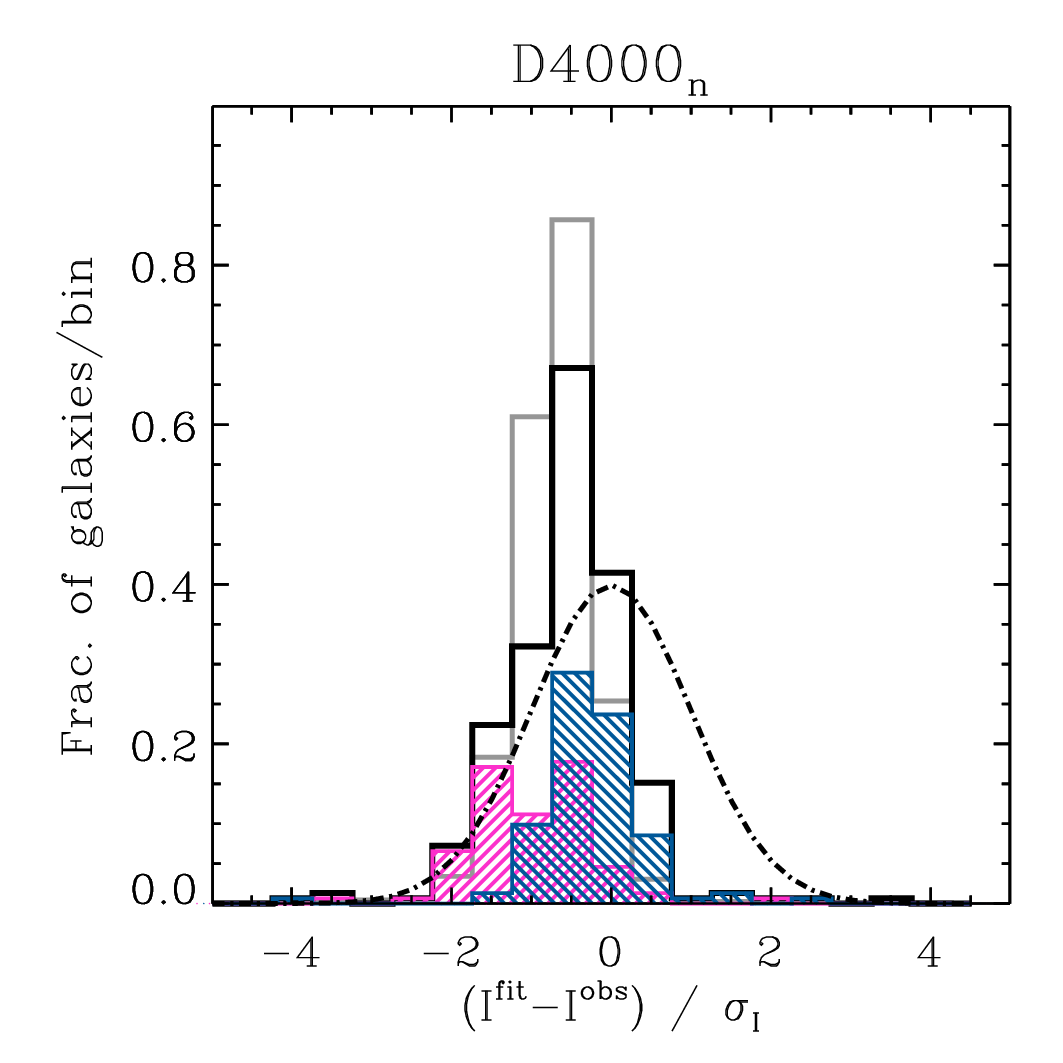}
\end{center}
\begin{center}    
\includegraphics[width=0.23\textwidth]{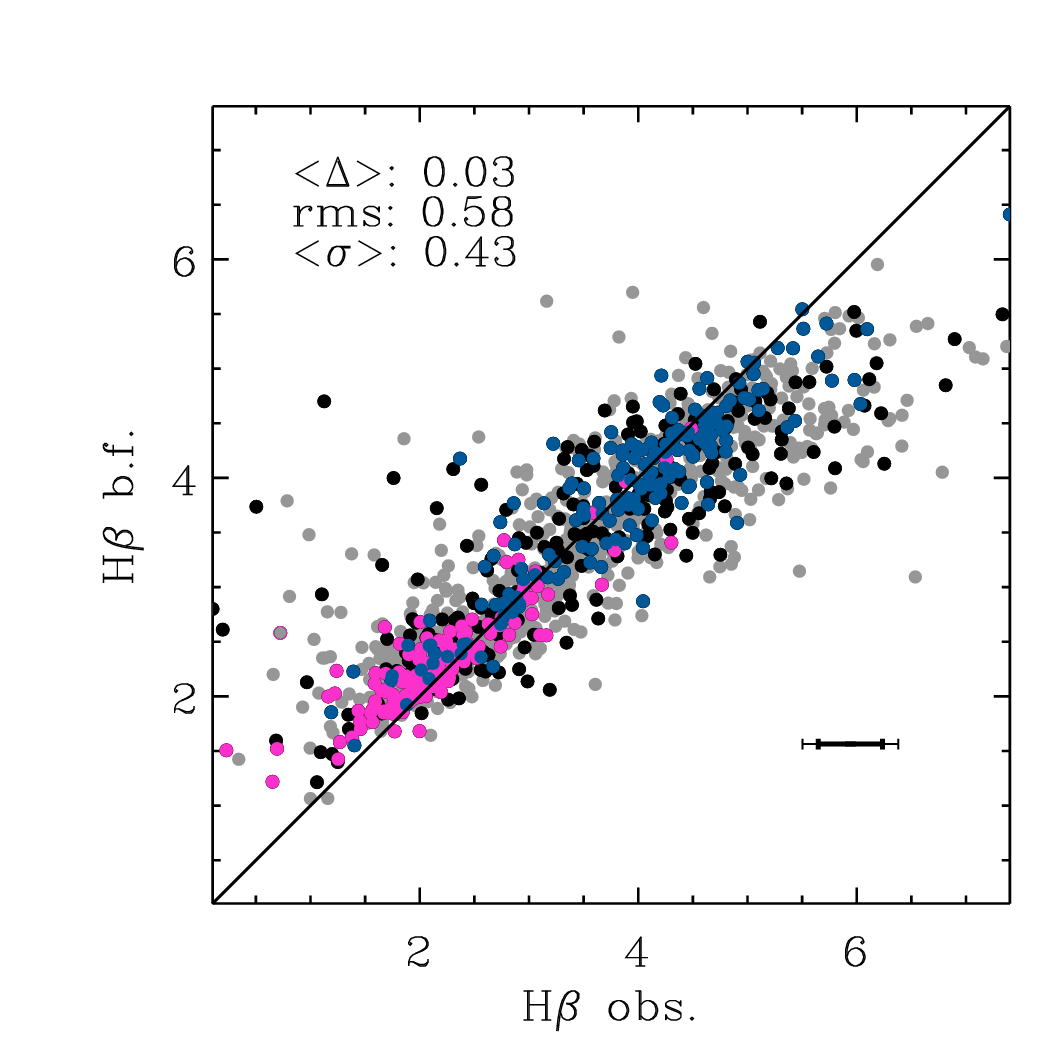}
\includegraphics[width=0.23\textwidth]{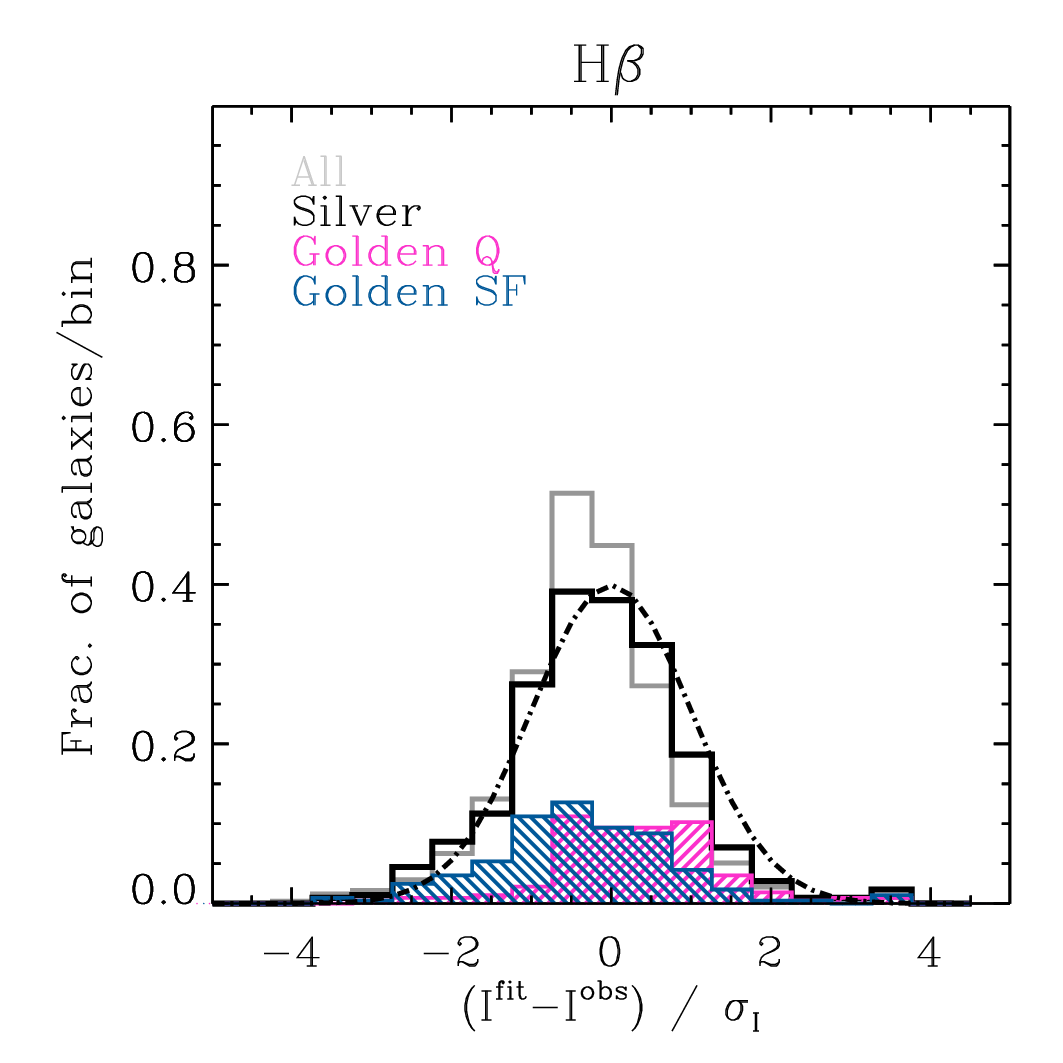}
\end{center}
\begin{center}
\includegraphics[width=0.23\textwidth]{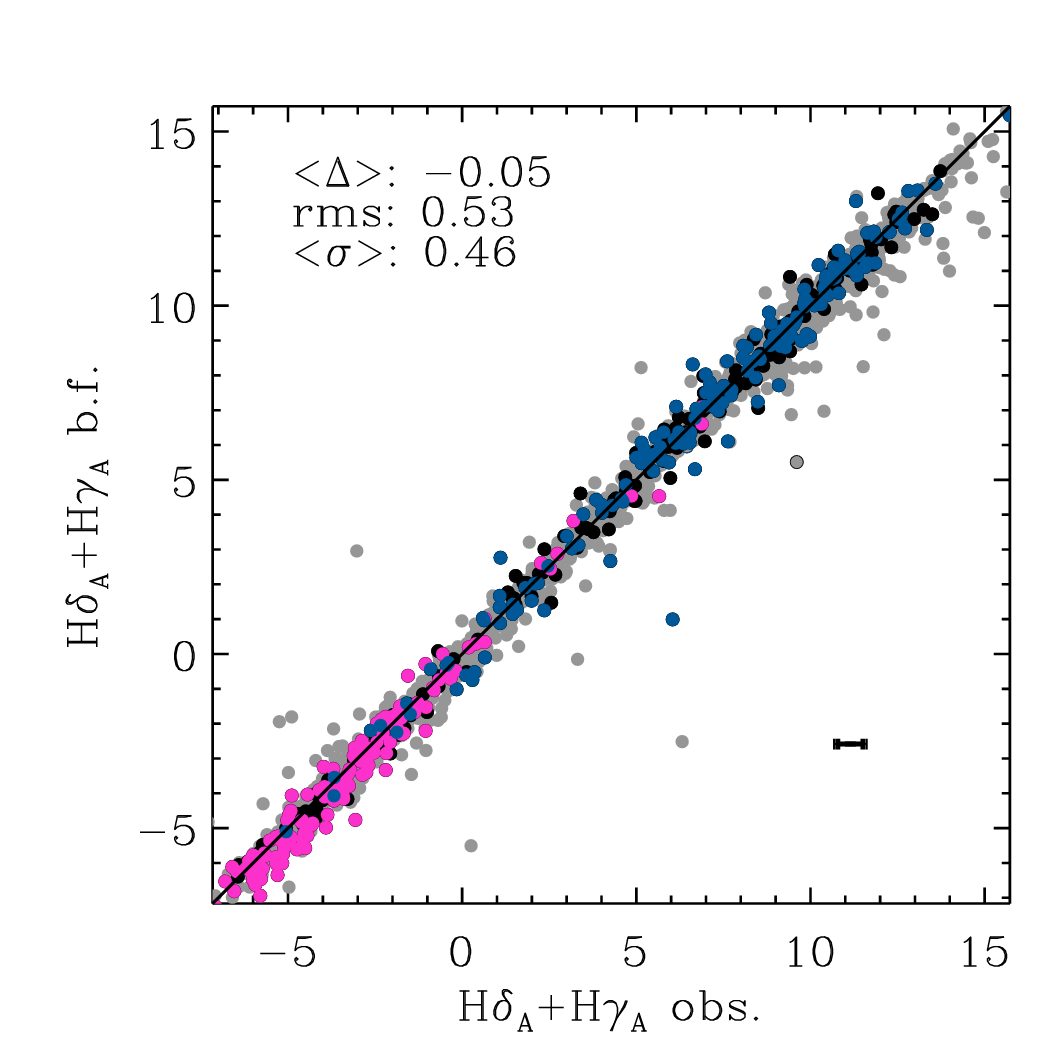}
\includegraphics[width=0.23\textwidth]{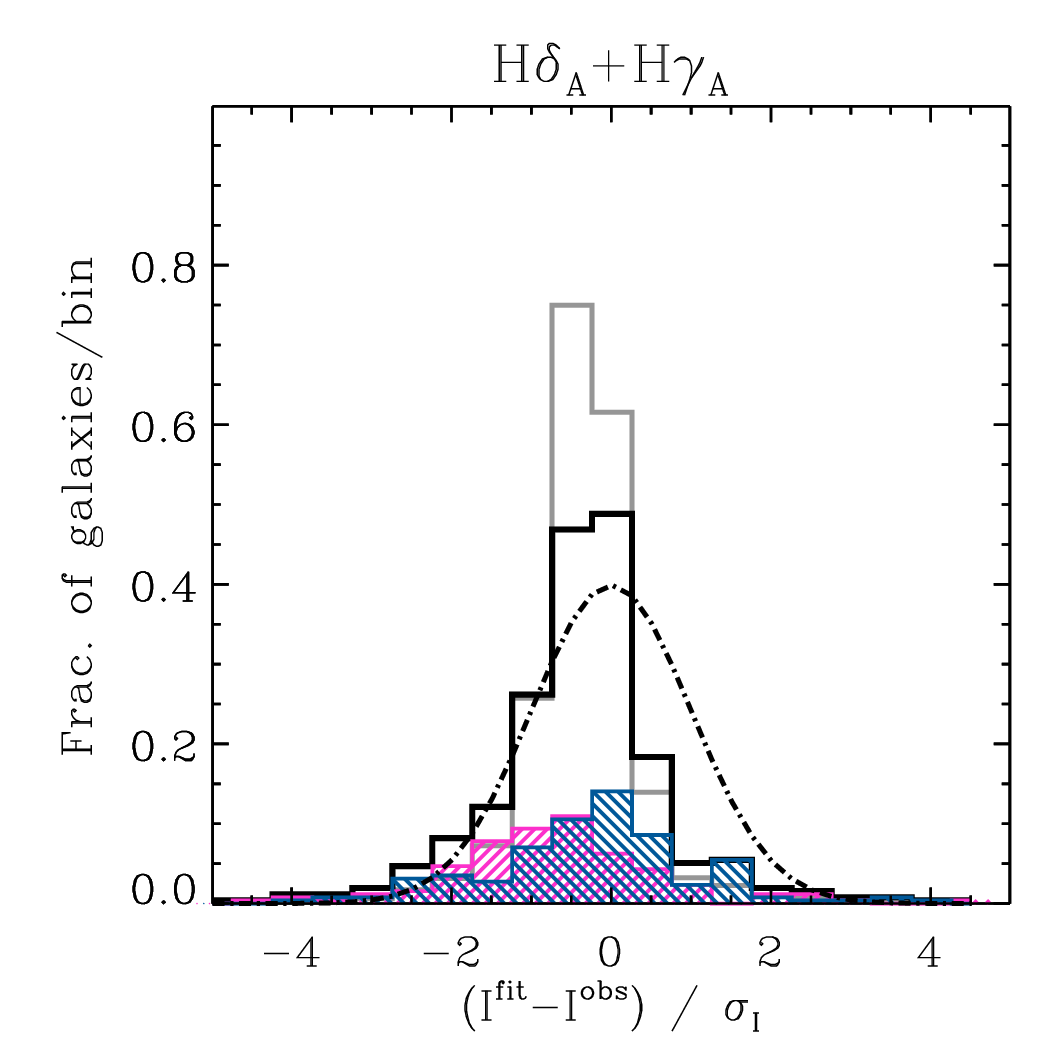}
\end{center}
\caption{Comparison between observed and best-fit predicted index strengths for the absorption features considered in the {\tt BaStA} fit as age-sensitive diagnostics. Grey filled circles show the full primary LEGA-C catalog for which the index is measured; black circles highlight galaxies in the \silver~sample; magenta/blue circles highlight quiescent/star-forming galaxies in the \golden~sample. Filled symbols indicate galaxies for which the index has been used in the fit. The thin/thick errorbars indicate the median observational error for \silver/\golden~galaxies. The median offset, rms scatter and the median uncertainty are reported in each panel for the \silver~sample. The right-hand panels show the distribution of the difference between the absorption index strength predicted by the best-fit model and the observed one, normalized by the observational error (regardless of whether the index was used in the fit). Grey histograms are for the whole LEGA-C sample (normalized to unit area); black/magenta/blue histograms refer to the \silver/\golden~Q/\golden~SF (normalized to the \silver~sample). A gaussian centered on zero and with unit standard deviation is shown for reference (dot-dashed curve).}
\label{fig:indx_obs_bf_age}
\end{figure}
\begin{figure}
\begin{center}
\includegraphics[width=0.23\textwidth]{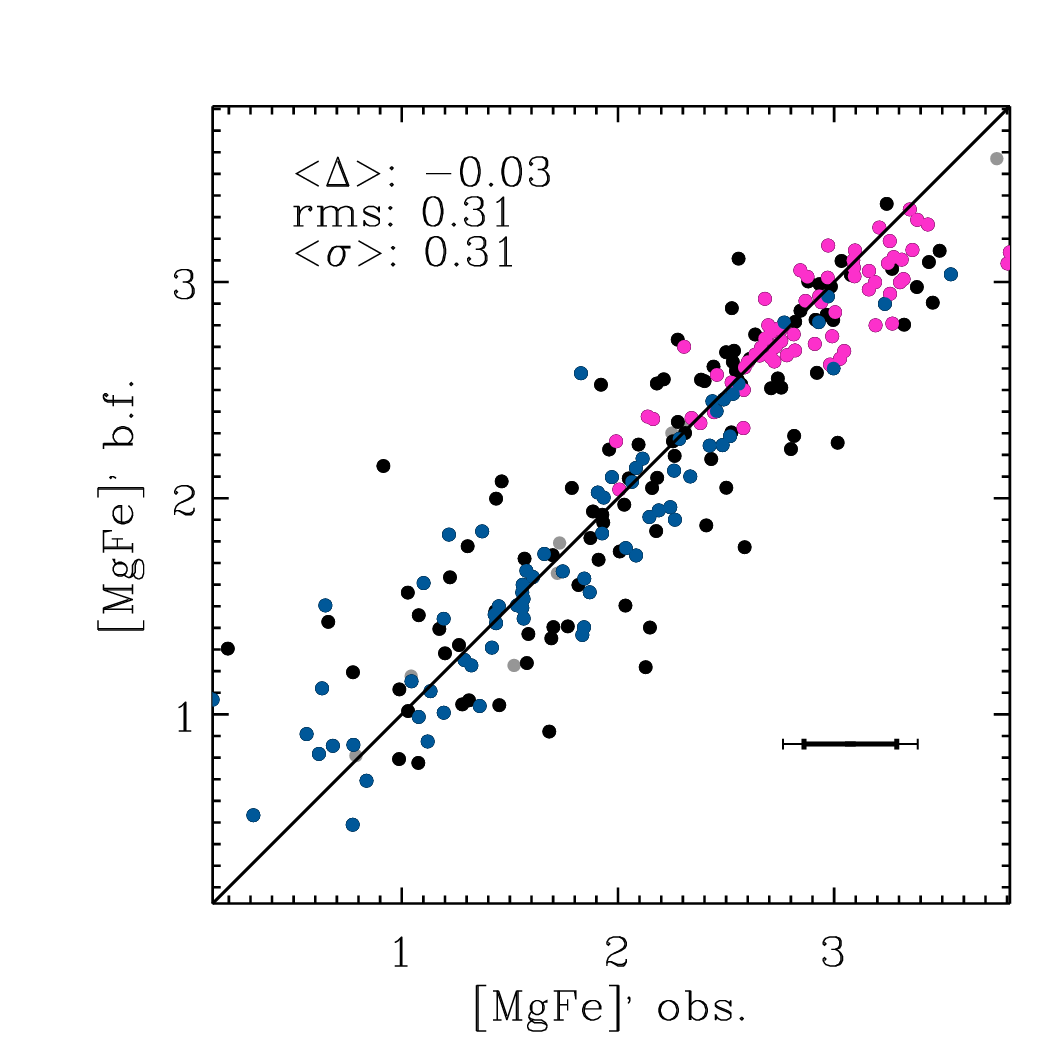}
\includegraphics[width=0.23\textwidth]{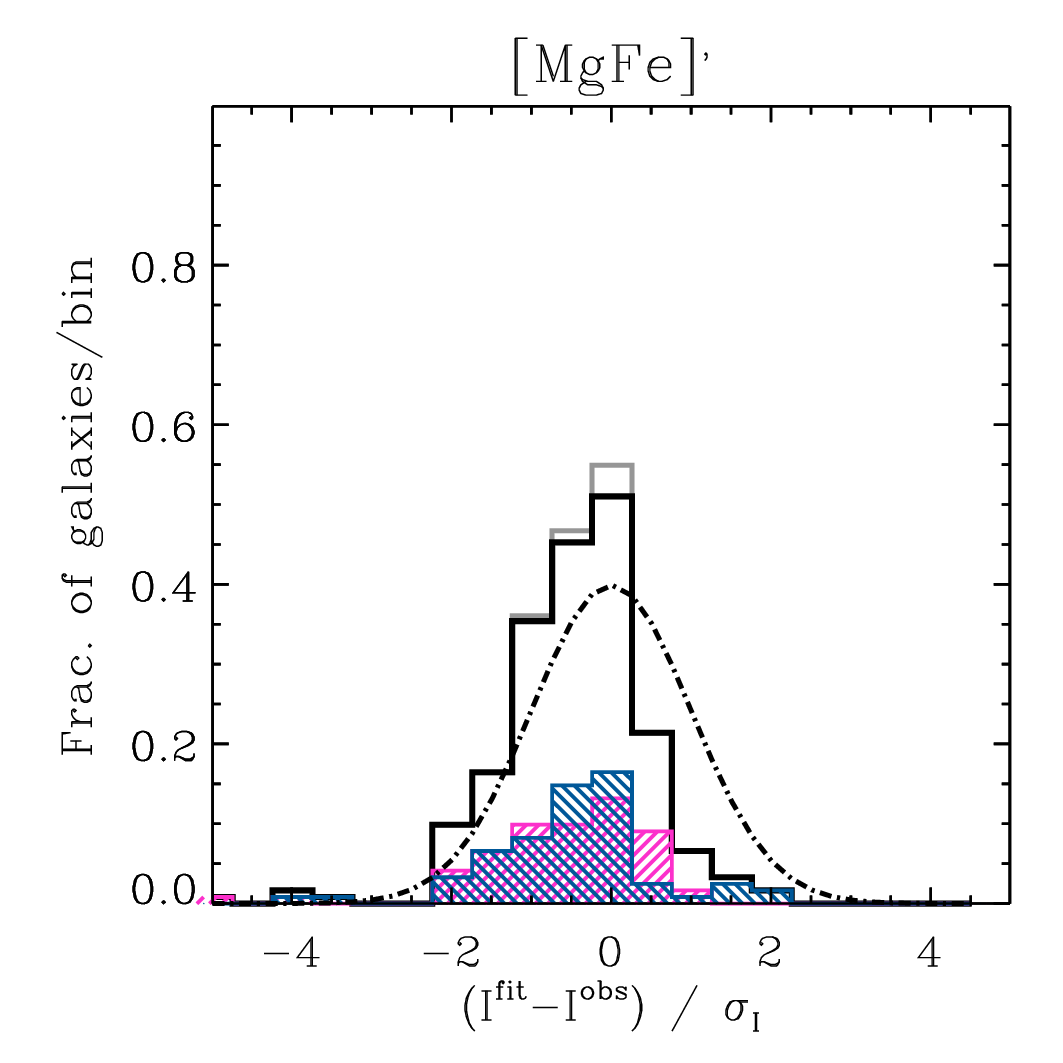}

\includegraphics[width=0.23\textwidth]{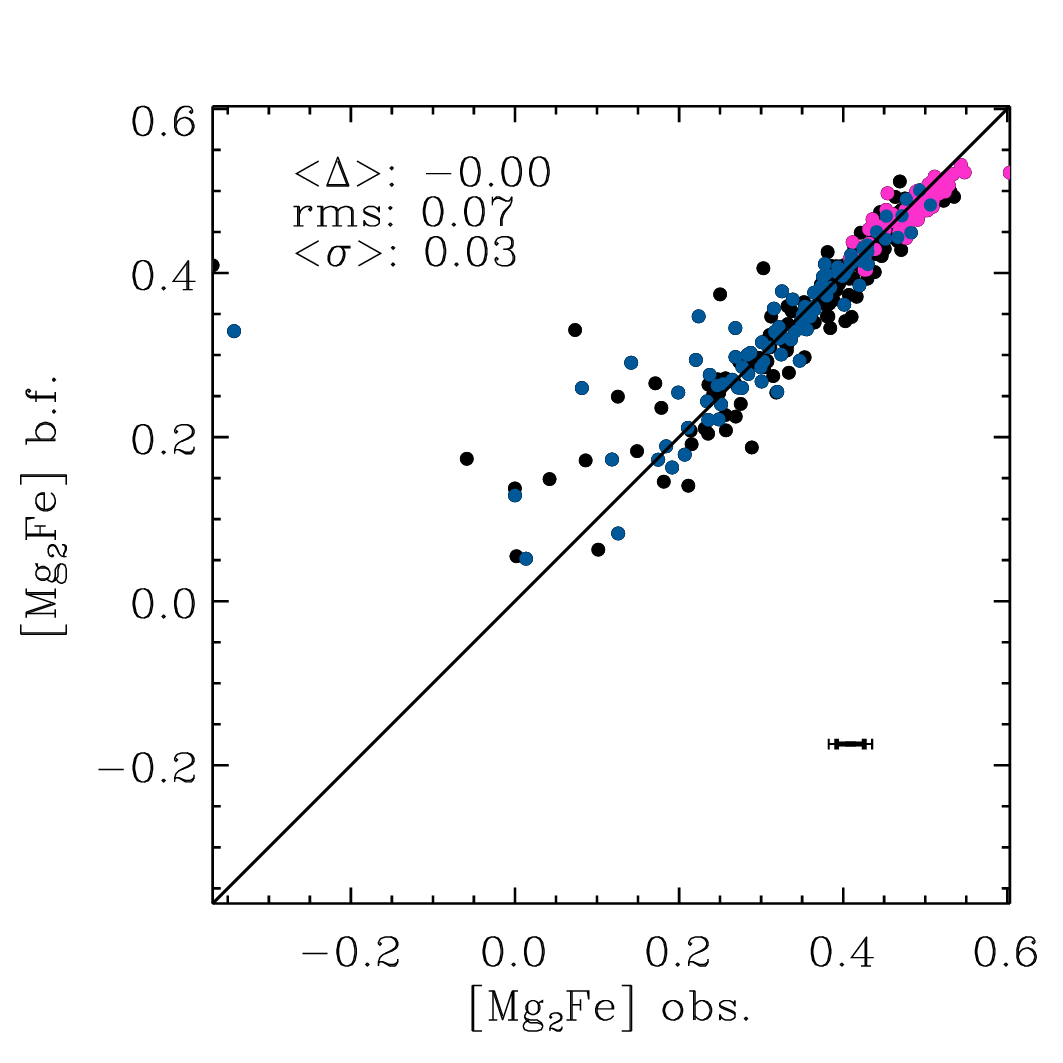}
\includegraphics[width=0.23\textwidth]{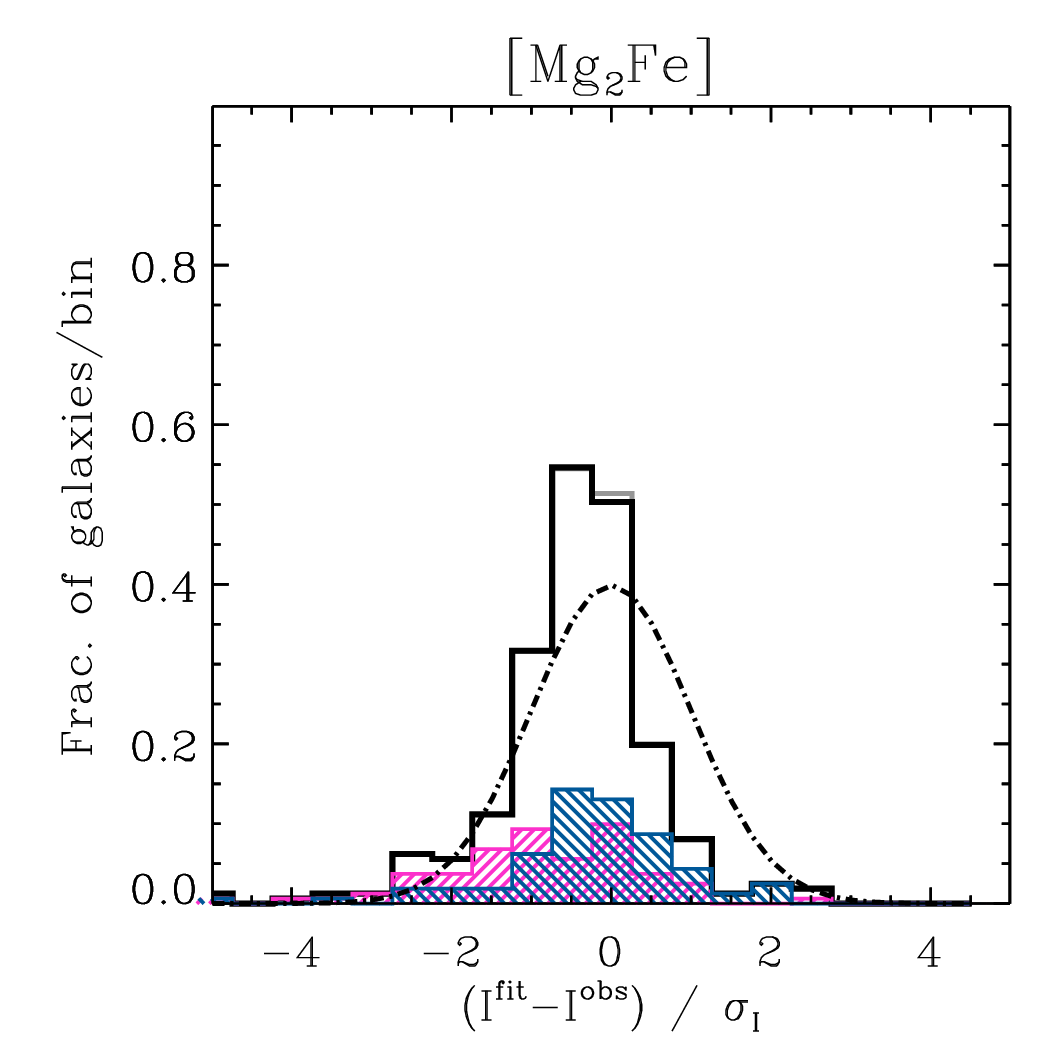}

\includegraphics[width=0.23\textwidth]{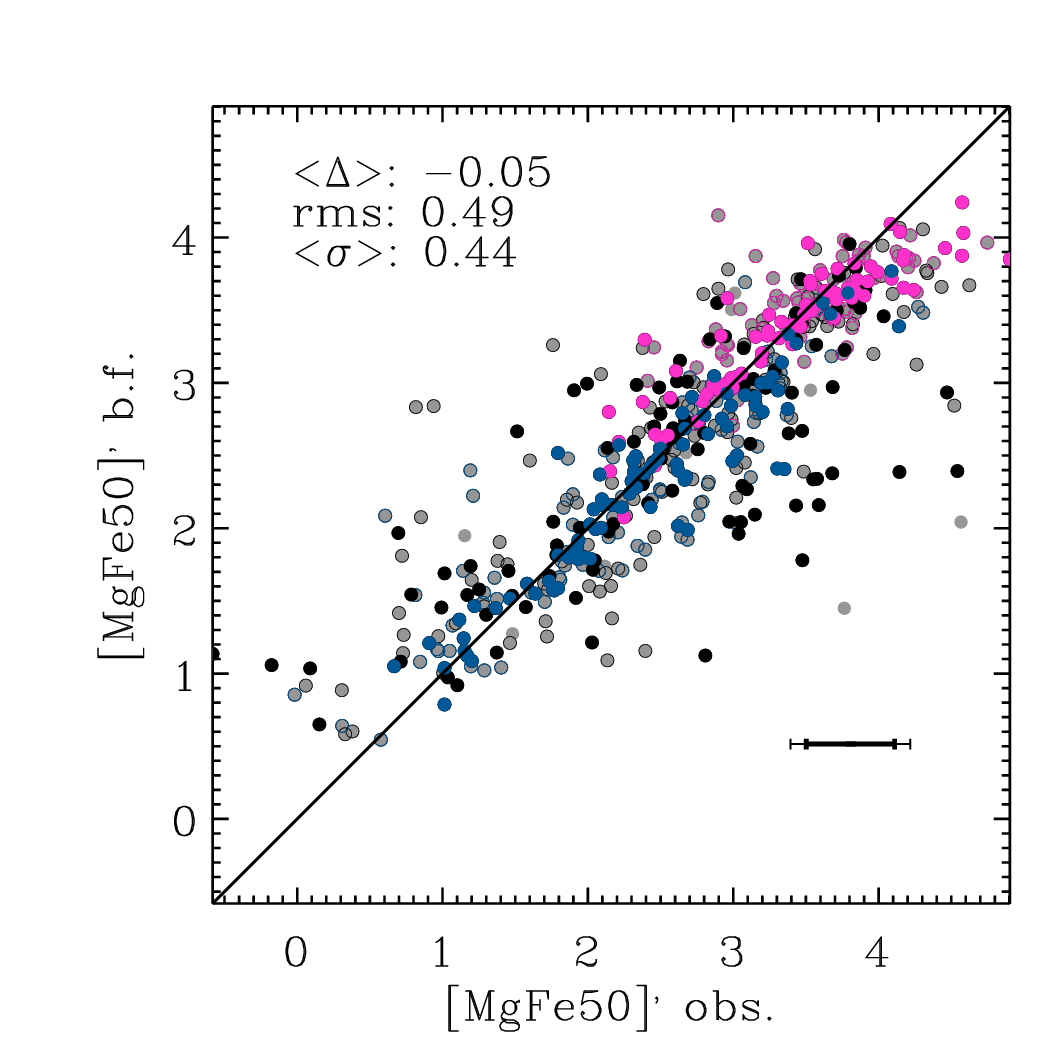}
\includegraphics[width=0.23\textwidth]{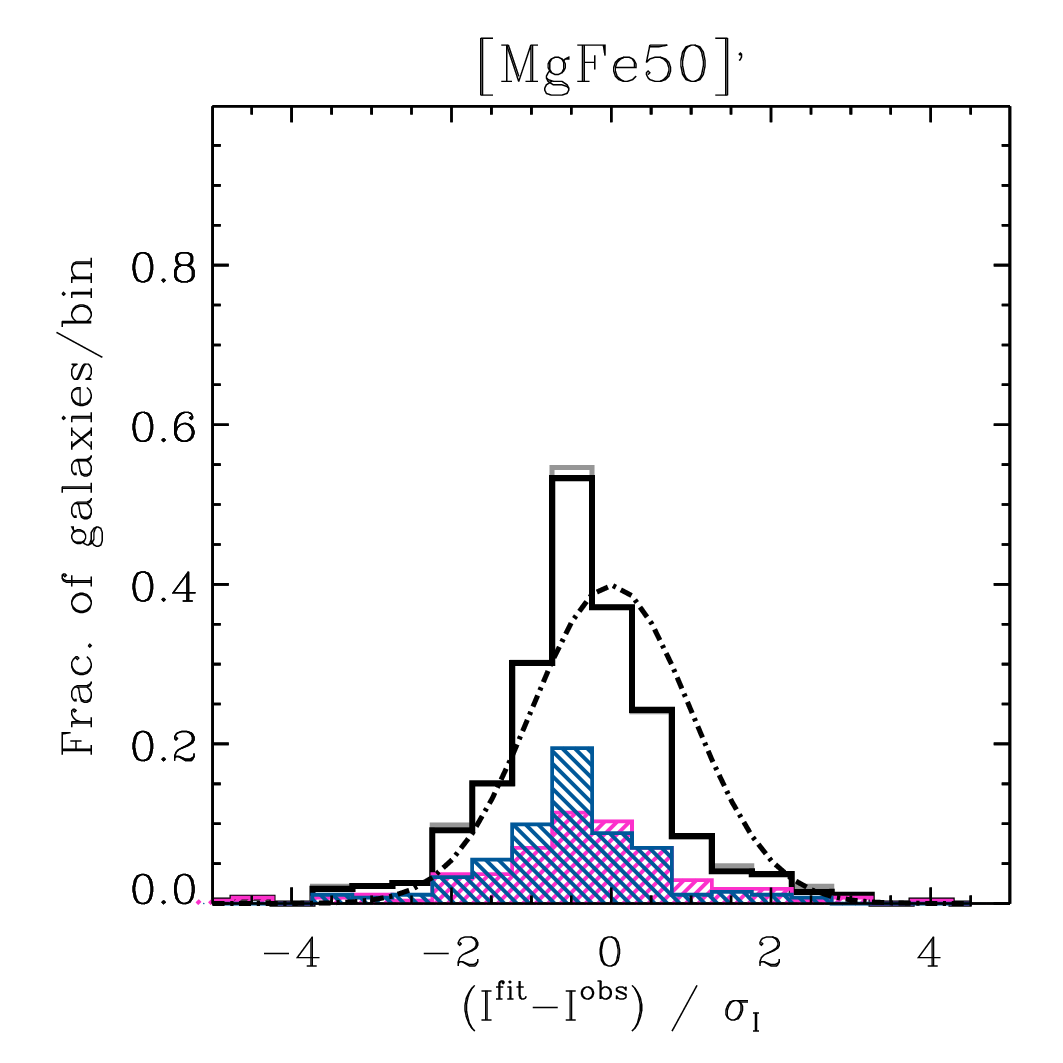}

\includegraphics[width=0.23\textwidth]{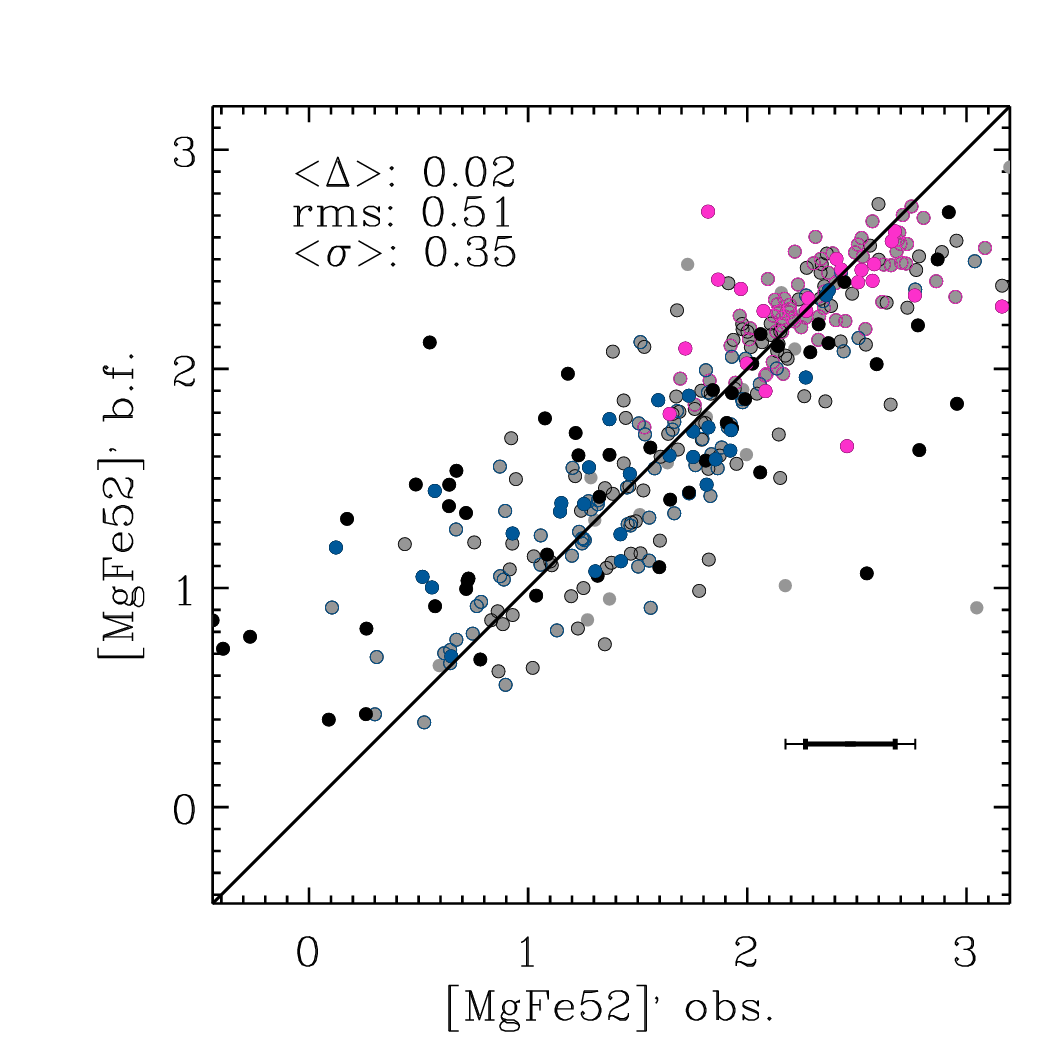}
\includegraphics[width=0.23\textwidth]{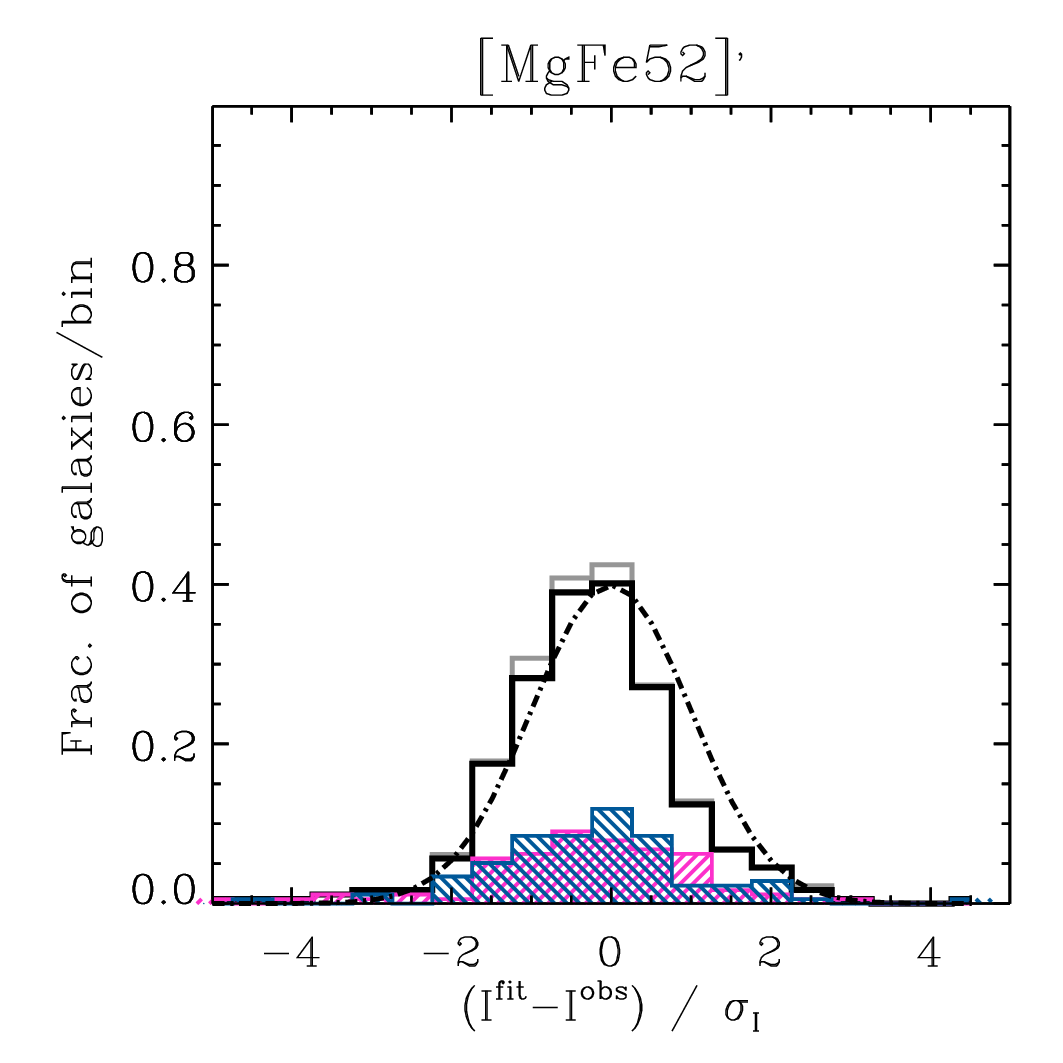}
\end{center}
\caption{Same as Fig.~\ref{fig:indx_obs_bf_age} but for the absorption features considered in the {\tt BaStA} fit as metallicity-sensitive diagnostics.}
\label{fig:indx_obs_bf_met}
\end{figure}

Figures~\ref{fig:indx_obs_bf_age} and~\ref{fig:indx_obs_bf_met} show, for the indices used for stellar populations constraints, how well the observed index strengths are reproduced by the models. The left panels compare the observed index strength with the one predicted by the best-fit model (i.e. the model of minimum $\chi^2$ considering both absorption indices and photometry), both for the whole LEGA-C primary sample (grey circles) and for the \silver/\golden~samples used in this work (black/magenta/blue circles). The right panels display the histograms of the difference between predicted and observed index strength normalized by the observational uncertainty. In all cases the agreement is very good and within the observational uncertainties. Nevertheless, we notice the tendency to underestimate \dn~and \hdg~and slightly overestimate \hb~for \golden~quiescent galaxies. We checked that the comparison between best-fit model and observations is slightly improved for the quiescent galaxies if we fit only absorption features (without photometry) to dust-poor models. This may signal some tension in the SPS models between different observational constraints (indices and photometry). As we discuss in Appendix~\ref{sec:appendix_model_prior}, we are overall confident that our treatment of dust, necessary for analysing the general galaxy population in a consistent way, does not introduce significant bias. It is worth noticing that a mismatch in \dn-$H\delta$ between LEGA-C and model predictions (IllustrisTNG) has been discussed in \cite{Wu21}, originating either from systematic uncertainties in SPS models or from the requirement of small recent bursts of star formation in otherwise old systems. 

As mentioned, in addition to spectral absorption indices, we add constraints from the spectral shape in order to infer simultaneously stellar masses and dust attenuation together with age and metallicity. We extract the photometric information from the UltraVISTA catalog of \cite{Muzzin13a}. We consider the fluxes in the photometric bands $r, i, z, Y, J$, which cover the reft-frame wavelength range between 3800 and 7500\AA~at the mean redshift of our sample. Following the conclusions in Appendix B of \cite{DR3}, we exclude photometric bands that sample redward of the rest-frame 8000\AA. 
We also exclude from the fit the B and V bands which sample the near-UV spectral range, motivated by the fact that at the rest-frame UV wavelengths the CB19 models are less reliable, manifesting also as a mismatch between the observed colors and the model colors at fixed absorption indices. We performed several tests changing both the model library (the base SSPs and the recipes for composite stellar populations) and the observational constraints. We found that fits that included the B and V photometry provided parameter estimates that varied significantly changing e.g. the SPS models or the chemical evolution assumptions. This affected in particular massive quiescent galaxies that were assigned younger ages and higher dust attenuations, inconsistent with their $J-24$\micron~colors.
Similarly as for the indices, we have checked that for magnitudes and colors in the bands used for the fit the mean deviation between observations and best-fit model is 0.1-0.2 times the observational uncertainty.

\section{Estimates of stellar populations physical parameters}\label{sec:stelpop_results}
The stellar populations physical parameters that we are primarily interested in are the light-weighted mean age and the light-weighted mean stellar metallicity. 
In addition to these, we also consistently extract stellar mass and dust attenuation estimates. The stellar mass is the mass surviving in stars and remnants and is obtained for each model in the library as the normalization factor that minimizes the $\chi^2$ of the $rizYJ$~photometry. The dust attenuation of each model is computed in the rest-frame g-band, directly comparing the broad-band flux of the attenuated model spectrum with that of the unattenuated model spectrum.\footnote{Notice that this differs from the approach used in our previous works \citep[see][]{gallazzi05,gallazzi14} where dust attenuation was estimated from the excess between the observed color and the best-fit dust-free model based on an index-only fit.} For each of these physical parameters we construct the posterior probability density function (PDF), marginalizing over all the nuisance parameters, as described in Sec.\ref{sec:stelpop_method}. Table~\ref{Tab:mean_errors} reports the mean uncertainties in physical parameter estimates for the \silver~and \golden~subsamples. In Appendix~\ref{sec:appendix_basta} we address systematic differences with respect to our previous modeling assumptions \citep{gallazzi05,gallazzi14}, while in Appendix~\ref{Appendix:compare_parameters} we compare our estimates with those derived with {\tt BAGPIPES} or {\tt Prospector} in \cite{kaushal24} and \cite{Nersesian25}, respectively. Although the scientific results presented in this paper (Sec.~\ref{sec:scaling_legac} and ~\ref{sec:discussion}) pertain to the general galaxy population, in this section and the Appendix we keep the separation between Q and SF galaxies in order to present the typical accuracy in parameter estimates that can be achieved for the two classes of objects. This is relevant because the ability to constrain ages and metallicities depend on the age of the stellar populations and the overall mix of stellar populations in a galaxy (i.e. the total duration of the SFH), affecting the strength of the features and the underlying degeneracies.

\begin{table*}
\caption{Samples used in the scientific analysis and mean uncertainties on the physical parameter estimates.}
\begin{center}

\begin{tabular}{|l|c|c|cc|cc|cc|cc|}
\hline
sample & $\rm N_Q$ & $\rm N_{SF}$ &  \multicolumn{2}{c|}{mean $\sigma_{\log<Age/yr>_r}$ [dex]} & \multicolumn{2}{c|}{mean $\sigma_{\log<Z_\ast/Z_\odot>_r}$ [dex]} & \multicolumn{2}{c|}{mean $\sigma_{\log(M_\ast/M_\odot)}$ [dex]} & \multicolumn{2}{c|}{mean $\sigma_{A_g}$ [mag]} \\ 
 &  &  & Q & SF & Q &SF  &Q &SF &Q & SF \\
\hline
\silver         & 232 & 320   & 0.16 & 0.16 & 0.19 & 0.35 & 0.12 & 0.13 & 0.24 & 0.38 \\
\golden       & 148 & 175    & 0.15 & 0.15 & 0.15 & 0.29 & 0.11 & 0.12 & 0.21 & 0.36\\
\hline
\end{tabular}
\end{center}
\label{Tab:mean_errors}
\tablefoot{1) Samples (\silver: all LEGA-C with ${\tt primary} = 1$, ${\tt flag\_spec} = 0$, ${\tt sigma\_star} > 0$, $0.55 < {\tt z\_spec} < 1.1$, and at least one of (\hb, \hd, \hg) and one of (\mgtwofe,\mgfef,\mgfeft,\mgfep); \golden: all \silver~galaxies with $S/N>20$); 2) and 3) number of unique galaxies classified as quiescent (Q) or as star-forming (SF) according to their specific SFR (i.e. lying below or above, respectively, of the dashed line in the left panel of Fig.~\ref{fig:UVJ_SSFR}); 4) mean error on light-weighted age (dex); 5) mean error on light-weighted stellar metallicity (dex); 6) mean error on stellar mass (dex); 7) mean error on dust attenuation (mag). The mean errors are given separately for Q/SF galaxies.}
\end{table*}

\subsection{Stellar metallicities and ages}
The derived estimates on $r$-band light-weighted stellar age and stellar metallicity are presented in the upper panels of Fig.\ref{fig:age_zstar_distributions}, along with their associated uncertainties given by the 84th-16th inter-percentile width of the PDF. The distributions are presented for the \silver~and \golden~galaxies, distinguished into quiescent and star-forming as in Fig.~\ref{fig:UVJ_SSFR}. For completeness we illustrate also the distribution of parameters and parameter uncertainties for the whole parent DR3 sample. We find that LEGA-C galaxies span a range of light-weighted ages between roughly $\log<Age/yr>=8.9$~and 9.8 (i.e. from $\sim600$~Myr to 6.3~Gyr), with Q galaxies dominating at ages older than $\log<Age/yr>=9.2$~ (1.6 Gyr). The light-weighted age is constrained typically to within 0.16 (0.15)~dex for all (high-S/N) spectra, with no significant difference between Q and SF nor as a function of age. 

LEGA-C galaxies are predominantly found to have super-solar metallicities up to $3\cdot Z_\odot$, including both Q and SF galaxies. The stellar metallicities of star-forming galaxies  extend down to $0.1\cdot Z_\odot$. We notice that the highest metallicity values ($\log(Z_\ast/Z_\odot)\gtrsim0.4$) that are close to the highest metallicity covered by the models may be biased low by $\sim0.1$~dex, and they tend to have asymmetric posterior distributions skewed toward lower metallicities. The saturation to high metallicity is however mitigated with respect to other works that are limited to an upper metallicity boundary of $\sim1.6\cdot Z_\odot$ ($\log Z/Z_\odot=0.2$), i.e. approximately a factor $2$ lower than ours \citep[e.g.][]{Bevacqua24,Nersesian25}. As opposed to the light-weighted age, the uncertainties on stellar metallicity depend on both spectral quality and galaxy type. Stellar metallicity is typically constrained to within 0.17~dex for \silver~quiescent galaxies and to within 0.15~dex for the subsample of \golden~quiescent galaxies. The accuracy on stellar metallicity decreases to 0.34~dex and to 0.28~dex for \silver~and~\golden~star-forming galaxies, respectively. These are the mean uncertainties, but we notice that uncertainties on stellar metallicity can extend up to 0.6~dex especially for low-S/N, low-metallicity star-forming galaxies. This highlights the intrinsic difficulty in constraining stellar metallicities for young/star-forming galaxies and the need for high-S/N spectroscopy covering the red metallic features.
\subsection{Stellar masses and dust attenuations}
\begin{figure*}
\begin{center}
\includegraphics[width=0.45\textwidth]{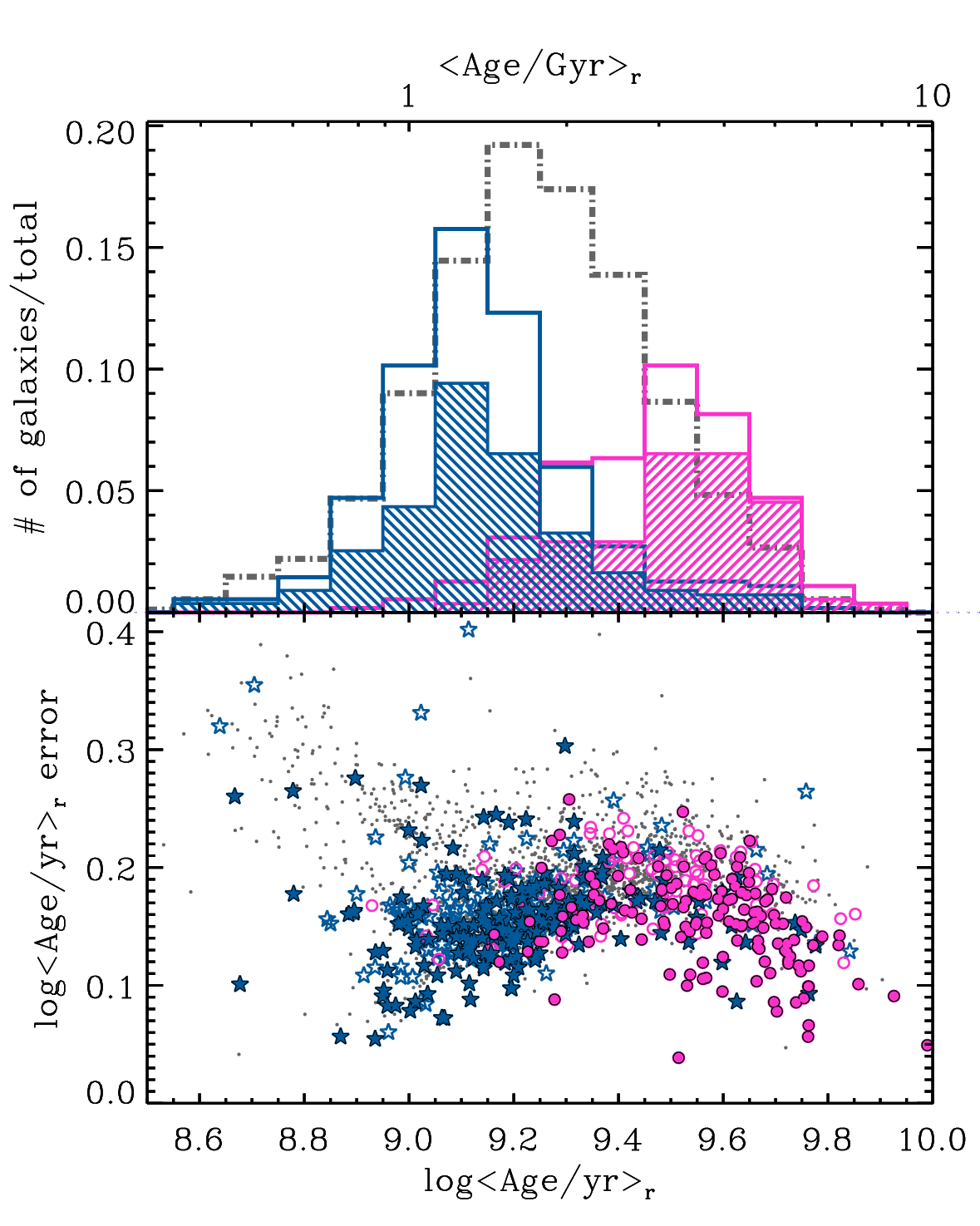}
\includegraphics[width=0.45\textwidth]{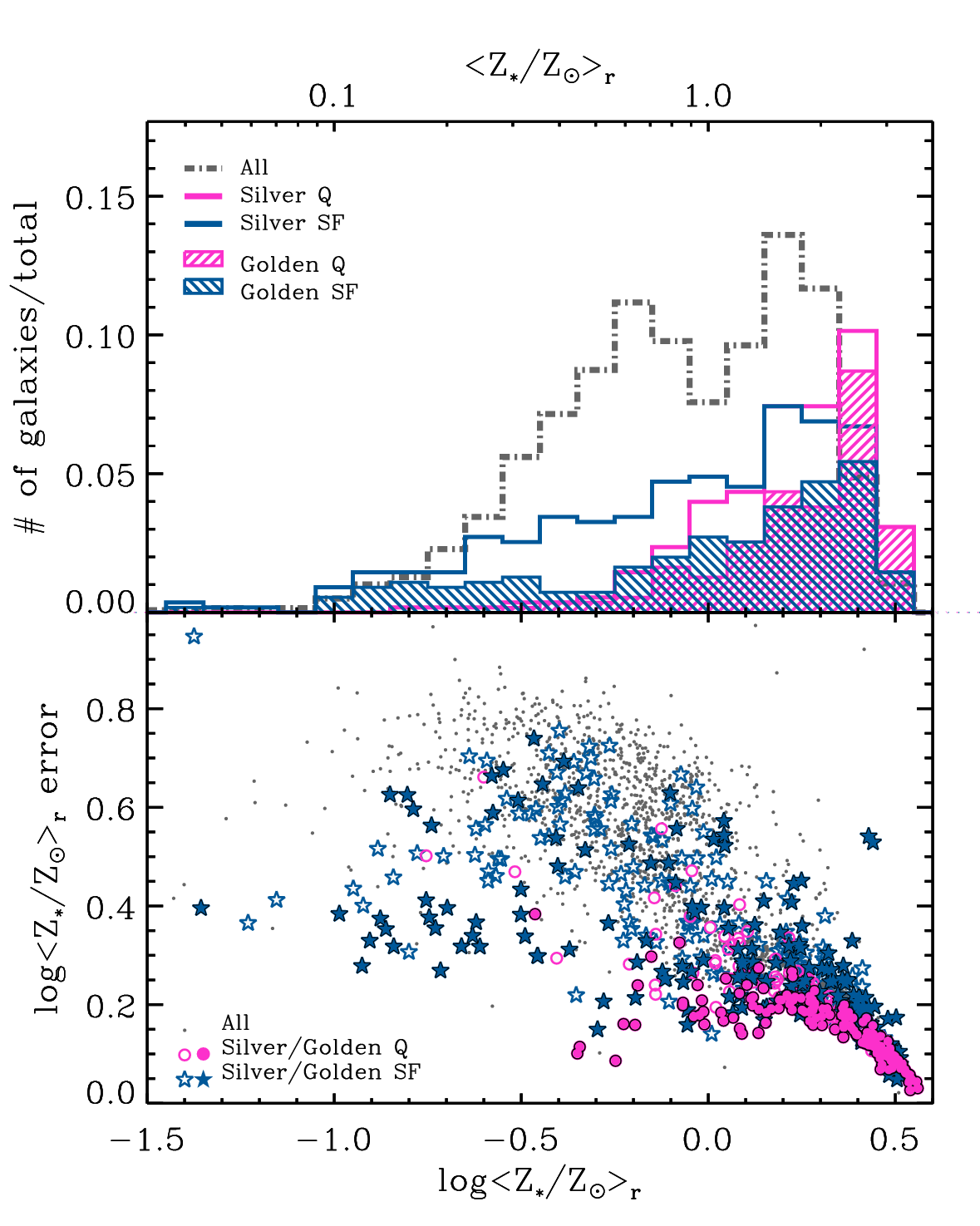}
\end{center}
\begin{center}
\includegraphics[width=0.45\textwidth]{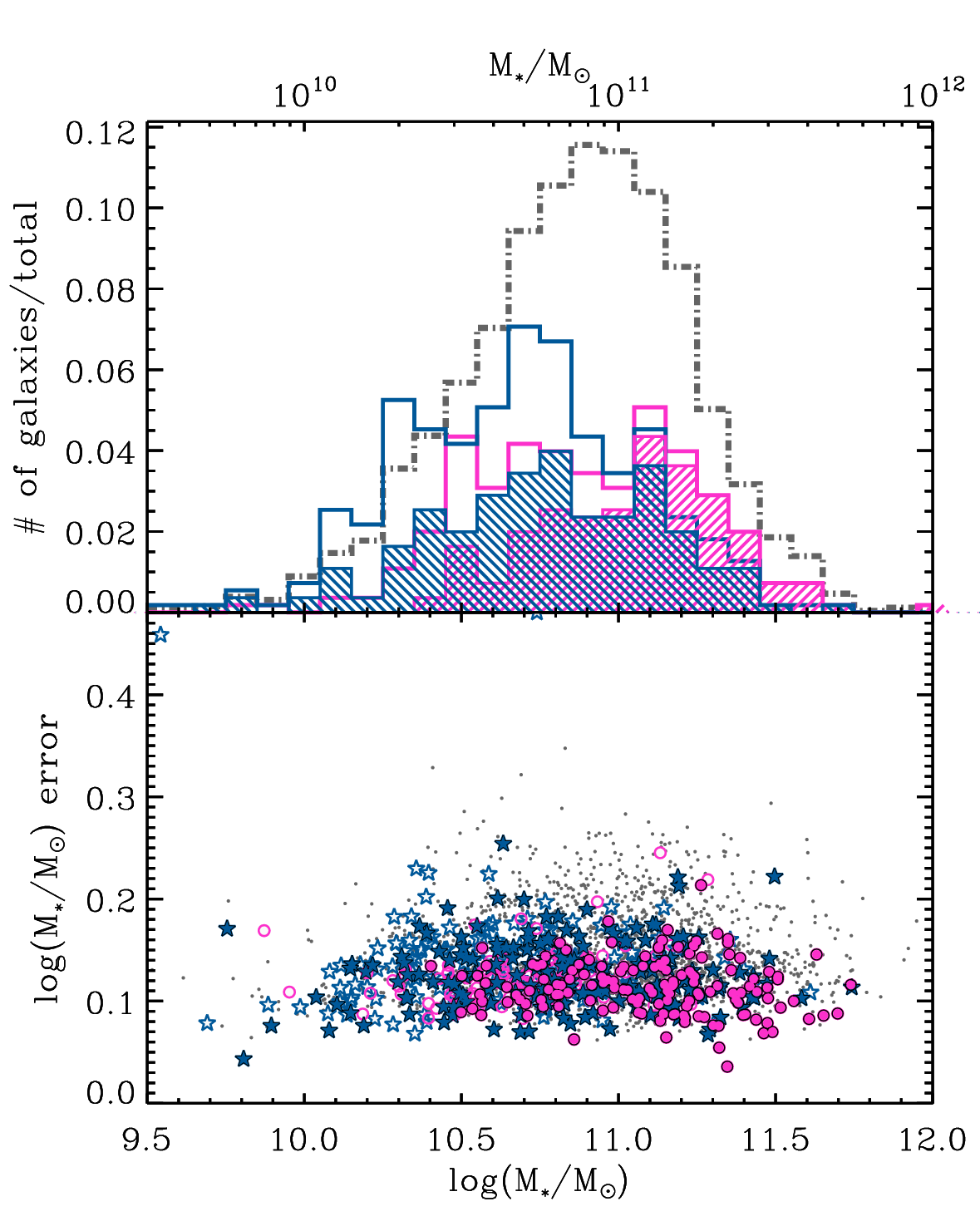}
\includegraphics[width=0.45\textwidth]{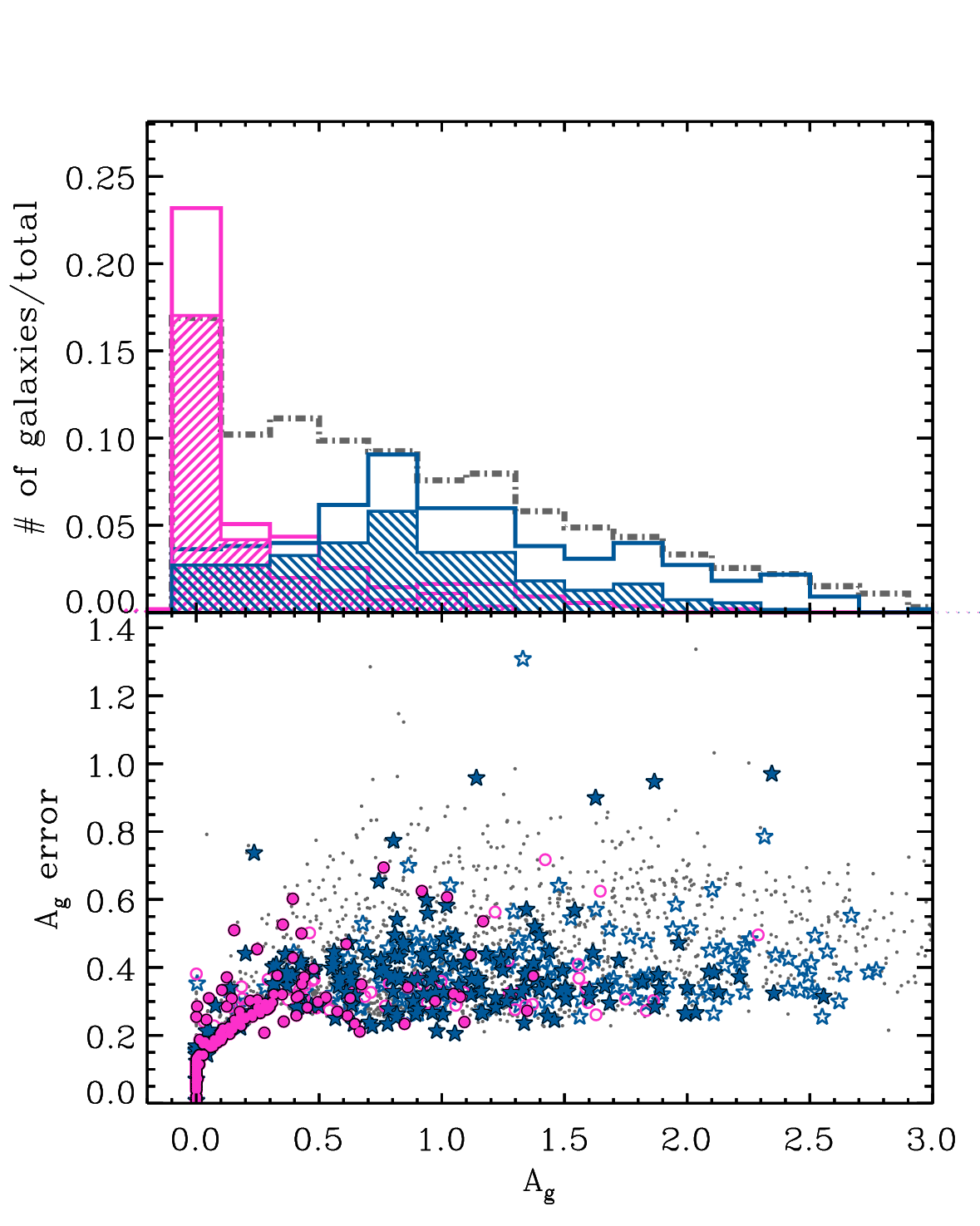}   
\caption{Distribution in $r$-band light-weighted mean age (upper-left), r-band light-weighted stellar metallicity (upper-right), stellar mass (lower-left) and dust attenuation in the g-band (lower-right) for the LEGA-C \silver~and \golden~samples analysed in this work. The upper panels display the distribution in the derived physical parameter, the lower panels show the uncertainties in the parameter estimates against the fiducial values (i.e. the median of the PDF). Galaxies belonging to the \silver~sample are distinguished into quiescent (magenta circles) and star-forming (blue stars) based on a cut in specific SFR. Galaxies in the \golden~subsample are highlighted by filled symbols. Grey dots and grey dot-dashed histograms (normalized to unit integral) display the whole LEGA-C parent sample regardless of the indices used in the fit. Magenta/blue histograms show the distribution for the Q/SF galaxies in the \silver~(hollow) and in the \golden~(hatched) subsamples, both normalized to the total \silver~sample.}\label{fig:age_zstar_distributions}
\end{center}
\end{figure*}

The lower panels of Fig.~\ref{fig:age_zstar_distributions} show the distributions in stellar mass and in rest-frame g-band dust attenuation ($A_g$), and their associated uncertainties. Our galaxy sample ranges between $10^{10}$~and $6\cdot10^{11} M_\odot$~in stellar mass, with rather homogeneous uncertainties of 0.12~dex and 0.13~dex for quiescent and star-forming galaxies respectively. We have checked that the stellar masses we derive are in excellent agreement with those in the DR3 catalog obtained from photometry-only {\tt Prospector} fit, with a mean offset of only 0.02~dex and scatter of 0.11/0.13~dex for quiescent/star-forming galaxies. 
The majority of our galaxy sample, and in particular quiescent galaxies, show negligible or low dust attenuation. However a small fraction of quiescent galaxies and the majority of star forming galaxies have $\rm A_g$~larger than 0.5 mag. Uncertainties are between 0.2 and 0.4 mag, but can be up to 0.6 mag for lower-quality spectroscopic data or young galaxies for which the looser constraints on stellar metallicity give more freedom to dust attenuation. For galaxies with a significant fraction of stars younger than $10^7$\,yr (the ``birth cloud'' timescale), uncertainties are also boosted by degeneracies between the two dust components assumed in the \cite{CF00} dust attenuation model.

\subsection{Comparison of light- and mass-weighted properties}\label{sec:light_vs_mass}

\begin{figure}
\includegraphics[width=0.45\textwidth]{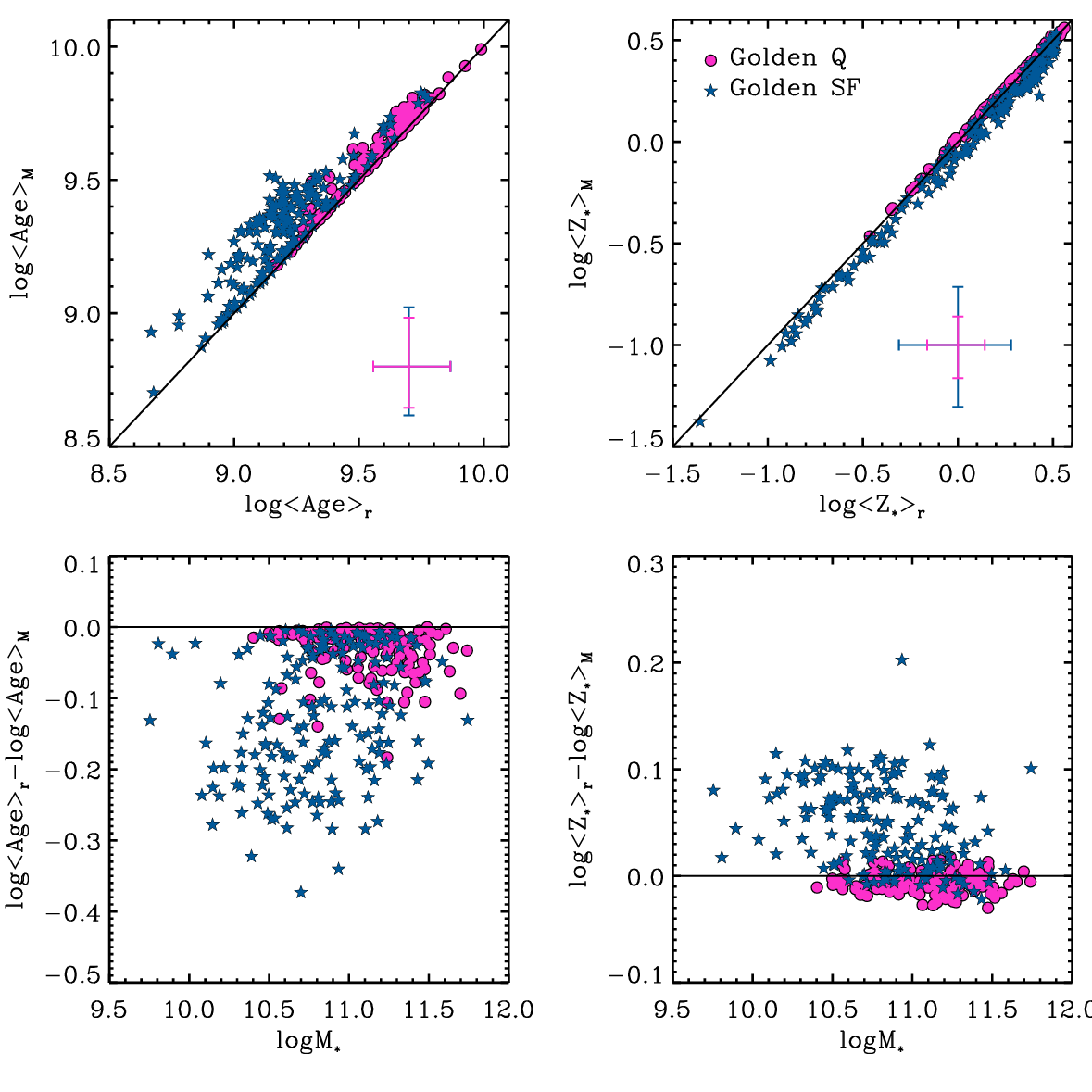}
\caption{Comparison between light-weighted and mass-weighted ages and stellar metallicities derived in this work for \golden~galaxies. Magenta circles indicate quiescent galaxies, blue stars indicate star-forming galaxies. The errorbars show the median uncertainties for the two subsamples. The upper panels show the 1:1 comparison, while the bottom panels show the difference between light- and mass-weighted quantities versus stellar mass.}\label{fig:wm_vs_wr}
\label{default}
\end{figure}
Together with light-weighted mean parameters, we also derive estimates of mass-weighted mean ages and metallicities. Note that these estimates are weighted on the \emph{present-day} stellar mass, corresponding to the formed stellar mass reduced by the fraction of mass returned to the ISM. While these parameters would be potentially more physically relevant, they are also intrinsically more subject to systematics associated to the adopted SFH modeling. Fig.~\ref{fig:wm_vs_wr} compares the light-weighted and the mass-weighted properties for \golden~quiescent and star-forming galaxies. We notice that for quiescent galaxies the two weights lead to very similar mean stellar population properties, with mass-weighted ages on average 0.05~dex older, and mass-weighted metallicities consistent or only marginally smaller than light-weighted metallicities. 
For star-forming galaxies, mass-weighted ages are also always larger than light-weighted ones with differences spanning between zero and 0.4 dex. Mass-weighted stellar metallicities are typically smaller than light-weighted ones, up to 0.1 dex. 
This trend is expected from our parametrization of the metal enrichment history (which is monotonically increasing with time) and because older, less metal-rich stellar generations weigh more in the mass-weighted metallicity with respect to the light-weighted one.

\section{The stellar metallicity and age scaling relations in LEGA-C}\label{sec:scaling_legac}
In this section we quantify the distribution of the general population of massive ($>10^{10}M_\odot$) $\left<z\right>\sim0.7$ galaxies from the LEGA-C survey in mean stellar metallicity and age as a function of stellar mass and velocity dispersion. The reference population statistical trends are those obtained from the volume- and completeness-weighted \silver~sample, but we also present the results for the unweighted \golden~sample as a check against spectral quality.
\subsection{Trends with stellar mass}
\begin{figure*}[h!]
    \centering
\includegraphics[width=\textwidth]{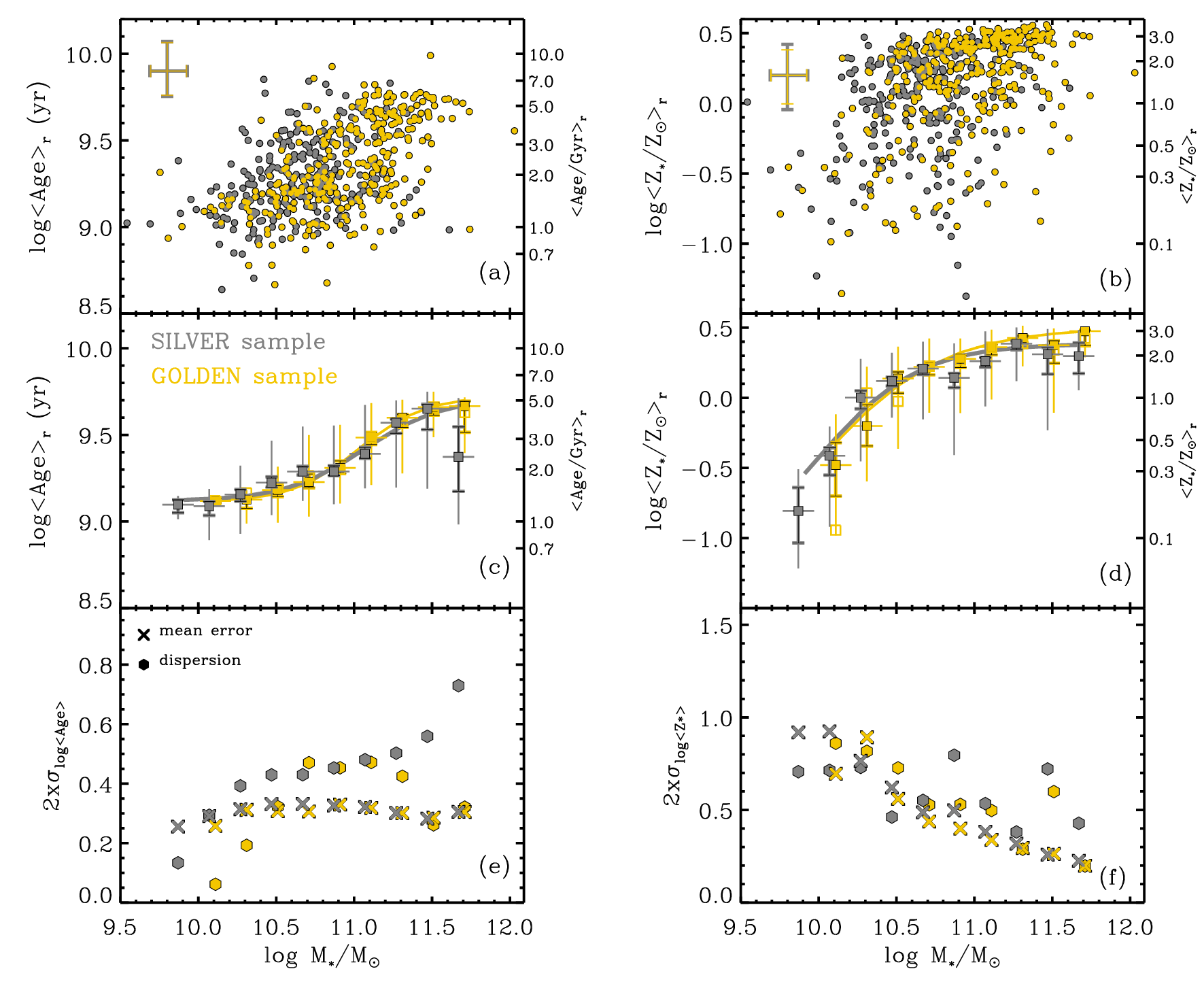}
\caption{Luminosity-weighted age ({\it left panels}) and luminosity-weighted stellar metallicity ({\it right panels}) versus stellar mass for LEGA-C galaxies. The {\it upper panels} show individual measurements: \silver~sample is shown with silver circles and \golden~galaxies are highlighted with golden circles. {\it Middle panels} show the median age (panel c) and median stellar metallicity (d) in bins of stellar mass 0.2~dex wide and with at least 5 galaxies. Thick vertical errorbars show the error on the median, while thin horizontal and vertical bars indicate the 16th-84th interpercentile range of the distribution. Silver filled squares refer to the \silver~sample weighted for completeness with ${\tt Tcor}\times {\tt w\_spec\_silver}$, while golden filled (empty) squares refer to the \golden~sample without (with) weights. The corresponding lines in panel c and d display the relation as in Eq.\ref{eqn:sigmoid} that best fits the median values of age/metallicity for the weighted \silver~and weighted \golden~sample respectively. The {\it bottom panels} compare the mean error on individual age and stellar metallicity estimates (crosses) with the dispersion, as given by the 16th-84th interpercentile range, at fixed stellar mass (hexagons), with silver/golden symbols for \silver/\golden~sample.}\label{fig:scaling_legac}
\end{figure*}

Figure~\ref{fig:scaling_legac} shows the distribution of our sample galaxies in luminosity-weighted mean age (left panels) and in luminosity-weighted mean stellar metallicity (right panels) versus stellar mass. The upper panels display the individual measurements for the reference \silver~sample and for the high-S/N \golden~subsample. As already seen in Fig.~\ref{fig:age_zstar_distributions}, the high-S/N requirement tends to select against metal-poor/low-mass galaxies.  
The middle panels show the median trends of age and metallicity in bins of stellar mass 0.2~dex wide and with at least 5 galaxies. The golden squares with errorbars show the median and dispersion in the parameter (expressed by the 84th-16th inter-percentile range of the distribution) limited to the \golden~sample, while the silver squares and errorbars refer to the broader \silver~sample weighted for completeness by ${\tt Tcor}\times {\tt w\_spec\_silver}$ (see Sec.~\ref{sec:stats_corr}). Finally, the bottom panels compare the dispersion in age and metallicity at fixed mass (hexagons) with the mean parameter uncertainties (crosses) for the \golden~sample distribution (golden symbols) and the weighted \silver~sample distribution (silver symbols). 

The distribution in light-weighted ages (Fig.\ref{fig:scaling_legac}a) indicates the presence of two sequences at ages younger/older than 3 Gyr ($\log\left<Age/yr\right>9.5$), particularly evident for the \golden~sample. There is a clear trend of increasing age with increasing stellar mass, from $\sim1.2$~to $\sim5$~Gyr ($\log\left<Age/yr\right>=9.1-9.7$) from $10^{10}$ to $10^{11.7}M_\odot$ (Fig.\ref{fig:scaling_legac}c). The larger and more complete statistics of the \silver~sample confirms the median trend of the robust, high-S/N \golden~sample, though with larger dispersion in some of the mass bins. The dispersion in light-weighted age is larger by $0.15-0.2$~dex than the mean parameter uncertainty, across almost the whole range of stellar mass probed and in particular around $10^{11}M_\odot$. We notice that the larger dispersion in the \silver~sample is due both to the inclusion of lower-S/N galaxies and to the application of completeness weights (except in the highest-mass bin, which, however, suffers from low number statistics).

For a convenient quantification of the mean trend, we fit a parametric function to the median $\log\left<Age\right>_r$ as a function of $\log(M_\ast)$\footnote{To fit scaling relations, we use the {\tt MPFIT} IDL routine which finds the bestfit parameters of the supplied function by minimizing the sum of the weighted squared differences between the model and data. The fits are performed on the median values in bins of stellar mass accounting for the error on the median.}. Following \cite{leja19} and \cite{Nersesian2020}, we find that the trend is well described by a sigmoidal functional form. Specifically, for a generic parameter $P$ we parametrize the relation with $M_\ast$ as:\\
\begin{equation}
%;  Y = Y0 + A*tanh(B*log(X/X0)
P(M_\ast) = \bar{P} + A*tanh\left(B*\log\frac{M_\ast}{\bar{M_\ast}}\right) - C \label{eqn:sigmoid}
\end{equation}

where $\bar{M_\ast}$~is the characteristic mass of inflection, A represents the dynamical range of the relation, B regulates the extent of the inflection range, and
\begin{eqnarray}
C = A*tanh\left(B*\log\frac{10^{11.5}}{\bar{M_\ast}}\right) \nonumber
\end{eqnarray}
normalizes the relation so that $\bar{P}$~represents the zero-point at $M_\ast=10^{11.5}M_\odot$.

For the median light-weighted age of the LEGA-C galaxies, the zero point at $10^{11.5}M_\odot$  corresponds to a value of $\log \left<Age\right>_r=9.62\pm0.04$ (4.2 Gyr). The inflection occurs at a stellar mass of $1.2\pm0.4\cdot 10^{11}M_\odot$. The parameters of the fitted relation for the high-SN \golden~sample are consistent, with smaller uncertainties, with those of the weighted \silver~sample.

Unlike age, the distribution in stellar metallicity does not highlight the presence of two sequences. LEGA-C galaxies have predominantly high stellar metallicities, above solar, with a spread to lower metallicities down to 10\% solar (Fig.\ref{fig:scaling_legac}b). Stellar metallicity increases on average from $\sim30$\% solar to $\sim3\cdot$~solar from $10^{10}$ to $10^{11.7}M_\odot$. The trend in $\log\left<Z_\ast/Z_\odot\right>$ is steep at low masses and flattens above $\sim6\cdot10^{10}M_\odot$ (Fig.\ref{fig:scaling_legac}d). The dispersion in stellar metallicity at fixed mass is slightly larger than the mean parameter uncertainty at stellar masses above $10^{10.5}M_\odot$, while below $10^{10.5}M_\odot$ the uncertainties in stellar metallicity for the \silver~sample prevent a robust statement about the intrinsic dispersion in stellar metallicity.

To better quantify the trends as a function of stellar mass, we fit a curve to the median points. We adopt the same sigmoidal function in Eq.\ref{eqn:sigmoid} as in the case of age, although the stellar metallicity$-$mass relation, in the mass range probed by our data, displays only the concave part above the inflection point. 
The stellar metallicity of the weighted \silver~sample reaches a value of $\log\left<Z_\ast/Z_\odot\right>=0.37\pm0.04$ at $10^{11.5}M_\odot$. The relation for the incomplete, higher-S/N \golden~sample saturates at a higher value of $0.46\pm0.01$.

The median trends and functional fits are reported in Tables~\ref{Tab:median_relations_all} and~\ref{Tab:functional_fit_all}. In Fig.~\ref{Fig:scaling_legac_extpb10} in Appendix~\ref{sec:appendix_basta}, we compare the LEGA-C scaling relations with those we obtained at similar redshift but for a more limited sample in G14: under the same SPS modeling assumptions as in G14, we find consistent results, with LEGA-C being crucial in increasing the statistics and extending the scaling relations to lower masses. We notice, though, that the differences in modeling impact the normalization of the scaling relations and the detailed shape of the galaxy distribution, in particular for the age bimodality.

\begin{table*}
\caption{Median trends of light-weighted age and stellar metallicity as a function of stellar mass for the \silver~and \golden~samples, as shown in Fig.\ref{fig:scaling_legac}.}\label{Tab:median_relations_all}
\begin{center}
\begin{tabular}{|c|ccc|c|ccc|c|c|}
\hline
\noalign{\vspace{3pt}}
\multicolumn{10}{|c|}{\silver~sample, weighted by ${\tt Tcor}\times {\tt w\_spec\_silver}$} \\
\hline
\noalign{\vspace{3pt}}
$\log(M_\ast/M_\odot$) & \multicolumn{3}{c|}{$\log<Age/yr>_r$} & $<2\times\sigma_{\log Age}>$  & \multicolumn{3}{c|}{$\log<Z_\ast/Z_\odot>_r$} &$<2\times\sigma_{\log Z_\ast}>$ & $\rm N_{bin}$\\
 & p50 & p16 & p84 &  & p50  & p16 & p84 & & \\
\hline
\noalign{\vspace{3pt}}
       9.91  &   $9.10^{+0.05}_{-0.03}$  &  9.01   &   9.15  &	   0.26 & $-0.81^{+0.23}_{-0.17}$ &   $-1.22$  &   $-0.51$  &  0.92 & 5 \\
      10.11  &   $9.09^{+0.05}_{-0.03}$  &  8.89   &   9.19  &	0.29 & $-0.41^{+0.14}_{-0.06}$ &   $-0.92$  &   $-0.20$  &	0.92 & 21 \\
      10.31  &   $9.16^{+0.04}_{-0.03}$  &  8.93   &   9.32  &	0.31 & $ 0.00^{+0.08}_{-0.05}$ &   $-0.45$  &   $ 0.28$  &	0.77 & 51 \\
      10.51  &   $9.22^{+0.03}_{-0.03}$  &  9.04   &   9.47  &	0.33 & $ 0.12^{+0.04}_{-0.03}$ &   $-0.14$  &   $ 0.32$  &	0.62 & 81 \\
      10.71  &   $9.29^{+0.02}_{-0.03}$  &  9.12   &   9.55  &	0.33 & $ 0.21^{+0.04}_{-0.02}$ &   $-0.15$  &   $ 0.40$  &	0.49 & 113 \\
      10.91  &   $9.29^{+0.02}_{-0.03}$  &  9.10   &   9.55  &	0.33 & $ 0.14^{+0.07}_{-0.03}$ &   $-0.41$  &   $ 0.39$  &	0.50 & 97 \\
      11.11  &   $9.39^{+0.03}_{-0.04}$  &  9.19   &   9.67  &	0.32 & $ 0.26^{+0.04}_{-0.03}$ &   $-0.06$  &   $ 0.47$  &	0.38 & 91 \\
      11.31  &   $9.57^{+0.06}_{-0.02}$  &  9.20   &   9.70  &	0.30 & $ 0.39^{+0.04}_{-0.02}$ &   $ 0.12$  &   $ 0.50$  &	0.32 & 58 \\
      11.51  &   $9.65^{+0.12}_{-0.03}$  &  9.19   &   9.75  &	0.28 & $ 0.31^{+0.14}_{-0.05}$ &   $-0.23$  &   $ 0.49$  &	0.26 & 23 \\
      11.71  &   $9.37^{+0.20}_{-0.17}$  &  8.98   &   9.71  &	0.31 & $ 0.30^{+0.12}_{-0.09}$ &   $ 0.05$  &   $ 0.48$  &	0.23 & 6 \\
\noalign{\vspace{3pt}}
\hline
\noalign{\vspace{3pt}}
\multicolumn{10}{|c|}{\golden~sample} \\
\hline
\noalign{\vspace{3pt}}
$\log(M_\ast/M_\odot$) & \multicolumn{3}{c|}{$\log<Age/yr>_r$} & $<2\times\sigma_{\log Age}>$  & \multicolumn{3}{c|}{$\log<Z_\ast/Z_\odot>_r$} &$<2\times\sigma_{\log Z_\ast}>$ & $\rm N_{bin}$\\
 & p50 & p16 & p84 &  & p50 & p16 & p84 & & \\
\hline
\noalign{\vspace{3pt}}
      10.11  &   $9.12^{+0.01}_{-0.01}$ &    9.09  &    9.15  &	 0.26& $-0.48^{+0.22}_{-0.16}$ &  $-0.98$  &   $-0.12$  &    0.70 & 8 \\
      10.31  &   $9.13^{+0.05}_{-0.02}$ &    8.99  &    9.18  &	 0.31& $-0.20^{+0.14}_{-0.15}$ &  $-0.60$  &   $ 0.22$  &    0.89 & 12 \\
      10.51  &   $9.18^{+0.04}_{-0.03}$ &    8.99  &    9.32  &	 0.31& $ 0.14^{+0.10}_{-0.05}$ &  $-0.36$  &   $ 0.36$  &    0.56 & 36\\
      10.71  &   $9.23^{+0.03}_{-0.05}$ &    9.03  &    9.50  &	 0.31& $ 0.22^{+0.06}_{-0.03}$ &  $-0.11$  &   $ 0.42$  &    0.44 & 53\\
      10.91  &   $9.31^{+0.03}_{-0.04}$ &    9.10  &    9.56  &	 0.33& $ 0.28^{+0.06}_{-0.02}$ &  $-0.11$  &   $ 0.42$  &    0.40 & 60\\
      11.11  &   $9.48^{+0.04}_{-0.03}$ &    9.21  &    9.68  &	 0.32& $ 0.36^{+0.05}_{-0.02}$ &  $-0.01$  &   $ 0.49$  &    0.34 & 72\\
      11.31  &   $9.60^{+0.06}_{-0.02}$ &    9.28  &    9.70  &	 0.30& $ 0.43^{+0.04}_{-0.02}$ &  $ 0.22$  &   $ 0.51$  &    0.29 & 51\\
      11.51  &   $9.66^{+0.05}_{-0.02}$ &    9.49  &    9.75  &	 0.29& $ 0.37^{+0.13}_{-0.03}$ &  $-0.11$  &   $ 0.49$  &    0.26 & 22\\
      11.71  &   $9.67^{+0.15}_{-0.03}$ &    9.40  &    9.72  &	 0.30& $ 0.47^{+0.10}_{-0.01}$ &  $ 0.29$  &   $ 0.49$  &    0.20 & 5\\
\noalign{\vspace{3pt}}
\hline
\end{tabular}
\tablefoot{(1) central value of stellar mass bin; (2-3-4) median (with its uncertainties) and percentiles of mean light-weighted age; (5) average 84-16 percentile range of log Age PDF; (6-7-8) median (with its uncertainties) and percentiles of mean light-weighted metallicity; (9) average 84-16 percentile range of log $Z_\ast$ PDF; (10) number of galaxies in stellar mass bin. Only bins with at least 5 galaxies are considered.}
\end{center}
\label{Tab:mean_relations}
\end{table*}

\begin{table*}
\centering
\caption{The relations fit to the median age and metallicity as a function of stellar mass for the \silver~and \golden~samples, using the functional form of Eq.\ref{eqn:sigmoid}, and shown in Fig.\ref{fig:scaling_legac}.}
\begin{tabular}{|l|l|c|c|c|c|}
\hline
\noalign{\vspace{3pt}}
parameter & sample  & $\bar{P}$ & A & B & $\bar{M_\ast} [M_\odot]$ \\
\hline
$\log<Age/yr>_r$ & LEGA-C \silver    & $9.62\pm0.04$  &   $0.3\pm0.1$  &  $2\pm1$ & $1.2\pm0.4 \cdot 10^{11}$ \\
                             & LEGA-C \golden & $9.66\pm0.02$  &   $0.30\pm0.02$  &   $2.4\pm0.3$ & $1.07\pm0.08\cdot 10^{11}$ \\
\hline
$\log<Z_\ast/Z_\odot>_r$ & LEGA-C \silver    & $0.37\pm0.04$ &  $0.8\pm1.4$ &   $1.6\pm1.3$ &  $1\pm2\cdot 10^{10}$ \\
                             & LEGA-C \golden & $0.46\pm0.01$ &   $1.0\pm1.7$ &    $1.2\pm0.6$ &  $0.9\pm2\cdot 10^{10}$ \\
\hline
\end{tabular}
\label{Tab:functional_fit_all}
\tablefoot{(1) characteristic parameter $\bar{P}$ at $10^{11.5}M_\odot$; (2) "A" parameter, which regulates the increase in age / metallicity over the inflection; (3) "B" parameter, which regulates the mass range of inflection; (4) characteristic mass $\bar{M_\ast}$ of the inflection point.}
\end{table*}

\subsection{Trends with stellar velocity dispersion}
\begin{figure*}[h!]
\centering
\includegraphics[width=\textwidth]{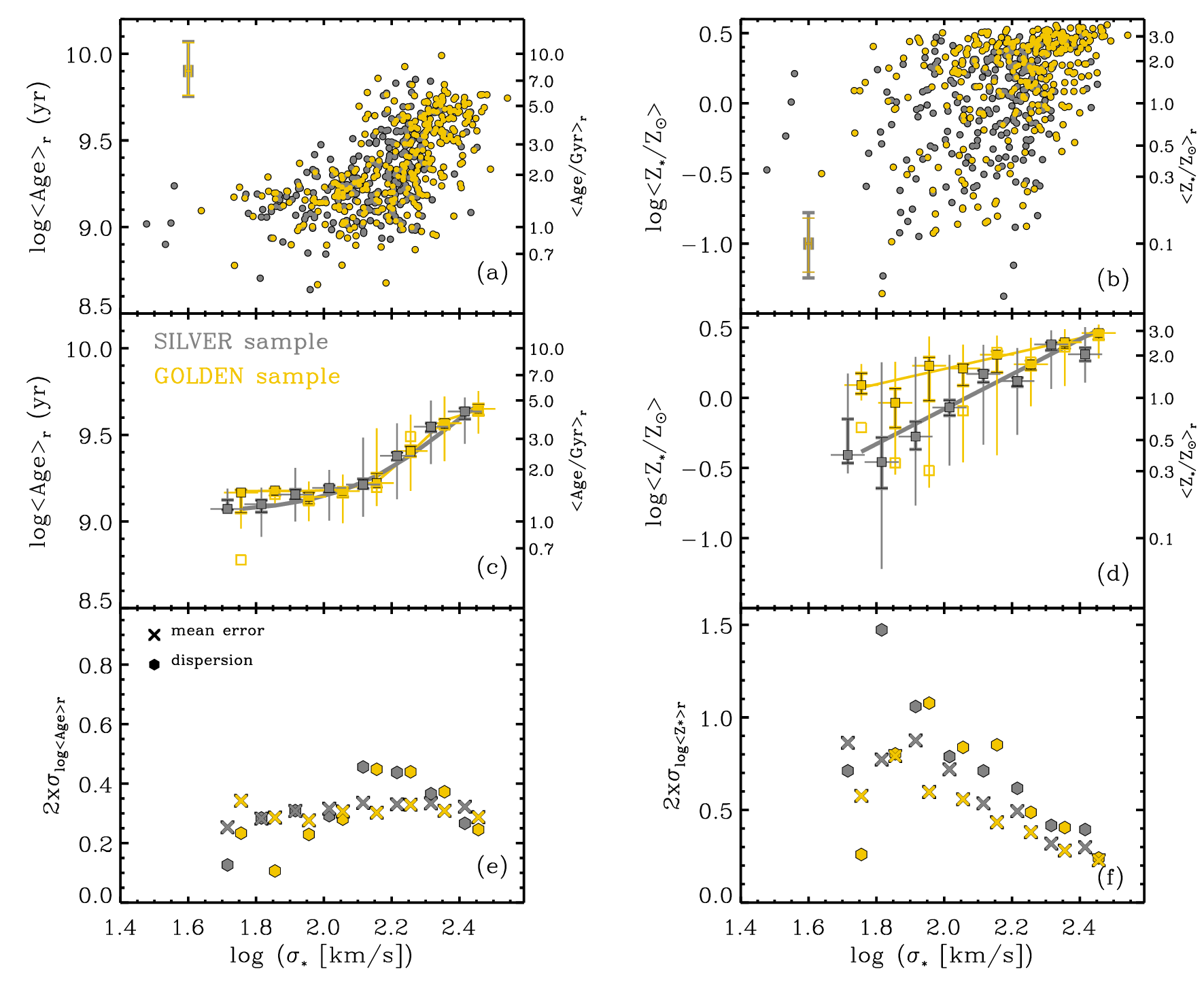}
\caption{Luminosity-weighted age ({\it left panels}) and luminosity-weighted stellar metallicity ({\it right panels}) versus stellar velocity dispersion for LEGA-C galaxies. Symbols and colors have the same meaning as in Fig.~\ref{fig:scaling_legac}. The lines in panel (c) show the bestfit $\rm age-\sigma_\ast$~relation using Eq.\ref{eqn:sigmoid}, while those in panel (d) show the bestfit linear $Z_\ast-\sigma_\ast$ relation using Eq.~\ref{eqn:linear}.}\label{fig:scaling_legac_sigma} 
\end{figure*}

In Fig.\ref{fig:scaling_legac_sigma} we explore how mean stellar ages and metallicities vary as a function of stellar velocity dispersion rather than stellar mass. The median trends and percentiles of the distributions are reported in Table~\ref{Tab:mean_relations_sigma}, while the functional fits are summarized in Table~\ref{Tab:fit_relations_sigma}. We find the same qualitative trends of increasing age and stellar metallicity with increasing $\sigma_\ast$, similar to those observed as a function of stellar mass. The median light-weighted age increases rapidly by 0.6~dex (from $\sim1$~to $\sim5$~Gyr) over 0.6~dex in velocity dispersion. The same increase in age occurs over 1.7~dex in stellar mass. The rapid increase appears to result from a sequence of young galaxies with ages $\sim1-2$~Gyr ($\log\left<Age/yr\right>=9.1-9.3$) spanning a relatively large range in $\sigma_\ast$, which transitions to a more compact distribution of old galaxies between $\log \sigma_\ast=2.1-2.3$. Across this transition regime the dispersion in light-weighted age is highest and larger than the typical uncertainty in age estimates. Above $\log \sigma_\ast>2.3$~instead galaxies are more uniformly old. 
For convenience, we fit the same sigmoidal function as in Eq.~\ref{eqn:sigmoid}, replacing $M_\ast$~with $\sigma_\ast$~and fixing the reference zero point at $\sigma_\ast=10^{2.4}$~km/s, which roughly corresponds to $M_\ast=10^{11.5}\,\mathrm{M_\odot}$ used as reference zero point for the relations with the stellar mass.
The fit to the weighted \silver~distribution provides a light-weighted age of $\rm log\left<Age\right>_r=9.59\pm0.03$ at $10^{2.4}$~km/s, with a transition at $\sigma_\ast=236\pm108$ km/s ($\rm \log \sigma_\ast=2.37_{-0.26}^{+0.16}$). The control \golden~sample confirms this trends with a more precise transition at $\sigma_\ast=184\pm4$ km/s (Table~\ref{Tab:fit_relations_sigma}).

The right-hand panels show that galaxies follow a rather tight sequence in stellar metallicity above $\log\sigma_\ast=2.3$. At lower velocity dispersions, the distribution in stellar metallicity broadens significantly. We measure a dispersion in stellar metallicity at fixed $\sigma_\ast$ larger than expected from the median uncertainty on metallicity estimates, both for the \silver~and \golden~sample, across almost the whole range in velocity dispersion. Unlike the trend with stellar mass, the increase in median stellar metallicity with $\sigma_\ast$ can be well described by a simple linear function:
\begin{equation}
    P = P_0 + \alpha \log(\sigma_\ast/10^{2.4} km/s) \label{eqn:linear} 
\end{equation}
The median stellar metallicity reaches the value of $\log\left<Z_\ast/Z_\odot\right>=0.42\pm0.03$ at $10^{2.4}$~km/s, and decreases with decreasing velocity dispersion with a slope of $1.2\pm0.2$. Unlike in the relation with stellar mass, we notice that the \golden~sample delivers a shallower slope for the $Z_\ast-\sigma_\ast$ relation. The discrepancy likely originates from the different ranking of high-S/N galaxies in stellar mass rather than in stellar velocity dispersion: at masses below $10^{10.8}M_\odot$ the scatter in the $M_\ast-\sigma_\ast$ relation increases and at fixed $\sigma_\ast$ lower-S/N galaxies tend to have lower $M_\ast$ than \golden~galaxies. Although the formal uncertainties are small, the slope of the relation could be subject to possible biases either in the \golden~sample due to incompleteness at lower $\sigma_\ast$ or in the \silver~sample due to larger uncertainties on stellar metallicity estimates. 
We tentatively apply statistical weight corrections to the medians of the \golden~sample (this is not done by default because such weights are very noisy due to the small number statistics of the \golden~sample). These are now shown as golden empty squares in Fig.\ref{fig:scaling_legac} and~\ref{fig:scaling_legac_sigma}. Despite the large scatter, the statistically corrected golden points are in agreement with the relation determined from the \silver~sample. This shows that the observed discrepancy in the distribution as a function of $\sigma_\ast$ between the uncorrected golden points and the \silver~sample is mainly due to the statistical incompleteness of the \golden~sample.

Comparing Fig.~\ref{fig:scaling_legac} with Fig.~\ref{fig:scaling_legac_sigma} it emerges that the dispersion in light-weighted age is higher at fixed $M_\ast$~than at fixed $\sigma_\ast$. In particular, the dispersion at fixed $\sigma_\ast$~ is comparable to the typical age uncertainties, and it is higher only across the transition from the young to the old sequence at $\sigma\sim10^{2.2} km/s$. This suggests that below and above the transition regime in $\sigma_\ast$, velocity dispersion is more predictive of the age of stellar populations, while stellar mass alone cannot predict the age of stellar populations. Stellar metallicity instead displays a similar dispersion with respect to the median uncertainties at fixed stellar mass (above $10^{10.7}M_\odot$) or at fixed velocity dispersion (above $\log\sigma_\ast=2.1$), suggesting that both parameters are equally predictive of the stellar metallicity in massive galaxies. In the lower-mass regime, the dispersion in metallicity is lower at fixed mass than at fixed sigma.

\begin{table*}
\caption{Median trends of light-weighted age and stellar metallicity as a function of stellar velocity dispersion for the \silver~and \golden~samples, as shown in Fig.\ref{fig:scaling_legac_sigma}.}
\begin{center}
\begin{tabular}{|c|ccc|c|ccc|c|c|}
\hline
\noalign{\vspace{3pt}}
\multicolumn{10}{|c|}{\silver~sample, weighted by ${\tt Tcor}\times {\tt w\_spec\_silver}$} \\
\hline
\noalign{\vspace{3pt}}
$\log(\sigma_\ast [km/s])$ & \multicolumn{3}{c|}{$\log<Age/yr>_r$} & $<2\times\sigma_{\log Age}>$  & \multicolumn{3}{c|}{$\log<Z_\ast/Z_\odot>_r$} &$<2\times\sigma_{\log Z_\ast}>$ & $\rm N_{bin}$\\
 & p50 & p16 & p84 &  & p50  & p16 & p84 & & \\
\hline
\noalign{\vspace{3pt}}
       1.76  &   $9.07^{+0.05}_{-0.00}$  &  9.06   &   9.19  &	0.25 & $-0.41^{+0.26}_{-0.06}$ &   $-0.54$  &   $0.17$  &  0.86  & 8 \\
       1.86  &   $9.10^{+0.02}_{-0.05}$  &  8.91   &   9.20  &	0.28 & $-0.46^{+0.17}_{-0.19}$ &   $-1.22$  &   $0.25$  &	0.77 & 26\\
       1.96  &   $9.16^{+0.03}_{-0.03}$  &  9.00   &   9.31  &	0.31 & $ -0.28^{+0.11}_{-0.09}$ &   $-0.77$  &   $ 0.29$  &	0.88 & 45\\
       2.06  &   $9.19^{+0.01}_{-0.03}$  &  9.01   &   9.30  &	0.32 & $ -0.07^{+0.05}_{-0.06}$ &   $-0.48$  &   $ 0.31$  &	0.72 & 82\\
       2.16  &   $9.21^{+0.03}_{-0.02}$  &  9.03   &   9.48  &	0.34 & $ 0.17^{+0.02}_{-0.06}$ &   $-0.33$  &   $ 0.38$  &	0.54 & 111\\
       2.26  &   $9.38^{+0.02}_{-0.02}$  &  9.13   &   9.57  &	0.33 & $ 0.12^{+0.02}_{-0.04}$ &   $-0.26$  &   $ 0.35$  &	0.49 & 158\\
       2.36  &   $9.55^{+0.02}_{-0.03}$  &  9.33   &   9.70  &	0.33 & $ 0.38^{+0.01}_{-0.04}$ &   $0.06$  &   $ 0.48$  &	0.32 & 89\\
       2.46  &   $9.64^{+0.02}_{-0.04}$  &  9.45   &   9.72  &	0.32 & $ 0.31^{+0.05}_{-0.05}$ &   $ 0.11$  &   $ 0.50$  &	0.30 & 28\\
\noalign{\vspace{3pt}}
\hline
\noalign{\vspace{3pt}}
\multicolumn{10}{|c|}{\golden~sample} \\
\hline
\noalign{\vspace{3pt}}
$\log(\sigma_\ast [km/s])$ & \multicolumn{3}{c|}{$\log<Age/yr>_r$} & $<2\times\sigma_{\log Age}>$  & \multicolumn{3}{c|}{$\log<Z_\ast/Z_\odot>_r$} &$<2\times\sigma_{\log Z_\ast}>$ & $\rm N_{bin}$\\
 & p50 & p16 & p84 &  & p50 & p16 & p84 & & \\
\hline
\noalign{\vspace{3pt}}
      1.76  &   $9.17^{+0.01}_{-0.12}$ &    8.96  &    9.19  &	 0.34& $0.09^{+0.08}_{-0.06}$ &  $-0.02$  &   $0.24$  &    0.58 & 5 \\
      1.86  &   $9.18^{+0.01}_{-0.03}$ &    9.10  &    9.21  &	 0.29& $-0.04^{+0.10}_{-0.18}$ &  $-0.55$  &   $ 0.26$  &    0.79 & 13\\
      1.96  &   $9.14^{+0.03}_{-0.04}$ &    9.00  &    9.23  &	 0.28& $ 0.23^{+0.06}_{-0.25}$ &  $-0.64$  &   $ 0.44$  &    0.60 & 19\\
      2.06  &   $9.18^{+0.02}_{-0.03}$ &    8.99  &    9.27  &	 0.31& $ 0.21^{+0.03}_{-0.12}$ &  $-0.46$  &   $ 0.38$  &    0.56 & 47\\
      2.16  &   $9.22^{+0.06}_{-0.02}$ &    9.09  &    9.54  &	 0.30& $ 0.31^{+0.02}_{-0.13}$ &  $-0.41$  &   $ 0.44$  &    0.43 & 50\\
      2.26  &   $9.41^{+0.03}_{-0.03}$ &    9.18  &    9.62  &	 0.33& $ 0.24^{+0.03}_{-0.04}$ &  $-0.06$  &   $ 0.43$  &    0.38 & 86\\
      2.36  &   $9.57^{+0.02}_{-0.03}$ &    9.35  &    9.72  &	 0.31& $ 0.40^{+0.01}_{-0.04}$ &  $ 0.08$  &   $ 0.49$  &    0.28 & 77\\
      2.46  &   $9.65^{+0.03}_{-0.04}$ &    9.51  &    9.75  &	 0.29& $ 0.46^{+0.02}_{-0.05}$ &  $0.28$  &   $ 0.52$  &    0.23 & 24\\
\noalign{\vspace{3pt}}
\hline
\end{tabular}
\end{center}
\tablefoot{(1) central value of stellar velocity dispersion bin; (2-3-4) median (with its uncertainties) and percentiles of mean light-weighted age; (5) average 84-16 percentile range of log Age PDF; (6-7-8) median (with its uncertainties) and percentiles of mean light-weighted metallicity; (9) average 84-16 percentile range of log $Z_\ast$ PDF; (10) number of galaxies in velocity dispersion bin. Only bins with at least 5 galaxies are considered.}
\label{Tab:mean_relations_sigma}
\end{table*}

\begin{table*}
\centering
\caption{The relations fit to the median age and metallicity as a function of stellar velocity dispersion for the \silver~and \golden~samples, and shown in Fig.\ref{fig:scaling_legac_sigma}.}
\begin{tabular}{|l|l|c|c|c|c|}
\hline
\noalign{\vspace{3pt}}
parameter & sample  & $\bar{P}$ & A & B & $\bar{\sigma_\ast} [km/s]$ \\
\hline
$\log<Age/yr>_r$ & LEGA-C \silver    & $9.59\pm0.03$  &   $0.5\pm0.3$  &  $3.0\pm1.2$ & $236\pm108$ \\
                             & LEGA-C \golden & $9.62\pm0.01$  &   $0.24\pm0.01$  &   $9.5\pm1.4$ & $184\pm4$ \\
\hline
\noalign{\vspace{3pt}}
parameter & sample  & \multicolumn{2}{c|}{$P_0$}& \multicolumn{2}{c|}{$\alpha$} \\
\hline
$\log<Z_\ast/Z_\odot>_r$ & LEGA-C \silver    & \multicolumn{2}{c|}{$0.42\pm0.03$} &  \multicolumn{2}{c|}{$1.2\pm0.2$}  \\
                             & LEGA-C \golden & \multicolumn{2}{c|}{$0.42\pm0.01$} &   \multicolumn{2}{c|}{$0.55\pm0.08$}  \\
\hline
\end{tabular}
\label{Tab:fit_relations_sigma}
\tablefoot{For light-weighted age we use Eq.\ref{eqn:sigmoid} and we report: (1) characteristic age at $10^{2.4}km/s$; (2) "A" parameter, which regulates the increase in age over the inflection; (3) "B" parameter, which regulates the $\sigma_\ast$ range of inflection; (4) characteristic $\sigma_\ast$ of the inflection point. For stellar metallicity we use the linear function in Eq.~\ref{eqn:linear} and we report: (1) the intercept $P_0$ at $\log\sigma_{\ast,0}=2.4$ km/s, and (2) the slope $\alpha$.}
\end{table*}

\section{Discussion}\label{sec:discussion}

\subsection{Downsizing and age bimodality}
We find a downsizing trend such that the light-weighted mean ages of more massive galaxies are older than those of lower-mass galaxies. 
The median relation between luminosity-weighted age and stellar mass displays a sigmoidal behavior which originates from the presence of an old and a young sequence of galaxies. The characteristic stellar mass across which the relation inflects (i.e. when it changes from being dominated by young galaxies to being dominated by old galaxies) is around $10^{11}M_\odot$. Such a non-linear, bimodal behavior of the age distribution in the general galaxy population has been observed by several studies also in the local Universe, both as a function of the global mass or surface mass density \citep[e.g.][]{kauffmann03b,Baldry04, gallazzi05, mateus06, Franx08, Williams10,Mattolini25} and as a function of the local surface brightness or stellar mass surface density \citep[e.g.][]{BdJ00,Rosa14,Zibetti17}.

\begin{figure}
     \centering
     \includegraphics[width=0.5\textwidth]{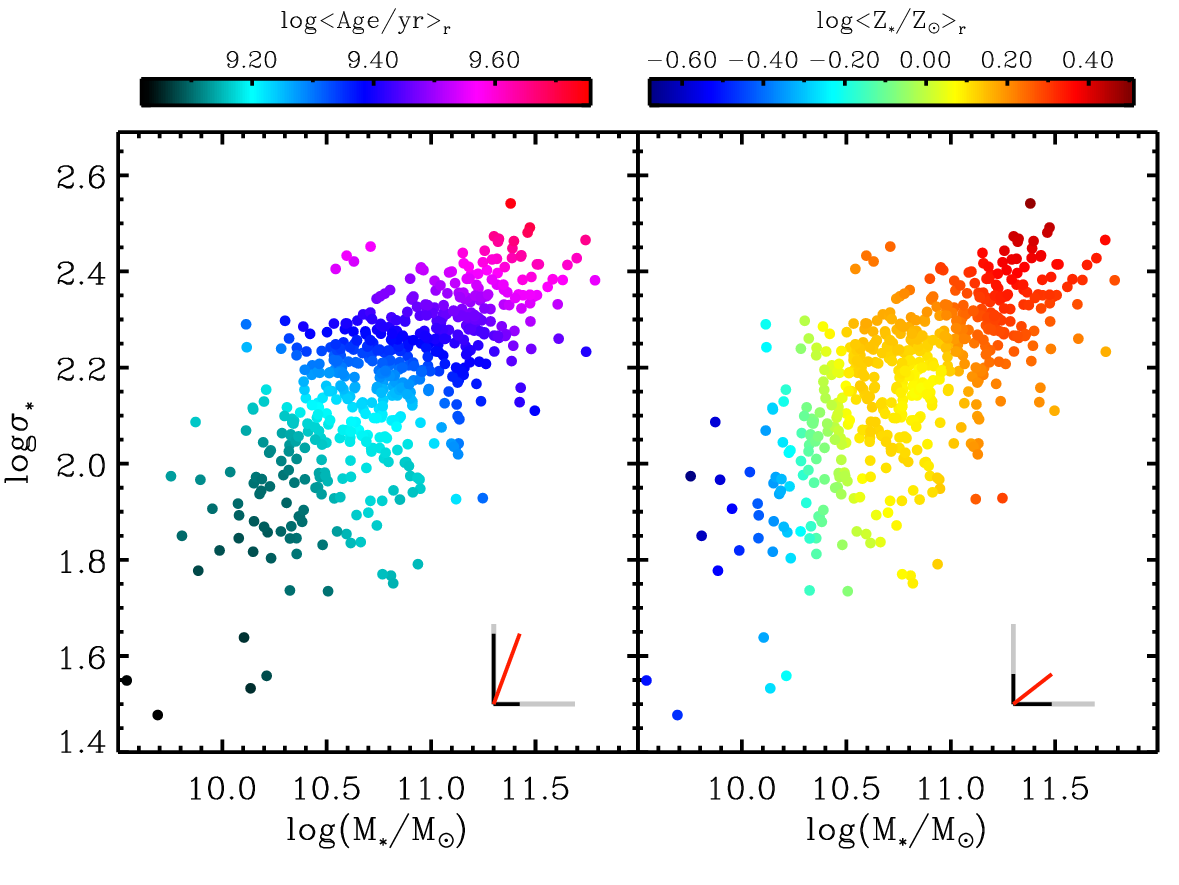}
     \caption{Stellar mass - velocity dispersion plane color coded by light-weighted age (left) and light-weighted stellar metallicity (right). For visualization purposes the age and metallicity are LOESS-smoothed. Vectors in the bottom right corner are proportional to the partial correlation coefficients, computed for the original (non-smoothed) physical parameters, following \cite{ScholzDiaz22,ScholzDiaz24}. The red line marks the direction of maximal increase in age/metallicity. The grey vectors refer to a correlation coefficient of 0.5 for comparison.}
     \label{fig:mstar_sigma}
 \end{figure}

 We observe a young and an old sequence also as a function of stellar velocity dispersion. The transition between the two regimes occurs around $\log \sigma_\ast=2.3$, more abruptly than the transition as a function of stellar mass, leading to a significant intrinsic dispersion in age at fixed $\sigma_\ast$~around the transition regime. This trend and the transition regime are very similar to those found for the full LEGA-C sample (thus extending to $z\sim1$) in \cite{Nersesian25} with ages estimated from {\tt Prospector} and in \cite{Cappellari23} with {\tt pPXF}. \cite{Cappellari23} discusses that the regime $\log\sigma_\ast\sim2.3$ corresponds to a `quenching boundary' such that the star formation histories of galaxies above this boundary have no young component ($\rm\lesssim1~Gyr$). 
 The smaller scatter in age at fixed velocity dispersion, below and above the transition regime, with respect to that at fixed stellar mass, as well as the sharper transition from the young to the old regime, suggest that stellar velocity dispersion (the depth of the potential well) is a better predictor of the main stellar formation epoch than stellar mass (the integral of the SFH). This is in line with the conclusion from other works in the local Universe \citep[e.g][]{vdW09,graves09a,Wake12,mcdermid15}. 
 Figure~\ref{fig:mstar_sigma} displays how light-weighted mean age maps in the stellar mass--velocity dispersion plane (left panel). A clear gradient of increasing age with increasing velocity dispersion is evident, especially for $\log \sigma_\ast>2.2$. Following \cite{ScholzDiaz24} we compute Spearman partial correlation coefficients to better quantify the dependence of age on mass and on $\sigma_\ast$ (black lines in Fig.~\ref{fig:mstar_sigma}). The coefficients indicate that light-weighted age primarily depends on $\sigma_\ast$~($\rho=0.44\pm0.06$) and secondarily on $M_\ast$~($\rho=0.16\pm0.08$) as also illustrated by the vector representation in Fig.\ref{fig:mstar_sigma}. \citep{Barone22} found a slightly stronger dependence of age on M/R (a proxy for the gravitational potential) than on M, for LEGA-C quiescent galaxies, although the correlations were weaker than what we find.
 
 We have checked that a similar sigmoidal behavior of age versus mass or velocity dispersion manifests also considering mass-weighted ages rather than light-weighted ones. A young and an old regime are still visible, with the young sequence shifted by 0.2~dex upward. However, the separation of the two sequences, hence the bimodality, is reduced, as well as the scatter in mass-weighted age at fixed mass. This suggests that both the main formation epoch (to which mass-weighted age  is most sensitive) and the duration of the star formation activity (which influences the difference between mass- and light-weighted ages) are a function of stellar mass, but a bimodality emerges mostly because of differences in the amount of the most recent star formation (affecting the light-weighted age).

 A downsizing trend, that manifests as an increase in age (or decrease in formation times) with increasing stellar mass or velocity dispersion, is also observed in \cite{kaushal24} from the analysis of the same LEGA-C sample, for the population as a whole with both {\tt Bagpipes} and {\tt Prospector}, but with a systematic difference between the two methods and SFH assumptions. \cite{Tacchella22} also describes two different regimes in the formation times for a sample of 160 massive galaxies in the redshift range $z=0.6-1$ from the HALO7D spectroscopic survey. They find uniformly old ages for masses above $2\cdot10^{11}M_\odot$, and a larger variation in both formation epoch and timescales at lower masses with a weaker mass dependence. Such a trend appears to extend beyond $z=1$ as found in \citep[][$1<z<1.6$]{Belli15} and in the VANDELS survey \citep[][$1<z<1.3$]{carnall19,Hamadouce22}. 

 \subsection{Chemical downsizing}
 We find a clear trend of increasing stellar metallicity with increasing galaxy mass, thus suggesting that higher-mass galaxies have been more efficient in reaching a higher degree of chemical enrichment. Lower-mass galaxies are still building up their metal content and/or have experienced a more prolonged history of gas flows altering the metal enrichment history. 
 
 Contrary to age, a bimodality is not apparent in the stellar metallicity-mass plane. The distribution in stellar metallicities is characterized by a concentration of galaxies in a high-metallicity sequence accompanied by an increasing scatter toward low stellar metallicities with decreasing stellar mass. This leads to a rather flat relation at masses $>10^{10.8}M_\odot$, and a gradual steepening below this regime.  
 This behavior resembles the one already observed in the local Universe for the general galaxy population \citep[][]{gallazzi05, panter08,Mattolini25} and for star-forming galaxies only \citep{Zahid17}. Interestingly, the same rapid increase at low masses and the flattening at higher masses is followed by the gas-phase metallicity, as observed at various redshifts \citep[e.g][]{Tremonti04,moustakas11,Curti20} and in the LEGA-C sample as well extending to higher masses than previous surveys at similar redshift \citep{Lewis24}. 
 It should be noted though that the stellar metallicity--mass relation is much more dispersed than the gas-phase relation. There are at least a couple of factors contributing to this difference: i) the stellar metallicity-mass relation displayed here pertains to the global population, while the gas-phase mass-metallicity relation only to star-forming galaxies; ii) the uncertainties on stellar metallicity become larger for lower-mass/younger galaxies, that are those for which gas-phase metallicities are better derived. The very broad similarity in the mass--metallicity relation for the stellar and the gas phase is non trivial, considering the different epochs of chemical enrichment that the two components trace. A combined analysis of the stellar and gas-phase metallicity scaling relations could constrain the efficiency of outflows in regulating star formation and metal enrichment timescales \citep[e.g.][]{Dalcanton07,Lu15}. However, systematic uncertainties in the different estimates of stellar and gas-phase metallicities should be treated with caution when making such comparison. 
 
 We find the median stellar metallicity to increase also as a function of stellar velocity dispersion. The trends we retrieve for the whole population are similar to those found by \cite{Cappellari23} with the FSPS or GALAXEV models and by \cite{Nersesian25}. Unlike with stellar mass, the median $\log Z_\ast$~follows a linear trend with $\log \sigma_\ast$~without an indication of a change in slope. This median linear trend results from an increasing scatter toward lower metallicities with decreasing $\sigma_\ast$. Such a distribution may suggest the existence of a mass-dependent lower limit in metallicity, referred to as MEtallicity-Mass-Exclusion zone in \cite{Bevacqua24}. In contrast to their work though, we observe such a behaviour for the global population rather than for quiescent galaxies only (as we show in Paper II).
 
 The statistics of our LEGA-C sample and the accuracy of the parameter estimates allow us to measure an intrinsic scatter in addition to the measurement uncertainties at least in the mass range $10^{10.5}-10^{11.5}M_\odot$ and in virtually the whole $\sigma_\ast$~range probed ($80-250~ \rm km/s$). This suggests that other parameters, such as the star formation activity or the galaxy structure, could regulate the metal enrichment. In Paper II we explore how the current star formation activity contributes to the scatter.
 
 In the right-hand panel of Fig.~\ref{fig:mstar_sigma} we show how the stellar metallicity varies in the $M_\ast-\sigma_\ast$~plane. A gradient in stellar metallicity almost parallel to the relation is evident, becoming more aligned with the mass axis below $\sim10^{10.8} M_\odot$. The Spearman partial correlation analysis indicates that stellar metallicity has a similarly strong dependence on stellar mass at fixed $\sigma_\ast$~as with velocity dispersion at fixed $M_\ast$ (partial correlation coefficients of $0.24\pm0.07$ and $0.19\pm0.08$ respectively, see also the vector representation in Fig.\ref{fig:mstar_sigma}). This suggests that both the stellar mass and the velocity dispersion determine the degree of metal enrichment. This partially agrees with what found in \citep{Barone18,Barone22} analysing the mass-size plane, although they retrieved a stronger dependence on M/R (which traces the depth of the potential well). Our finding reinforces the idea that the metallicity is connected with the ability of a galaxy to retain metals, hence the global and local potential well, modulated by galaxy mass \citep[e.g.][]{Zibetti22}.

\section{Summary}\label{sec:summary}
We have analysed the LEGA-C spectroscopic data to derive stellar population parameters and their scaling relations for volume-representative galaxy samples at redshift $0.6<z<0.77$. We use our {\tt BaStA} fitting code \citep{gallazzi05,Zibetti17} to interpret, in a Bayesian framework, optimally selected sets of stellar absorption features, in combination with rest-frame optical photometry, with a library of spectral models based on complex star formation and metal enrichment histories, and dust attenuations. We select a \silver~subset of the LEGA-C sample, based on the availability of key absorption features sensitive to both age and metallicity, comprising 552 unique galaxies, 323 of which have $S/N>20$~and constitute our \golden~sample. This work extends both in modeling assumptions, sample size and parameters coverage, the analysis we performed in \cite{gallazzi14}.

We analyse various sources of systematics that can affect the physical parameter estimates within our framework (Appendix~\ref{sec:appendix_basta}), as well as in comparison to other spectral fitting methods performed on the LEGA-C sample (Appendix~\ref{Appendix:compare_parameters}). We find that stellar metallicities are intrinsically more uncertain for younger/less-massive galaxies owing to larger degeneracies in the underlying SFHs. Deep spectroscopy and complex modeling of key absorption features remain necessary to push metallicity studies to the lower-mass star-forming population. Assumptions on SFHs are the primary drivers of systematic differences in inferred ages, while the choice of SPS models and the treatment of chemical enrichment histories have a significant impact on the normalization of stellar metallicities. We find that parameter estimates on an individual galaxy basis are subject to significant scatter between different estimates. The uncertainties retrieved with {\tt BaStA} reflect to a large extent this scatter and are thus representative of the underlying parameter degeneracies. This stresses the importance of applying consistent analysis when working on different datasets. Nevertheless, we find that the statistical trends obtained in this work are robust against different assumptions, but differ in the detailed shape (see Appendix~\ref{Appendix:compare_parameters}).

We provide the catalog of stellar population physical parameters derived in this work with our {\tt BaStA} code for the whole LEGA-C DR3, together with those estimated in previous works from the LEGA-C team using {\tt Prospector} \citep{Nersesian25} and {\tt Bagpipes} \citep{kaushal24}. We also provide a catalog of revised absorption index measurements for both individual spectra and duplicate-combined observations.

With the well-defined \silver~and \golden~samples, we characterize the relations of luminosity-weighted age and stellar metallicity as a function of stellar mass and velocity dispersion for the general $\left<z\right>=0.7$ galaxy population at masses $>10^{10} M_\odot$, by accounting for volume and survey completeness.
We find that the archaeological downsizing trends of increasing age and increasing stellar metallicity with increasing mass, known in the local Universe, were already in place when the Universe was 7~Gyr old. Similar trends in stellar age and metallicity are observed as a function of stellar velocity dispersion. A comparison of the scatter in the relations, as well as of the gradient in age and metallicity in the $M_\ast-\sigma_\ast$~plane, suggests that age is primarily predicted by stellar velocity dispersion being below or above the characteristic transition regime of $\log\sigma_\ast\sim2.3$, while both the stellar mass (the integral of the SFH) and the velocity dispersion (the depth of the potential well) determine the degree of chemical enrichment of galaxies' stellar populations. In Paper II we quantify how these relations differentiate for quiescent and star-forming galaxies, how they evolve from $z=0.7$ to $z=0.1$, and discuss the implications for the population evolution. The shape and scatter in these relations at intermediate redshifts can provide important constraints to galaxy evolution models in cosmological context and on the implementation of feedback and quenching mechanisms. This type of studies represents a benchmark for the archaeological analysis of galaxies over a continuous span of cosmic time at $z<2$, as will be obtained from WEAVE \citep{weave-steps}, 4MOST \citep{4most-steps, 4most-waves}, MOONS \citep{moonrise} surveys for large representative galaxy samples, as well as for increasing numbers of massive quiescent galaxies at even higher redshift from JWST \citep[e.g.][]{DeepDive_prop,Slob24,Carnall24}.

\section*{Acknowledgements}
We thank the referee for their suggestions to improve the clarity of the results and discussion.
We are grateful to Stephane Charlot and Gustavo Bruzual for assistance with the CB19 SPS models.\\ ARG and LSD acknowledge support from the INAF-Minigrant-2022 "LEGA-C" 1.05.12.04.01. SZ acknowledges support from the INAF-Minigrant-2023 "Enabling the study of galaxy evolution through unresolved stellar population analysis" 1.05.23.04.01. PFW acknowledges funding through the National Science and Technology Council grants 113-2112-M-002-027-MY2. LSD is supported by the "Prometeus" project PID2021-123313NA-I00 of
MICIN/AEI/10.13039/501100011033/FEDER, UE. This paper and related research have been conducted during and with the support of the Italian national inter-university PhD programme in Space Science and Technology.

\bibliographystyle{aa}
\bibliography{paper}

\begin{appendix}
    \onecolumn
\section{New catalog of absorption indices and quality of measurements}\label{Appendix:index_measures}
After the DR3 was released we discovered that a fraction of $\sim15$\% of LEGA-C spectra did not receive a measurement of the absorption indices, despite the good quality of the spectrum and the lack of artifacts. This was tracked down to be related to a silent bug in the run of {\tt\string pPXF} associated with the rebinning of the spectra. We thus re-processed the emission-line subtracted spectra for cleaning and index measurement. We then measure absorption indices with our routine {\tt\string BaStA\_index} and provide the revised catalog of index measurements. Measurements are provided for all the LEGA-C spectra, but we consider the measurement reliable only if less than 1/3 of the pixels in any of the central or side bands are flagged. 
We thus recover index measurements for the $\sim15$\% of LEGA-C spectra that were affected by the bug. The exact fraction varies from index to index because of the varying wavelength coverage. The revised index measurements are fully consistent with those previously published in the DR3 catalog for those galaxies that already had a measurement, with a scatter well within the measurement uncertainties. We notice that the index measurement errors are instead systematically lower by a factor of 1.3 with respect to those already published, due to an erroneous scaling of the noise spectrum during index measurement in the previous release.

\begin{figure}
\centering
\includegraphics[width=\hsize]{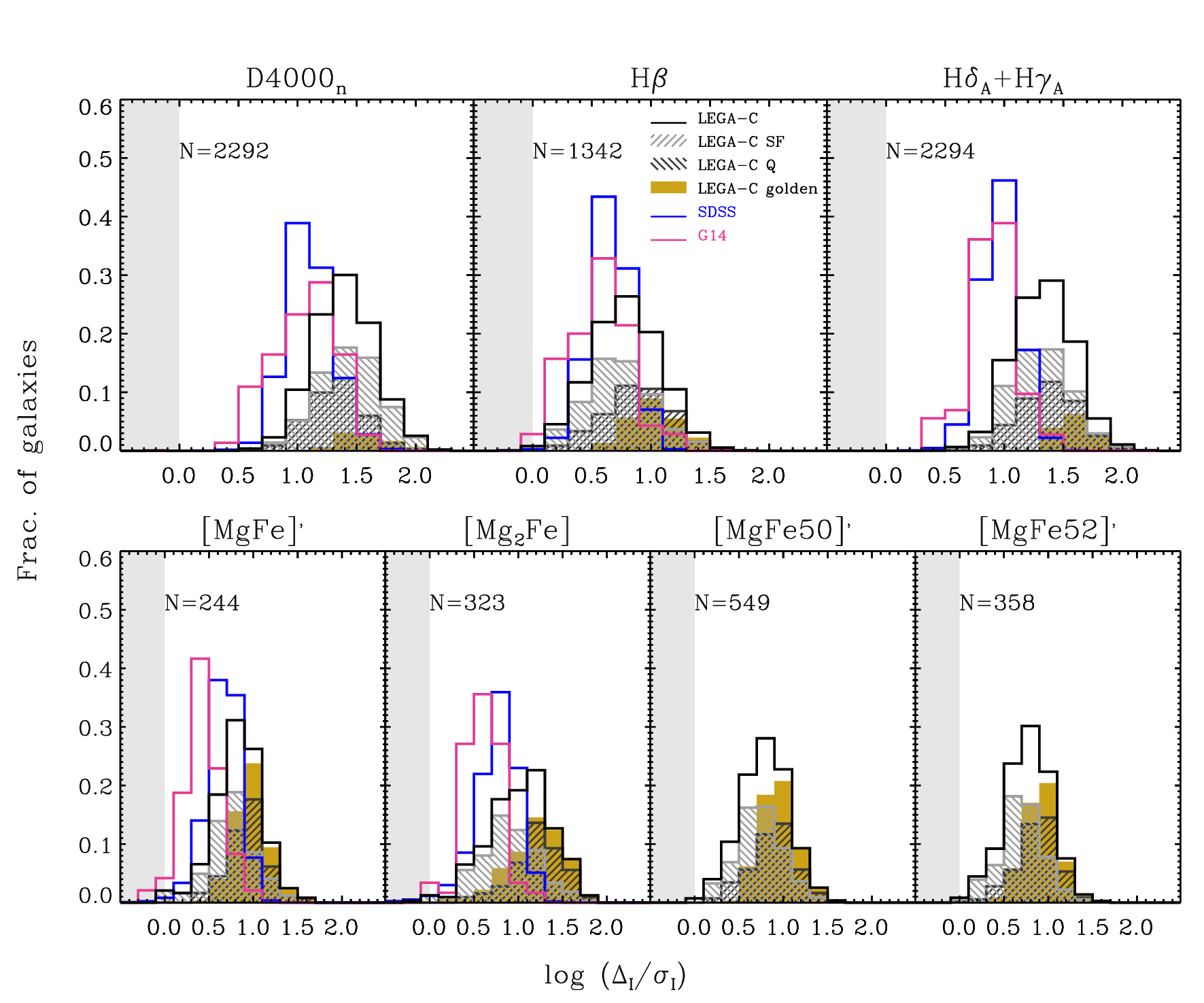}
\caption{Resolving power of the absorption indices used in this work. The resolving power is defined as the index dynamic range of the observed sample divided by the observational error on each index measure. The solid black histograms include all the galaxies with a valid measurement of the given absorption index among the 2588 unique galaxies in the LEGA-C primary sample with a measurement of stellar velocity dispersion. Each histogram is normalized by the total number of galaxies used, $N$, which is reported in each panel. The dark and light grey hatched histograms refer to the quiescent and star-forming LEGA-C galaxies, respectively. The golden histograms refer to the \golden~subsample used in this work. All these histograms are normalized by $N$. The distributions for LEGA-C galaxies are compared, when available, to those of SDSS DR7 (blue) and of G14 (magenta) samples, each normalized by their total number of galaxies.}
\label{fig:res_power}
\end{figure}

The quality of the index measurements can be expressed in terms of "resolving power", defined as the $5^{th}-95^{th}$ inter-percentile range of the index strength in the data divided by the individual observational error. In Fig.\ref{fig:res_power} we show the resolving power of the key absorption features used in this work for the LEGA-C sample analysed. For comparison the distributions of resolving power for SDSS DR7 galaxies and for the z=0.7 \cite{gallazzi14} sample are shown\footnote{except for the indices \mgfef~and \mgfeft~that were not measured for the SDSS DR7 and G14 sample.}.

We notice that the quality of index measurements in our LEGA-C sample is typically higher than that of SDSS DR7 for most indices, and higher than in G14 across the board. Among the Balmer indices, \hb~has lower resolving power because of the larger uncertainties in corrections for emission line infill. Among the metal-sensitive indices, \mgtwofe~has higher resolving power because of the use of more spectral information. We also notice that, with the exception of \dn, the resolving power is higher for quiescent galaxies than for star-forming ones.

As a further quality check of the index measurement and their fit, in Fig.\ref{fig:indx_obs_bf_age} and Fig.\ref{fig:indx_obs_bf_met}, we compare the value of the observed absorption index with that of the best-fit model for the indices considered in the {\tt BaStA} fit. The agreement is generally very good, with negligible mean offsets between observed and predicted index strengths, and scatter consistent with or only slightly larger than the mean observational uncertainties. The largest deviations are typically associated to galaxies for which the index has not been used in the fit. We notice small systematic offsets for \golden~quiescent galaxies for which the models tend to predict weaker \dn~and stronger \hb. The best-fit model reproduces well, with scatter consistent with observational errors, the other absorption indices that are not used in the fit. The exceptions are Mg indices and red Fe indices for quiescent galaxies, a possible indication of $\alpha$-enhancment which is not included in the modeling. We checked that these offsets are present to an equal or slightly larger extent for models based on exponential SFH and BC03 models.

We release the catalog of revised index measurements for the whole LEGA-C DR3, both measured from individual emission-line subtracted spectra and obtained by combining duplicate observations when available as described in Sec.~\ref{sec:data}. We provide measurements for all the Lick indices, \dn~and the composite Mg-Fe indices, together with their errors, resolving power, warning flags, as well as flags for duplicate treatment.

\section{Comparison of parameters from duplicate observations}\label{Appendix:duplicates}
We take advantage of duplicate observations in LEGA-C to test the reliability of the uncertianties on stellar population parameters.
In order to quantify how much the observational set-up can affect our stellar population parameter estimates, we compare the parameters derived from duplicate observations of the same galaxies, in a similar way as it has been done for observational quantities such as absorption indices. We consider galaxies belonging to the {\it bronze} sample and that have one or two duplicate observations. The difference in luminosity-weighted ages between duplicates has a standard deviation of $0.18$~dex to be compared with the mean uncertainty on individual age estimates of $0.16$~dex. The difference in luminosity-weighted stellar metallicity has a standard deviation of $0.33$~dex which is fully consistent with the mean uncertainty on individual metallicity estimates for star-forming galaxies. We note that duplicate observations can cover slightly different wavelength ranges and sample a different portion of the galaxy light. Moreover, duplicate observations would differ also because of uncertainties on the stellar continuum modeling and decoupling of emission lines. This comparison shows that these effects do not impact significantly the parameter estimates of individual galaxies, whose uncertainties are well represented by those estimated with {\tt BaStA}.

\begin{figure}
\centering
\includegraphics[width=0.45\textwidth]{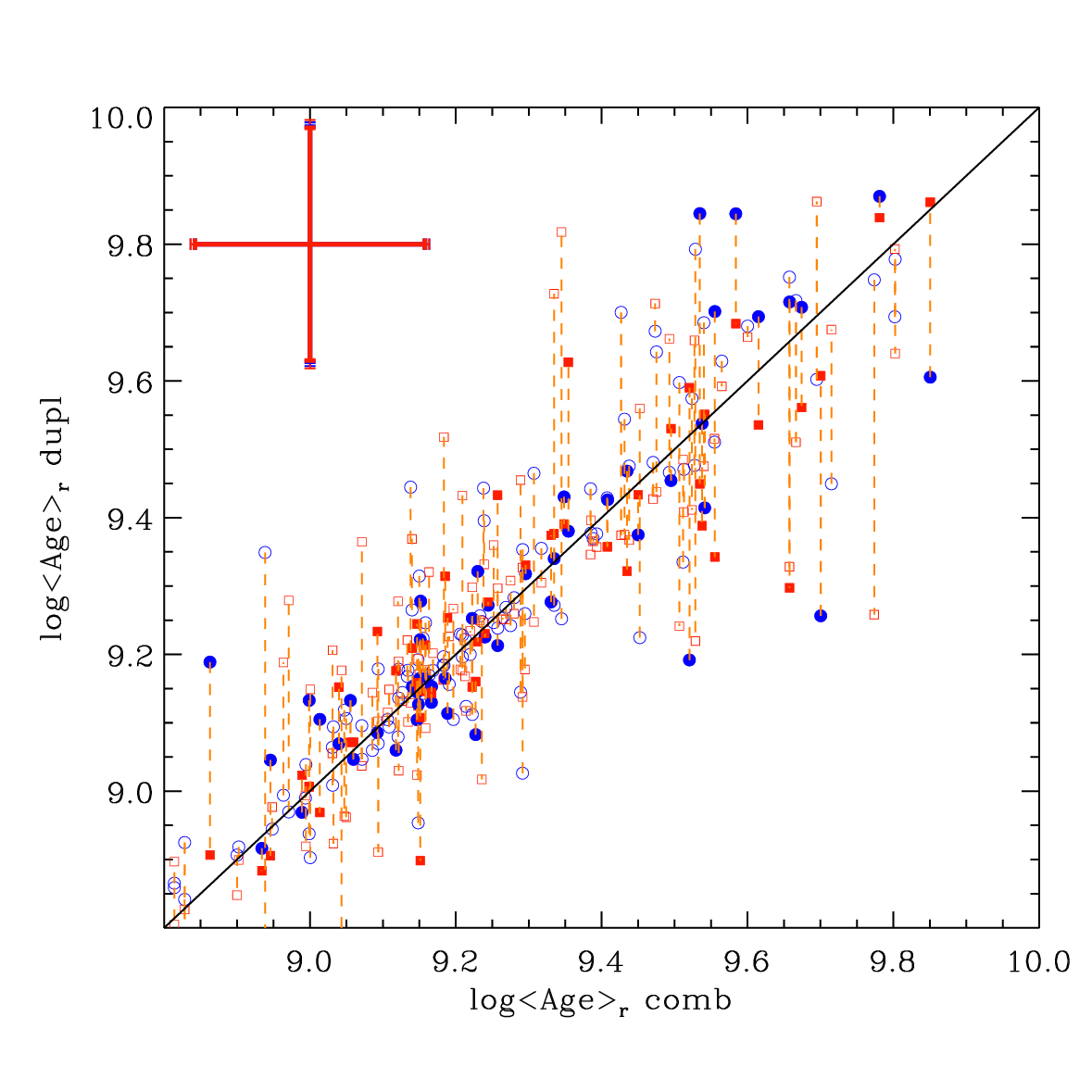}
\includegraphics[width=0.45\textwidth]{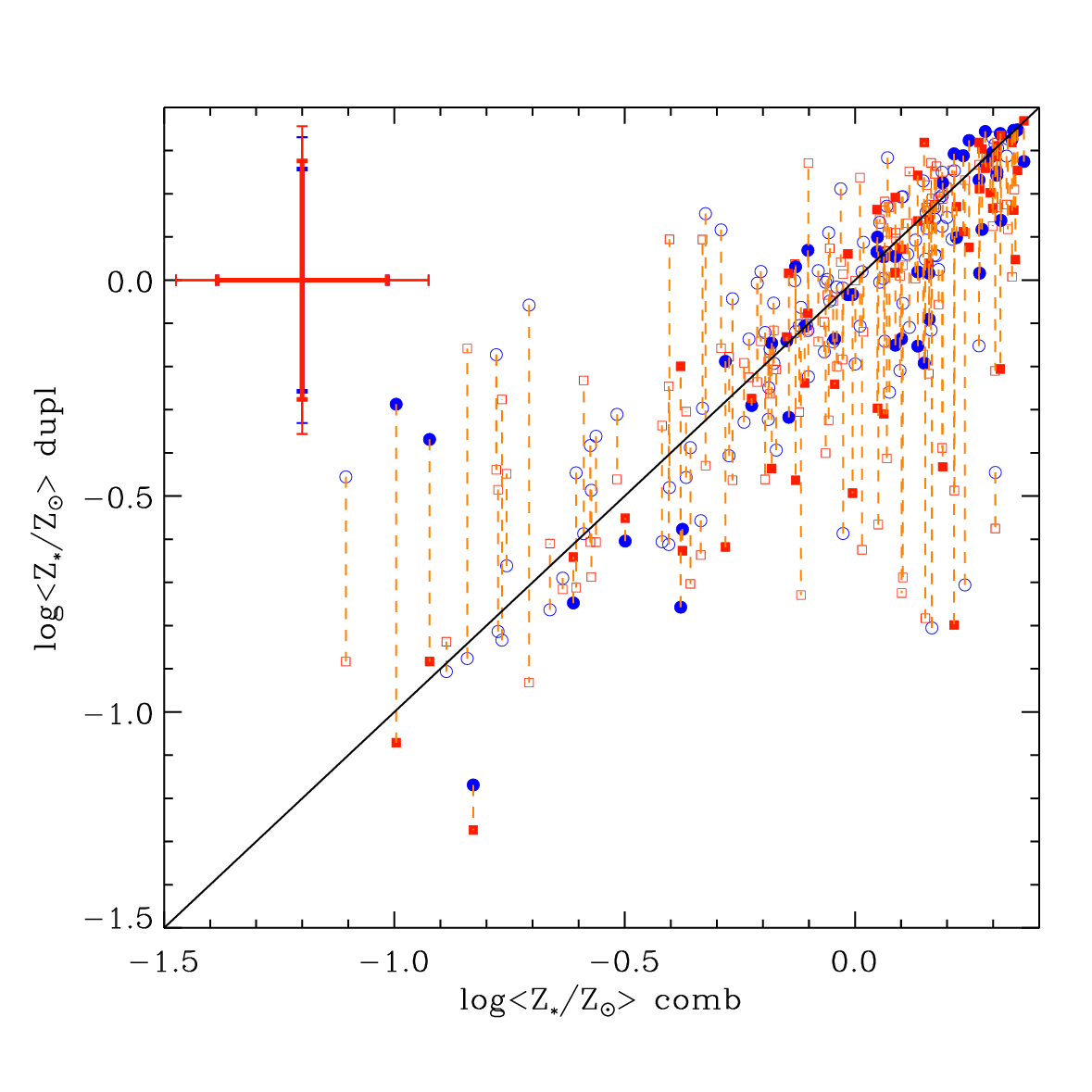}
\caption{Comparison of light-weighted ages (left) and light-weighted stellar metallicities (right) obtained for individual duplicate observations (y-axis; duplicate measurements for each galaxy are connected with a dashed line) with those obtained from the duplicate-combined indices (x-axis). Filled symbols refer to those galaxies whose fit is constrained from the full set of 5 optimal indices (\dn, \hb, \hdg, \mgtwofe, \mgfep) when duplicate observations are combined. The errorbars indicate the mean uncertainties in each estimate, with the thicker errorbars referring to estimates based on the 5 optimal indices. }
\label{fig:param_dupl_comp}
\end{figure}

As explained in Sec.\ref{Sec:sample_definition}, in order to increase the sample of galaxies with spectroscopic coverage of the key absorption features, we combine duplicate observations when available. Fig.\ref{fig:param_dupl_comp} compares the parameter estimates obtained from the combined observations (i.e. using weighted mean absorption indices) with those obtained on the individual observations for the same object. This figure shows a good correlation between the two sets of estimates, with those based on the combined duplicate observations lying typically in between those based on individual observations. The mean uncertainties on parameters based on the combined observations are smaller than those based on the individual observations (0.16~dex compared to 0.18~dex for age, 0.27~dex compared to 0.34~dex for metallicity). The availability of a larger number of indices, in particular the 5 optimal indices, has a larger impact on stellar metallicity estimates, with mean uncertainties reducing to 0.18~dex. This comparison reassures us that our procedure of combining duplicate observations is reliable and effective.

\section{Impact of dust modeling}\label{sec:appendix_model_prior}
\begin{figure}
\begin{center}
\includegraphics[width=0.5\textwidth]{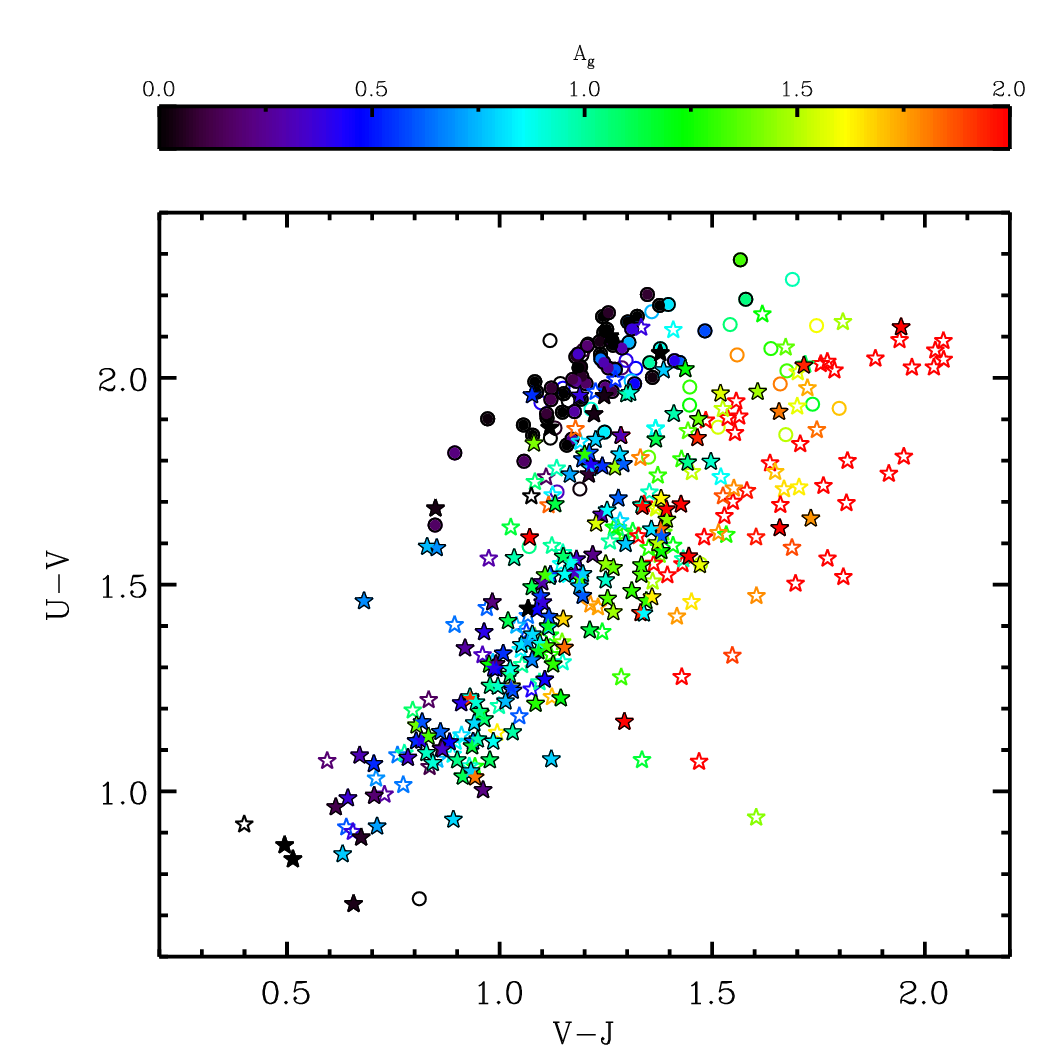}
\caption{Distribution in rest-frame U-V, V-J color for the galaxies in the LEGA-C sample analysed in this work. Filled/empty symbols refer to the \golden/\silver~sample, while circles/stars distinguish quiescent/star-forming galaxies. The color coding indicate the dust attenuation $\rm A_g$ estimated from our Bayesian fitting of spectral indices and $rizYJ$~photometry.}
\label{Fig:UVJ_dust}
\end{center}
\end{figure}

\begin{figure}
\begin{center}
\includegraphics[width=0.5\textwidth]{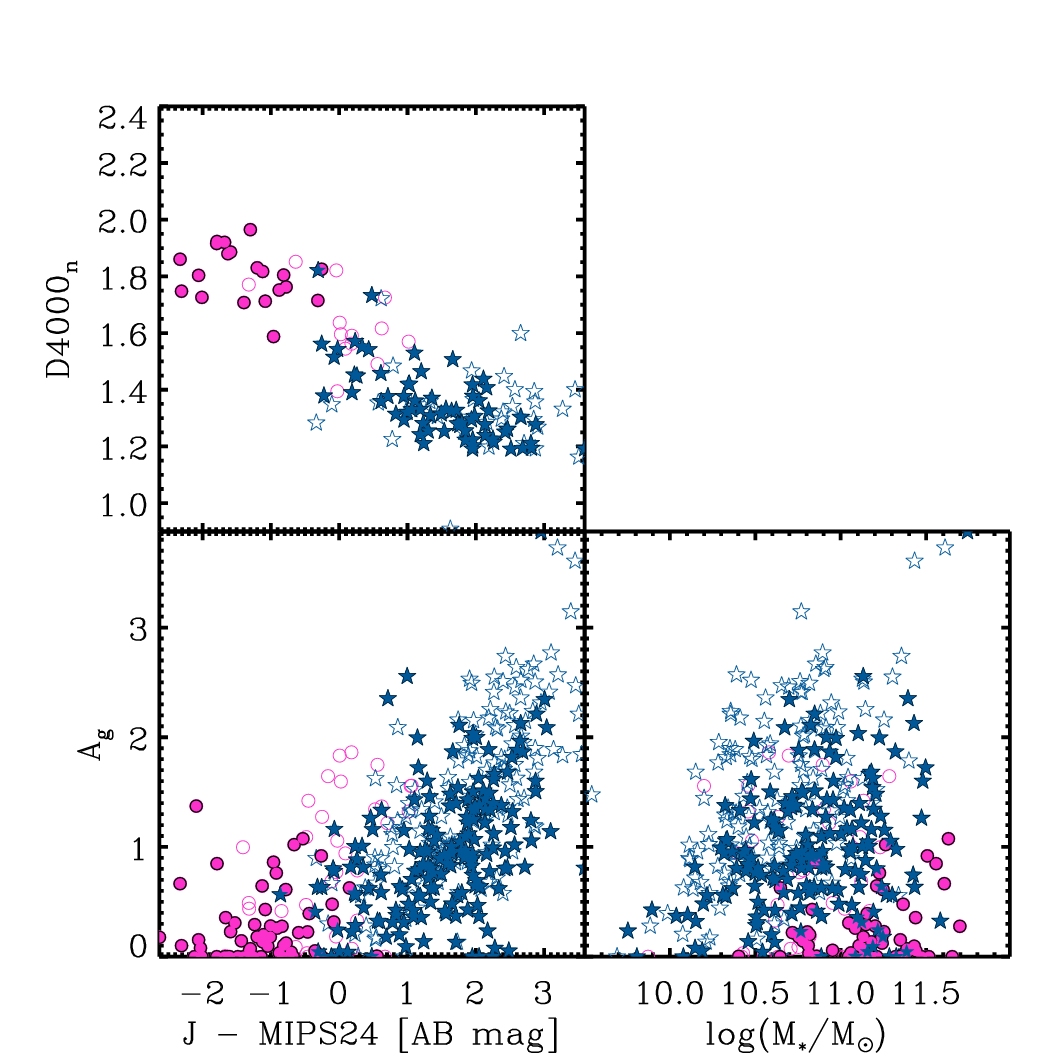}
\caption{Upper panel: 4000\AA-break versus $J-24$\micron~color; lower left panel: estimated dust attenuation $A_g$ versus $J-24$\micron~color; lower right panel: estimated $A_g$ versus stellar mass. Quiescent galaxies are indicated with magenta circles, star-forming galaxies with blue star symbols. Filled/empty symbols refer to the \golden/\silver~subsamples. This figure and Fig.\ref{Fig:UVJ_dust} show that the estimated dust attenuation correlates well with expectations based on the location of galaxies in the UVJ diagram and on their 24\micron~flux normalized to their optical emission (as traced by the observed J-band magnitude). In particular, quiescent galaxies are characterized by low dust attenuation as expected. This reassures that the treatment of dust does not significantly bias our estimates of stellar population parameters.}
\label{Fig:Ag_mips}
\end{center}
\end{figure}

We allow for a large flexibility of dust attenuation in our model library, setting a broad prior on the dust parameters, with the purpose of avoiding the fit to be driven by a too limited range in dust attenuation while at the same time being able to constrain dust and stellar mass as well.
A concern may be that this large range of dust attenuations in the model library may bias the estimates for quiescent galaxies to larger attenuations and consequently to lower ages/metallicities. We have checked that the combination of indices and photometry is able to isolate old from dusty models and reduce the degeneracy between age/metallicity and dust. In particular, we have checked that the apparently young ages of quiescent galaxies are not driven by an overestimation of the dust attenuation and the age-dust degeneracy. Indeed, the estimated dust attenuations for LEGA-C galaxies, based on indices and photometry, correlate well with the expectations based on the location in the $UVJ$ diagram (Fig.~\ref{Fig:UVJ_dust}) and on the 24\micron~emission (Fig.~\ref{Fig:Ag_mips}). 
We have repeated the fit using only models with a dust attenuation $A_g<0.2$~mag, to quantify the impact on the stellar populations of quiescent galaxies: the metallicity of quiescent galaxies would be systematically lower by only 0.05 dex on average with scatter consistent with the uncertainties; the majority of Q galaxies have ages consistent between the two fits, with a median/mean offset of 0.07/0.1~dex, while only 8\% of the quiescent galaxies, mainly of $S/N<20$, have ages older by more than 0.3~dex (which corresponds to $2\sigma$) in the case of low dust. Moreover, using only absorption indices as constraints to models with $A_g<0.2$~mag gives stellar metallicity and light-weighted age estimates for quiescent galaxies consistent with those obtained using both indices and photometry as constraints to the full suite of (dust-attenuated) models, with negligible bias and with scatter lower than the uncertainties. This reinforces the confidence on the ability of our approach in separating age-metallicity-dust effects.

\section{Comparison of parameter estimates based on different assumptions in the {\tt BaStA} method}\label{sec:appendix_basta}
Here we quantify the systematics introduced by different modelling assumptions between the model library adopted in this work and the one used in our previous works, with which the LEGA-C results are compared \citep{gallazzi05,gallazzi14}. 
The model library adopted in this work and the one adopted in our previous series of works differ under several aspects: 1) the SPS code and stellar library adopted as base models (CB19 with MILES instead of BC03 with Stelib stellar library), 2) the star formation histories (delayed gaussian instead of exponentially declining for the smooth component, and different parametrization of the bursts), 3) the chemical enrichment history (metallicity increasing with the mass formed instead of metallicity constant along the SFH). These changes add more realistic complexity in the interpretation of galaxy spectra. We quantify the contribution of each ingredient to the overall difference between the estimated parameters in the LEGA-C sample. A full description of the systematic differences in the SDSS analysis is presented in \cite{Mattolini25}. This is not meant to be a comprehensive analysis of the systematics associated to different models, but to provide both a quantitative assessment of the differences between our previous results and a sense of the impact of different ingredients.

We focus on the \silver~sample and perform the indices+photometry BaStA fit as described in Sec.~\ref{sec:stelpop_method} adopting model libraries that differ from the default model library used in this work by one ingredient at a time among the three mentioned above. Specifically, we consider: i) a library that differs from the default one in the base SPS models used, namely the original BC03 SSPs instead of the CB19, which differ both in the stellar library (Stelib vs MILES) and in the evolutionary tracks (Padova94 vs Parsec) - for this we consider also an intermediate step and check the difference between CB16 (Padova94) and CB19 (Parsec) both based on the MILES stellar library; ii) a library that differs from the default one in the assumed shape of the continuous SFH, namely an exponentially-declining form instead of a delayed-gaussian model; iii) a library in which the stellar metallicity is kept constant along the SFH instead of being a function of the stellar mass formed.

We find that:
\begin{itemize}
    \item The latest version of the SPS models CB19 tends to result in younger ages, with a larger spread toward younger ages for quiescent galaxies around $\log(Age/yr)=9.5$, in higher metallicities with an offset roughly constant with metallicity below $\log(Z/Z_\odot)=0.2$, in slightly larger dust attenuations and in higher stellar masses. These differences in age and metallicity are mostly associated to the change from Stelib to MILES stellar library, with milder (or counteracting) effects associated to the evolutionary tracks, at least in the range of parameters of LEGA-C galaxies.
    \item Assuming a delayed-gaussian SFH instead of an exponentially declining one impacts mostly on the light-weighted age, leading to younger values for galaxies with intermediate ages $\log(Age/yr)=9-9.5$, but also leads to $<0.1$~dex higher stellar metallicities. This is driven by the intrinsic shift to later epochs of the peak of the SFH. 
    \item Introducing a chemical evolution history, rather than having the metallicity constant, impacts mostly the stellar metallicity estimates with a roughly constant offset of less than 0.1~dex, leaving the other parameters almost unaffected. This is understood by the fact that at fixed mass-weighted average metallicity, a SFH with increasing metallicity produces a weaker absorption index with respect to the case of a fixed metallicity. At fixed index strength, the metallicity is thus lower if interpreted with a constant metallicity. 
\end{itemize}
 In summary, we find an overall systematic difference in stellar metallicity of 0.28/0.34~dex for Q/SF galaxies associated in large part to the change in SPS models, followed with equal share by the change in SFH and chemical enrichment history assumptions. Comparing the fits with the new set-up with respect to the previous one, we find that the overall systematic difference in light-weighted age is $-0.18/-0.13$~dex for Q/SF galaxies, mostly driven by changes in SFH assumptions, followed by the change in SPS models. We notice that while the systematic differences in mass and dust attenuation are lower than their typical uncertainty and with scatter comparable to the uncertainties, the offsets in age and metallicity are comparable to or larger than their typical uncertainty. A simple comparison of the best-fit $\chi^2$ indicates that the new modeling set-up provides a better quality fits to the data (0.6/0.5 with respect to 1/0.7 on average for Q/SF galaxies). The results are summarized in Table~\ref{Tab:compare_old_new}.

\begin{table}
\caption{Systematic differences in physical parameter estimates due to different modeling assumptions within our {\tt BaStA} code.}
\begin{center}
\begin{tabular}{|l|l|cc|cc|cc|cc|cc|c|}
\hline
 & & \multicolumn{2}{c|}{CB19 - CB16} & \multicolumn{2}{c|}{CB16 - BC03} & \multicolumn{2}{c|}{Sandage - Exponential} & \multicolumn{2}{c|}{var Z - fix Z} & \multicolumn{2}{c|}{overall new - old} & \\
  & & $<\Delta>$ & rms & $<\Delta>$ & rms  & $<\Delta>$ & rms  & $<\Delta>$ & rms   & $<\Delta>$ & rms  & $<error>$ \\
\hline
$\log(Z_\ast/Z_\odot)$ &SF &0.01 & 0.14  & 0.16 &    0.14  & $0.08$ & 0.13 & 0.09 & 0.05 & $0.34$ & 0.21 & 0.34 \\
                                     &Q  &0.01 & 0.05   & 0.12 &    0.10  & $0.09$ & 0.13  & 0.07 & 0.04 & $0.29$ & 0.16 & 0.17\\
\hline
$\log(Age_r/yr)$           &SF  & 0.04 & 0.06  & $-0.05$ &   0.13  & $-0.11$ & 0.20 & 0.00 & 0.02 & $-0.13$ & 0.13 & 0.16 \\ 
                                     &Q    & 0.05 & 0.06 & $-0.12$ &   0.09  & $-0.13$ & 0.10 &  0.01 & 0.02 &$-0.18$ & 0.12 & 0.16 \\
\hline
$\log(M_\ast/M_\odot)$ &SF  & 0.06 & 0.08  &0.07 &     0.04 & $-0.06$ & 0.13 & $-0.03$ & 0.03 & $0.03$ & 0.09 & 0.13 \\
                                      &Q    & 0.05 & 0.05 &0.04 &     0.07 & $-0.10$ & 0.07 & 0.00 & 0.02 & $-0.01$ & 0.10 & 0.12\\
\hline
$A_g$                           &SF   & 0.12 & 0.19  & 0.08 &   0.12 & $0.06$ & 0.30 & $-0.05$ & 0.06 & $0.21$ & 0.35 & 0.37 \\
                                     &Q     & 0.08 & 0.10 & 0.09 &   0.09 &  $-0.03$ & 0.19 & $-0.03$ & 0.12 & $0.10$ & 0.22 & 0.23\\
\hline                                          
\end{tabular}
\end{center}
\label{Tab:compare_old_new}
\tablefoot{We consider the effect of assumptions on the following ingredients: i) stellar library and evolutionary tracks in SPS models, done comparing CB19 with CB16 (evolutionary tracks) and CB16 with BC03 (stellar library); ii) parameters of star formation history (sandage - exponential); iii) metallicity evolution versus constant metallicity (var Z - fix Z); iv) overall difference between new and old model library. For each ingredient we report the mean and rms scatter of the differences in the derived physical parameters for quiescent and star-forming galaxies separately in the \silver~sample. These are compared to the average parameter uncertainty in our default fit (last column).}
\end{table}

\begin{figure}
\centering
\includegraphics[width=\textwidth]{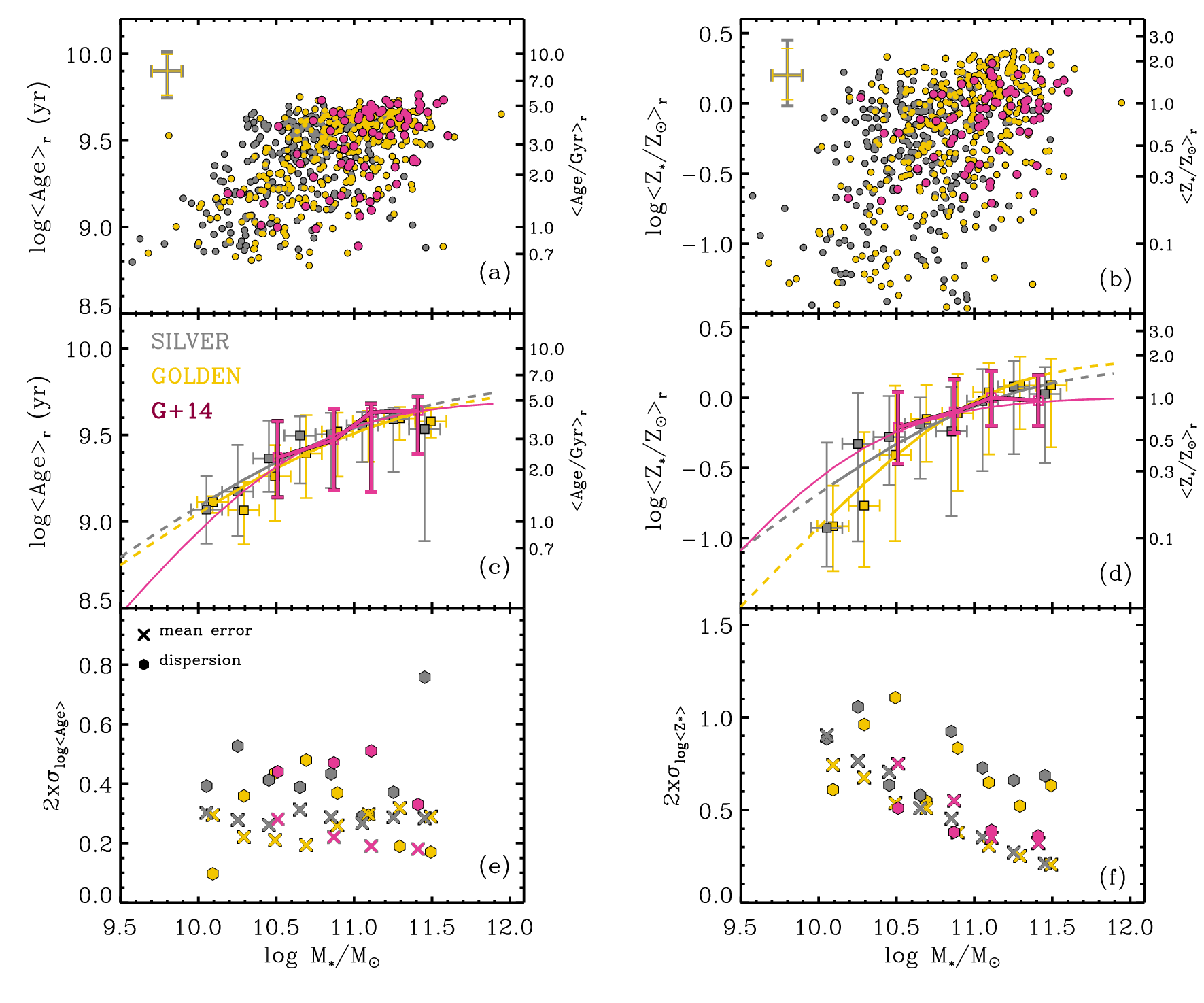}
\caption{The age-stellar mass and stellar metallicity-stellar mass distribution of LEGA-C galaxies (\silver~and \golden~samples) as in Fig.~\ref{fig:scaling_legac}, but using parameter estimates obtained with the modeling assumptions as in \cite{gallazzi14}. These are compared with the dataset and scaling relations fit obtained in \cite{gallazzi14} (magenta points and curves). The errorbars on the median points indicate the 84th-16th interpercentile of the distributions. The median ages and metallicities as a function of stellar mass are here fit adopting the functional form used in \cite{gallazzi14} (their Eq.1) for a direct comparison.}
\label{Fig:scaling_legac_extpb10}
\end{figure}

In Fig.\ref{Fig:scaling_legac_extpb10} we plot the age-mass and metallicity-mass relations obtained for the LEGA-C sample using the same model library and modeling assumptions as in our previous works, to make a direct comparison in particular with the results obtained in \cite{gallazzi14} with a more limited dataset in terms of number of galaxies and spectral quality. This plot shows that, under consistent assumptions, the LEGA-C dataset agrees with the results obtained in \cite{gallazzi14} at similar redshift. At the same time, it highlights the improved sampling of the age and stellar metallicity scaling relations thanks to the larger statistics in LEGA-C and the higher spectral quality which allow a better coverage of the lower mass and younger galaxies. This is essential to robustly characterize the scaling relations across the transition between quiescent and star-forming galaxies and down to masses $<10^{10.5}M_\odot$. The LEGA-C sample also reveals a larger dispersion in metallicity with respect to the mean uncertainties that could not be detected in G14.

\section{Catalog of stellar population parameters and comparison of estimates from different methods}\label{Appendix:compare_parameters}
We release a catalog of the stellar population parameters estimated in this work with {\tt BaStA}. While we select a robust sample for the analysis of the age and stellar metallicity scaling relations, parameter estimates are provided for the whole LEGA-C DR3. Parameters include: stellar mass, r-band-light-weighted and mass-weighted mean ages and stellar metallicities, g-band dust attenuation. We provide information about the indices used in each fit, as well as flags to select \silver~and \golden~samples. Estimates from {\tt BaStA} are based on duplicate-combined indices, when available.
The catalog also provides stellar populations and SFH parameters obtained from complementary spectral fitting performed within the LEGA-C team, specifically using: i) {\tt Prospector} as described in \cite{Nersesian25}, which adopts Simple Stellar Populations constructed from FSPS with MILES stellar
library and MIST isochrones, non-parametric SFHs and constant metallicity along the SFH (see their Table 1);  and ii) {\tt BAGPIPES} as described in \cite{kaushal24}  which adopts the 2016 version of \cite{BC03} models with MILES stellar library, double power-law SFH and constant metallicity (see their Table 1). 
These estimates are obtained from the individual spectra, in combination to photometry. Parameters from these other methods include stellar metallicity (constant along the SFH), light- and mass-weighted ages, stellar mass, V-band dust attenuation and current SFR.

\begin{figure*}
\centering
\includegraphics[width=\hsize]{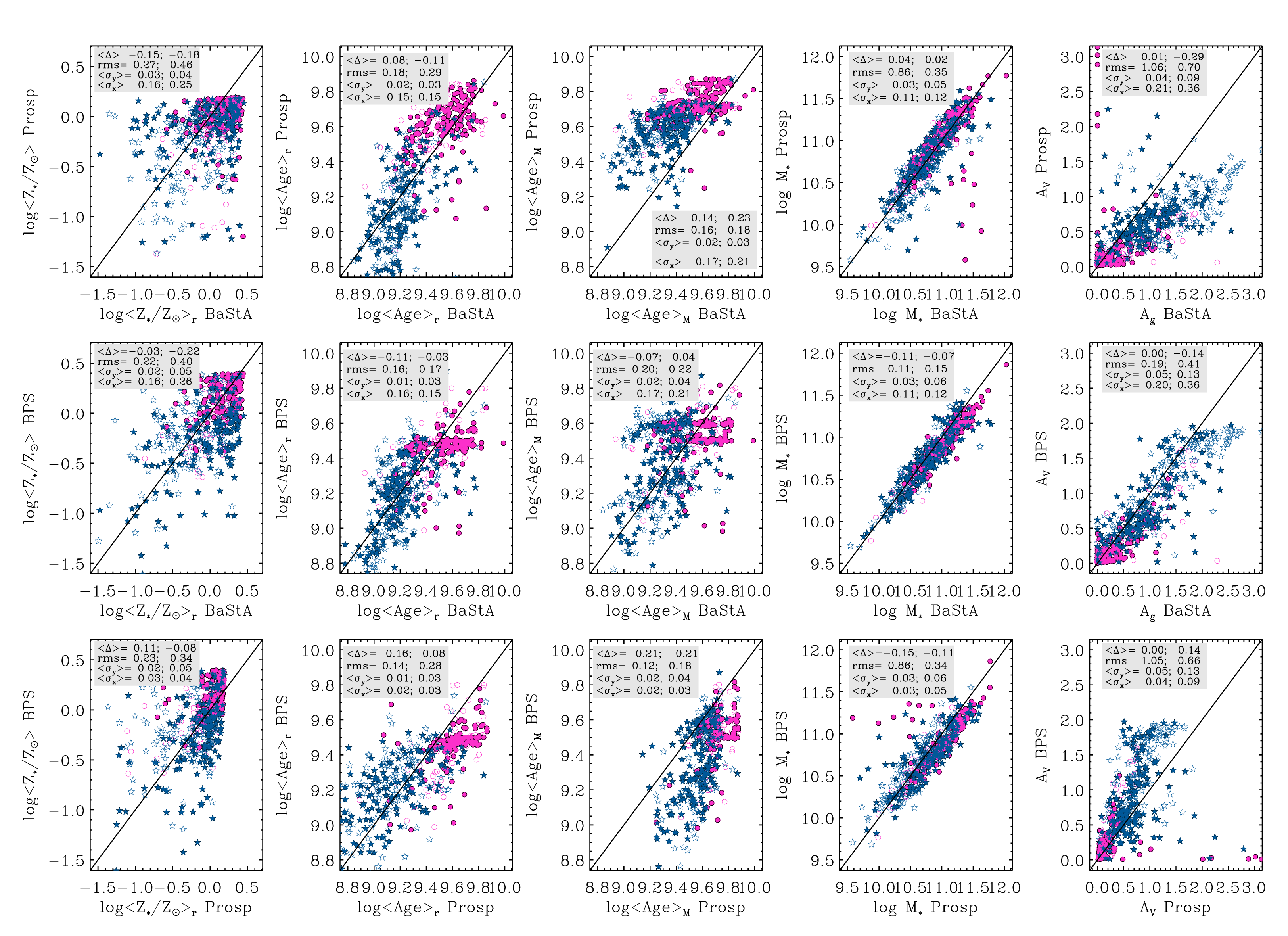}
\caption{Comparison between parameter estimates obtained with our default {\tt BaStA} code and those obtained with full-spectral fitting codes: {\tt Prospector} as in \cite{Nersesian25}, and {\tt Bagpipes} as in \cite{kaushal24}. We consider only \silver~and \golden~galaxies (empty and filled symbols), distinguished into quiescent (magenta circles) and star-forming (blue stars). In each panel we report the mean difference between the two estimates and its scatter, as well as the median uncertainty from each method, for quiescent and star-forming (left/right). The 1:1 line is also shown for guidance. Stellar metallicities have all been reported to a common scale of $Z_\odot=0.02$.}
\label{Fig:compare_methods}
\end{figure*}

In order to quantify potential systematic uncertainties originating from different spectral fitting methods, we compare the results obtained with {\tt BaStA} and with the SFH library adopted in this work with those obtained in \cite{Nersesian25} and in \cite{kaushal24}. 
Figure~\ref{Fig:compare_methods} shows the 1:1 comparison between the different estimates of the main physical parameters: light-weighted stellar metallicity, light-weighted and mass-weighted age, stellar mass and dust attenuation. For this comparison, all stellar metallicities have been reported to a common solar scale of $Z_\odot=0.02$. We show all \silver~galaxies but rely only on the \golden~sample for statistics, i.e. those galaxies for which we can most reliably constrain ages and stellar metallicities given the wavelength coverage and S/N of their spectra. We further distinguish galaxies into quiescent (magenta circles) and star-forming (blue stars) to highlight any systematics in the two populations. In Table~\ref{Tab:compare_methods} we summarize the differences between the three results on the physical parameters. We report the mean and the r.m.s. of the difference between each pair and the median uncertainty on the parameters from each method. In all cases, the reference estimated value is the median of the PDF, while the uncertainty is given by half of the $84^{th}-16^{th}$ percentile width.
This comparison incorporates differences associated to the observational diagnostics used (indices+photometry as opposed to full spectrum and photometry), the prior assumptions on SFH, metallicity and dust, and the base SPS models adopted.

We notice that stellar metallicity is the parameter on which there is poorer agreement. Quiescent galaxies span the high-metallicity end in all cases, while we see a larger scatter in the stellar metallicity estimates of star-forming galaxies, in accordance with their larger uncertainties. Notice that the models adopted in this work extend over a larger range in metallicity that those used with {\tt Bagpipes} and {\tt Prospector}. Moreover while the metallicities estimated with {\tt BaStA} are light-weighted averages,  {\tt Bagpipes} and {\tt Prospector} assume a constant stellar metallicity along the SFH. The scatter in the different estimates is $\sim0.2$~dex and $\sim0.4$~dex for quiescent and star-forming galaxies, respectively. This is significantly larger than the nominal uncertainties estimated with {\tt Bagpipes} and {\tt Prospector}, but closer to the uncertainty from {\tt BaStA} ($0.16$~dex and $0.25$~dex).  

The luminosity-weighted ages of quiescent galaxies from {\tt BaStA} and from {\tt Prospector} correlate with each other but with a systematic offset of $0.08$~dex, in the sense of older ages from {\tt Prospector}. The comparison with {\tt Bagpipes} is hampered by the fact that the majority of \golden~quiescent galaxies are assigned a light-weighted age that cluster around $\log(Age/yr)\sim9.5$, a feature noticed also in \cite{kaushal24}. The age estimates for star-forming galaxies are in better agreement between the three methods considered. The overall scatter in the age differences is comparable to the combined uncertainties for {\tt BaStA} and {\tt Prospector} or {\tt Bagpipes}. Likewise for metallicity, the uncertainties from {\tt Prospector} or {\tt Bagpipes} however underestimate the scatter between the two age estimates.

The comparison of mass-weighted ages shows in general larger scatter, as a consequence of the typically larger uncertainties and the stronger dependence on the assumed SFHs. We notice that {\tt Prospector} mass-weighted ages, that rely on non-parametric SFHs, are significantly older than those from {\tt BaStA} or {\tt Bagpipes} \citep[see also the discussion in][]{kaushal24}. 
We also find that mass-weighted ages compare to light-weighted ages in a similar way in the three methods considered (see Sec.~\ref{sec:light_vs_mass}). We find quantitatively more similar trends with {\tt BAGPIPES}, while the mass-weighted ages from {\tt Prospector} are systematically and significantly older, deviating more and more from the light-weighted ages for younger galaxies (see discussion in \cite{kaushal24}).

Stellar mass estimates correlate well with each other for both quiescent and star-forming galaxies, although {\tt Bagpipes} masses of quiescent galaxies tend to be smaller than both {\tt BaStA} and {\tt Prospector}. The scatter is typically comparable or slightly larger than the uncertainties estimated from {\tt BaStA}. We notice a large scatter in the comparison with {\tt Prospector} driven by few galaxies for which significantly low masses are estimated. 

Finally, the three methods agree in assigning small dust attenuation for quiescent galaxies below 0.5~mag, but with a tail to larger values for {\tt BaStA} and {\tt Bagpipes}. Star-forming galaxies span a larger range in dust attenuation, reaching higher values for {\tt BaStA} and {\tt Bagpipes}, which agree well with each other with a scatter consistent with the uncertainties.

It is worth noticing that in general the formal uncertainties associated to both {\tt Bagpipes} and {\tt Prospector} are significantly smaller than those estimated with {\tt BaStA} and than the scatter in the parameter estimates. The uncertainties quantified with {\tt BaStA} typically account for more than 60\% of the scatter between estimates from different methods, compared to $\sim10$\% from {\tt Prospector} or {\tt Bagpipes}. We thus consider
uncertainties from {\tt BaStA}, which we have checked to be well calibrated and consistent with the scatter between input and retrieved parameters from mock spectra \citep{Rossi25}, to be more representative of the underlying degeneracies.

\begin{figure}
\centering
\includegraphics[width=0.5\textwidth]{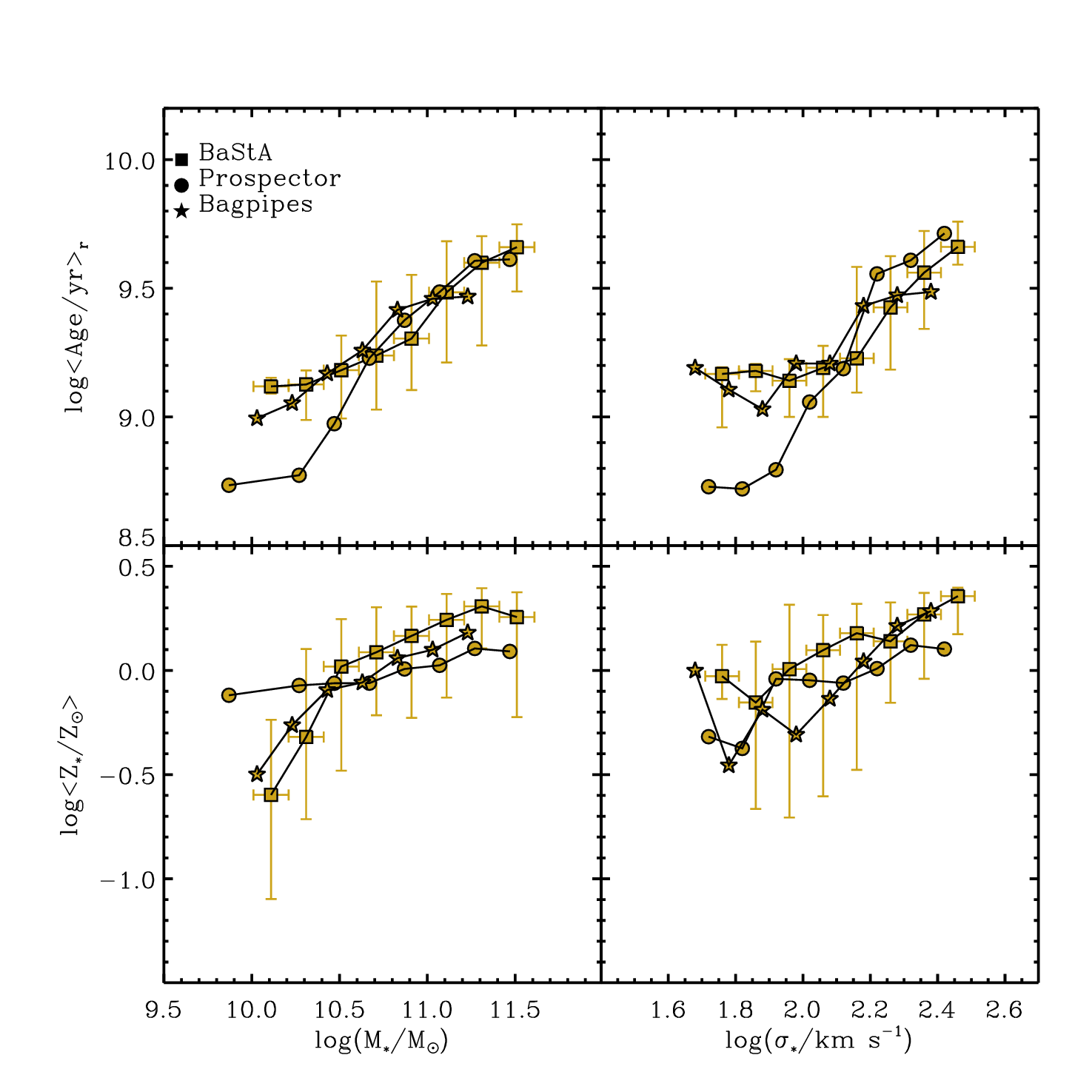}
\caption{Median age--mass and metallicty--mass relations for the \golden~sample, as obtained in this work (squares) compared to those obtained with {\tt Prospector} (circles) or {\tt Bagpipes} (stars) parameter estimates. Ages are light-weighted average in all cases, stellar metallicities are light-weighted in {\tt BaStA} and constant along the SFH in the other cases. For readability, the dispersion in the distribution, indicated by the errorbars, is shown only for {\tt BaStA} estimates.}
\label{Fig:compare_relations}
\end{figure}

One may wonder how much the scatter between these different estimates on an individual galaxy basis impact on the median scaling relations. Fig.~\ref{Fig:compare_relations} compares the median trends of age and metallicity versus stellar mass and velocity dispersion obtained with the three different estimates. We find the median relations, for the \golden~sample, to agree significantly better than the 1:1 comparison would suggest. In particular, the median light-weighted age as a function of stellar mass or velocity dispersion follows a similar sigmoidal behavior, with a young and an old regime, with all three codes. {\tt Prospector} however results in a steeper decrease in age at lower masses or velocity dispersion. {\tt BaStA} and {\tt Bagpipes} deliver consistent stellar metallicity scaling relations, while {\tt Prospector} results in a flatter relation. We stress that these differences include differences in spectral modeling as well as in observational diagnostics. In particular, our approach with {\tt BaStA} gives more weight to metal-sensitive features than full spectral fitting approaches do.

\begin{figure}
\centering
\includegraphics[width=0.5\textwidth]{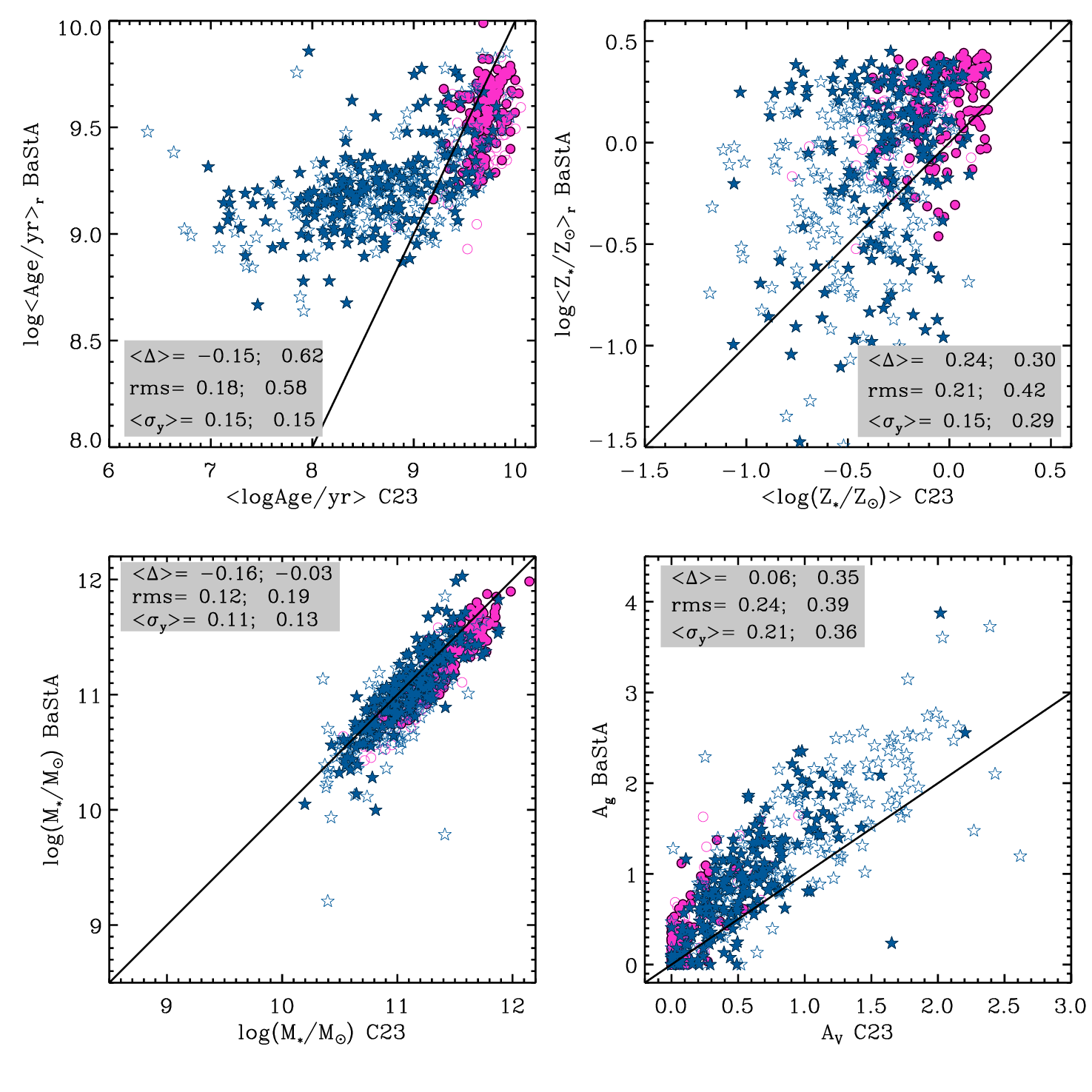}
\caption{Comparison between parameter estimates obtained in this work with those obtained with {\tt pPXF} and the FSPS population synthesis code published in \cite{Cappellari23}. We consider only galaxies in the \silver/\golden~samples (empty/filled symbols) distinguished into quiescent and star-forming (magenta circles and blue stars, respectively). The mean difference between {\tt BaStA} and {\tt pPXF} estimates, the rms scatter and the mean uncertainties on {\tt BaStA} estimates (those from {\tt pPXF} are not provided) are reported in each panel for quiescent and star-forming (left/right). Stellar metallicities have been reported to a common $Z_\sun=0.02$ scale, and our stellar masses have been rescaled up by 1.75 to match the Salpeter IMF used by \cite{Cappellari23}.}
\label{Fig:compare_cappellari}
\end{figure}

For completeness, in Fig.~\ref{Fig:compare_cappellari} we also show the comparison between our estimates and those from the other published catalog of stellar population parameters for the whole LEGA-C DR3 by \cite{Cappellari23} based on full spectral fitting with {\tt pPXF} \citep[see][for a similar comparison with {\tt Prospector} results]{Nersesian25}. We notice that {\tt pPXF} retrieves systematically older ages for the quiescent galaxies (exceeding the Universe' age at the LEGA-C redshift) and significantly younger ages for the star-forming galaxies. Stellar metallicities are systematically lower, partly because of the smaller range in the adopted SPS models, and show a larger separation between quiescent and star-forming galaxies than our estimates. Notice that for both age and metallicity \cite{Cappellari23} adopts a different definition, computing the average log age and log metallicity rather than averaging the linear quantities. This gives significanly more weight to lower ages and metallicities and it explains to a large extent the disagreement. While stellar masses correlate well, there is a systematic offset especially for quiescent galaxies. Dust attenuations are also systematically lower than ours, especially for star-forming galaxies.

\begin{table}
\caption{Comparison of physical parameter estimates obtained with our default {\tt BaStA} run with those obtained with {\tt Bagpipes} \citep{kaushal24} and {\tt Prospector} \citep{Nersesian25}. }
\begin{center}
\begin{tabular}{|l|cc|cc|cc|c|c|c|}
\hline
 & \multicolumn{2}{c|} {Prospector - BaStA} & \multicolumn{2}{c|} {Bagpipes - BaStA} & \multicolumn{2}{c|} {Bagpipes - Prospector} & & & \\
Parameter & $<\Delta>$ & rms  & $<\Delta>$ & rms & $<\Delta>$ & rms  & $\rm <\sigma>_{BaStA}$ & $\rm <\sigma>_{Prosp}$ & $\rm <\sigma>_{BPS}$ \\
\hline
\multicolumn{10}{|c|}{Quiescent} \\
\hline
$\log<Z_\ast/Z_\odot> $  &  $-0.15$ & 0.27 & $-0.03$ & 0.22 & 0.11 & 0.23 & 0.16 & 0.03 & 0.02 \\  
$\log<Age_r/yr> $        &  0.08  & 0.18 & $-0.11$ & 0.16 & $-0.16$& 0.14 & 0.15 & 0.02 & 0.01 \\
$\log<Age_M/yr> $        &  0.14  & 0.16 & $-0.07$ & 0.20 & $-0.21$& 0.12 & 0.17 & 0.02 & 0.02 \\
$\log(M_\ast/M_\odot) $  &  0.04  & 0.86 & $-0.11$ & 0.11 & $-0.15$& 0.86 & 0.11 & 0.03 & 0.03 \\
$A_V$                    &  $0.01$  & 1.06 &  0.0 & 0.19 &  0.00& 1.05 & 0.21 & 0.04 & 0.05 \\
\hline
\multicolumn{10}{|c|}{Star-forming} \\
\hline
$\log<Z_\ast/Z_\odot> $  & $-0.18$ & 0.46 & $-0.22$  & 0.40 & $-0.08$  & 0.34 & 0.25 & 0.04 & 0.05 \\  
$\log<Age_r/yr> $        &  $-0.11$ & 0.29 & $-0.03$  & 0.17 &  0.08  & 0.28 & 0.15 & 0.03 & 0.03 \\
$\log<Age_M/yr> $        &  0.23 & 0.18 &  0.04  & 0.22 & $-0.21$  & 0.18 & 0.21 & 0.03 & 0.04 \\
$\log(M_\ast/M_\odot) $  &  0.02 & 0.35 & $-0.07$  & 0.15 & $-0.11$  & 0.34 & 0.12 & 0.05 & 0.06 \\
$A_V$                    & $-0.29$ & 0.70 & $-0.14$  & 0.41 &  0.14  & 0.66 & 0.36 & 0.09 & 0.13 \\
\hline
  
\end{tabular}
\end{center}
\label{Tab:compare_methods}
\tablefoot{We consider light-weighted stellar metallicities, light-weighted and mass-weighted ages, stellar masses and dust attenuations. We report the mean difference, the scatter as well as the mean uncertainty for each estimate. The statistics are based only on \golden~galaxies, separately for quiescent and star-forming.}
\end{table}%


\end{appendix}

\end{document}